\documentclass[MSNbibl,nameyear,dvips]{arxstspdf}
\usepackage{flushend}
\usepackage{stfloats}
\usepackage{graphicx}

\volume{28}
\issue{4}
\pubyear{2013}
\firstpage{586}
\lastpage{615}
\doi{10.1214/13-STS451} 

\makeatletter
\setattribute{abstract}    {skip} {20\p@}
\setattribute{keyword}     {skip} {8\p@}

\setattribute{frontmatter} {skip} {6\p@ plus 3\p@ minus 3\p@}

  \fnbelowfloat
\renewcommand{\citep}[1]{(\citeauthor{#1}, \citeyear{#1})}
\newcommand{\eqref}[1]{(\ref{#1})}
\makeatother

\begin{document}
\begin{frontmatter}

\title{Wildfire Prediction to Inform Fire Management: Statistical
Science Challenges}
\runtitle{Wildfire Prediction for Fire Management}
\pdftitle{Wildfire Prediction to Inform Fire Management: Statistical Science Challenges}

\begin{aug}
\author[a]{\fnms{S.~W.} \snm{Taylor}\corref{}\ead[label=e1]{staylor@nrcan.gc.ca}},
\author[b]{\fnms{Douglas~G.} \snm{Woolford}\ead[label=e2]{dwoolford@wlu.ca}},
\author[c]{\fnms{C.~B.} \snm{Dean}\ead[label=e3]{cbdean@uwo.ca}}
\and
\author[d]{\fnms{David~L.} \snm{Martell}\ead[label=e4]{david.martell@utoronto.ca}}
\runauthor{Taylor, Woolford, Dean and Martell}

\affiliation{Natural Resources Canada/Simon Fraser University,
Wilfrid Laurier University, University of Western Ontario and
University of Toronto}

\address[a]{S.~W. Taylor is Research Scientist,
Natural Resources Canada, Pacific Forestry Centre, 506 West Burnside
Rd., Victoria, BC, Canada V8Z 1M5 \printead{e1}.}

\address[b]{Douglas~G. Woolford is Assistant Professor,
Wilfrid Laurier University, Department of Mathematics, Waterloo, ON, Canada
N2L 3C5 \printead{e2}.}

\address[c]{C.~B. Dean is Professor,
University of Western Ontario, London, ON, Canada, N6A 3K7 \printead{e3}.}

\address[d]{David~L. Martell is Professor,
University of Toronto, Faculty of Forestry, Toronto, ON, Canada, M5S
3B3 \printead{e4}.}

\end{aug}


\begin{abstract}
Wildfire is an important system process of the earth that occurs across
a wide range of spatial and temporal scales. A variety of methods have
been used to predict wildfire phenomena during the past century to
better our understanding of fire processes and to inform fire and land
management decision-making. Statistical methods have an important role
in wildfire prediction due to the inherent stochastic nature of fire
phenomena at all scales.

Predictive models have exploited several sources of data describing
fire phenomena. Experimental data are scarce; observational data are
dominated by statistics compiled by government fire management
agencies, primarily for administrative purposes and increasingly from
remote sensing observations. Fires are rare events at many scales. The
data describing fire phenomena can be zero-heavy and nonstationary over
both space and time. Users of fire modeling methodologies are mainly
fire management agencies often working under great time constraints,
thus, complex models have to be efficiently estimated.

We focus on providing an understanding of some of the information
needed for fire management decision-making and of the challenges
involved in predicting fire occurrence, growth and frequency at
regional, national and global scales.
\end{abstract}

%
\pdfkeywords{Environmetrics, forest fire, prediction, review, wildland fire}

\end{frontmatter}

\section{Introduction}\label{sec1}

``Predicting the behavior of wildland fires---among nature's
most potent forces---can save lives, money, and natural resources.''

{\hfill{Frank Albini (\citeyear{Al84})}}\vspace*{6pt}

Wildfires have likely occurred on the earth since the appearance of
terrestrial vegetation in the Silurian era, 420 million years B.P.
\citep{BowmanEtAl-2009-Science}, and are an important ecosystem process
on all continents except Antarctica, influencing the composition and
structure of plant and animal communities, as well as carbon and other
biogeochemical cycles. Emissions of $\mbox{CO}_2$, other trace gasses
and particulates from biomass burning contribute to annual and
inter-annual variation in atmospheric chemistry \citep
{AndreaeEtAl-2001-Biogeochem}, including the formation of cloud
condensation nuclei that influence global radiation and precipitation
budgets, and in the case of black carbon, accelerate the melting of ice
and snow \citep{BondEtAl-2013-GeophysicalResearch}. Wildfires also have
significant social and economic impacts, sometimes resulting in the
evacuation of communities, fatalities, smoke impacts on human health
\citep{FinlayEtAl-2012-PLoS}, property loss and the destruction of
forest resources.

Instrumental records suggest that the average\break global temperature
increased 0.8$^\circ\mbox{C}$ in the last century \citep
{HansenEtAl-2006-Proceedings}. However, because  global annual burned
area data have only been available for about the past 15 years from
satellite observations,\footnote{Using MODIS satellite data, \citet{Giglio-2009-Biogeosci}
estimated that the global annual burned area was between 3.31 and 4.31
million km$^2$ during 1997--2008 (Figure~\ref{Subfig:SuppFig1d}).} it has only been possible to examine the
effects of changes in climate during the past century on fire activity
in a few regions with long-term administrative records. For example,
area burned increased significantly in Canada as a whole, the province
of Ontario, Canada, and in northwestern Ontario in the latter compared
to the earlier half of the period 1918--2000 \citep
{PodurEtAl-2002-CJFR}; and, the fire season has been lengthening in the
provinces of Alberta and Ontario, Canada (\mbox{Albert-Green}
et~al., \citeyear{AlbertGreenEtAl-2013-CJFR}). Increases in the area burned in the
western US during the 1970--2005 period were associated with earlier
spring snowmelt \citep{WesterlingEtAl-2006-Science}. However, at a
regional scale, decreases in area burned in many ecological zones in
the province of British Columbia, Canada, were associated with
increases in precipitation during 1920--2000 (Meyn et~al., \citeyear{MeynEtAl-2010-IJWF}).
Climate warming scenarios of 2.5--3.5$^\circ\mbox{C}$ over the next
century are expected to result in increases in global wildfire activity
\citep{FlanniganEtAl-2009-IJWF}, but such changes are expected to vary
by region due to projected changes in the amount and distribution of
precipitation \citep{KrawchukEtAl-2009-PlosOne}.


Since wildfire management is likely to become increasingly challenging
under a changing climate, better predictive tools will be needed. We
believe that statistical science can make important contributions to
improving wildfire prediction at local to global scales.

\subsection{Prediction in Wildfire Management}\label{sec1.1}

Most wildfire management organizations in North America and elsewhere
have developed the capacity to respond rapidly to wildfires that
threaten communities and other values with highly-mobile fire
management resources (fire fighters, equipment and aircraft) in order
to contain and extinguish fires while they are small. Minimizing the
time intervals between when a fire is ignited, detected and actioned is
key to successful initial attack. While this approach is effective for
most fires, a small number (typically less than 5\% in Canada) escape
initial attack and continue to spread, requiring additional resources
as fire size and complexity increase.\footnote{The Incident Command System (ICS) is used by many wildfire
management organizations. It provides a flexible organizational
structure that can be expanded depending on the complexity of the
incident \citep{BigleyEtAl-2001-ManagementJournal}. The five incident
complexity classes (Type 5--1) recognized in ICS are associated with
an increasing need for resources for longer periods of time. For
example, a Type 5 wildfire that is less than a few hectares in size may
be controlled by 3--5 fire fighters, which may be supported by
helicopters or airtankers for up to one or two days, while a larger
Type~1 incident of thousands of hectares in size that threatens a
community will require a much more significant response, including a
specialized incident management team (IMT) and hundreds, perhaps even
thousands of firefighters and other resources that can be sustained for
many days to weeks.}

The number, severity and sizes of fires vary substantially within and
between regions, as well as within and between years, due in part to
variation in weather, climate, other environmental conditions and
demographic and human behavioral factors.\break  Much early fire research in
North America focused on the development of fire danger rating systems
that were designed to capture the cumulative effects of weather in
numerical measures of daily fire potential (\cite{TaylorEtAl-2006-IJWF};
\cite{HardyEtAl-2007-IJWF}). The fire danger systems developed and used in
Australia, Canada and the United States, for example, are based
primarily on empirical models of weather effects on the moisture
content and flammability of various organic layers (e.g., the moss
layers and dead pine needles on the forest floor) 
(Fujioka et~al., \citeyear{FujiokaEtAl-2008-EnvSci}). Fire danger measures are connected to fire
activity in many environments \citep{ViegasEtAl-1999-IJWF}. Thus, when
fire occurrence and fire behavior models were later developed, they
often included fire danger measures as covariates \citep
{Wotton-2009-EnvAndEcoStats}. Computer-based fire management
information systems have subsequently been developed to collect,
process, interpolate and distribute weather, fire danger measures and
model predictions throughout fire organizations, many in almost
real-time (\cite{DoanEtAl-1974-ForestryChron}; \cite{LeeEtAl-2002-CompAndElec}).

One important feature of many fire regimes is the sharp peaks in fire
activity that are often associated with high pressure systems,
lightning storms or other severe synoptic-scale weather events.
Although fire management organizations collaborate and often share
resources on regional, national and even continental scales, they are
not always able to respond fully to some peaks in fire activity, which
subsequently place significant stress on the system and increase the
likelihood of elevated costs and losses. In addition to limits on
resources, fire suppression effectiveness varies with fire size and
inten\-sity---direct fire suppression methods cannot be used when the
fire intensity exceeds safe working conditions for ground crews, or
when high winds or smoke ground aircraft or render their drops
ineffective. Thus, there is increasing interest in mitigating the risk
of extreme fire behavior by manipulating fuel conditions (vegetation),
in reducing the vulnerability of communities, and in choosing to
monitor rather than fully suppress some fires that pose little or no
threat to public safety, property or forest resources.

Fire activity varies substantially, and often rapidly, from local to
national scales; spatio-temporal variability is one of the main
challenges in wildfire management. Because resources are limited, both
for mitigating and responding to wildfire risks, predictive models are
needed to support planning and deci\-sion-making (\cite
{AndrewsEtAl-2007-SciAmer}; \cite{PreislerEtAl-2013-EncOfEnviron}). \citet
{Martell-1982-CJFR} described many of the strategic, tactical and
operational deci\-sion-making problems faced by fire managers, and of
early efforts to bring operations research methods to bear on them.
These include:

\begin{longlist}[1.]
\item[1.] Strategic decisions about the long-term requirements for
resources (e.g., number and type of airtankers) in large regions, such
as states or provinces, and where they should be home-based, depending
on the expected number, variation and distribution of incidents.
\item[2.] At the tactical level, the number and size of fires that are
expected to be ignited, detected and reported over shorter periods of
days to weeks influences decisions concerning the state of preparedness
or organizational readiness, the allocation of resources within a
region, and the acquisition (or release) of additional resources from
outside the region through mutual aid resource sharing agreements. The
expected daily fire occurrence is important for prepositioning fire
crews and routing aircraft for fire detection. The expected growth of
individual fires over days or weeks informs decisions concerning the
evacuation of communities in the path of a fire or whether some fires
burning in remote areas can be simply monitored and allowed to burn
relatively free\-ly without threatening public safety, resources or
infrastructure.
\item[3.] Because conditions can change rapidly, operational decisions are
typically made over minutes and hours during a day. Airtankers and
other resources may be re-deployed and dispatched to fires as each day
progresses. The expected behavior and growth of an individual fire over
the daily burning period is important for planning the dispatch and
safe deployment of firefighters and other resources on fires.
\end{longlist}

\begin{figure}

\includegraphics{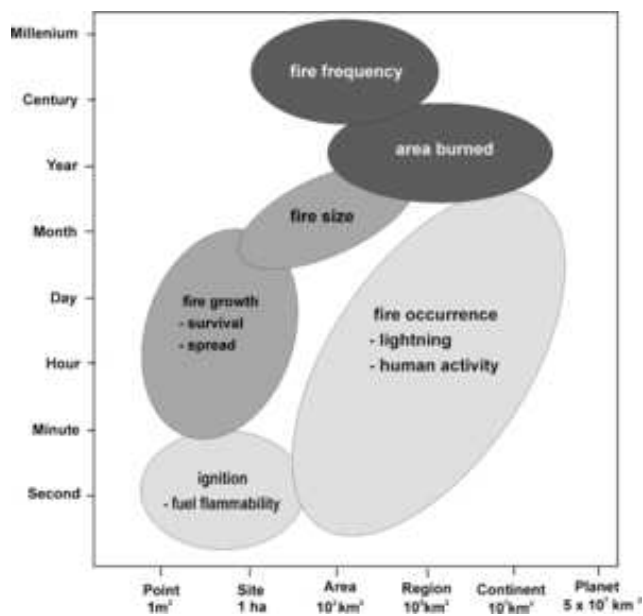}

\caption{Wildland fire risk elements are compounded over a range of
spatial and temporal scales. Reinterpreted from Simard (\citeyear{Simard-1991-IJWF}).}\label{Fig:Simard1991}
\end{figure}

In this paper we review some of the models that have been developed to
predict fire occurrence,\break  growth and frequency, and how they are linked
across multiple scales (Figure~\ref{Fig:Simard1991}). While there have
been important contributions from many regions, we have focused on the
North American fire literature because that is the region in which we
have carried out most of our fire-related research. Section~\ref{sec2} discusses
tools for ignition and fire occurrence prediction, with connections to
point processes and case--control methods. Section~\ref{sec3} discusses fire
spread/\break growth and fire size models. Section~\ref{sec4} reviews models for
estimating burned area and fire frequency. The \hyperref[app]{Appendix} provides an
overview of the sources---and limitations---of various types of
wildfire data that have been used in predictive models.

\begin{figure*}

\includegraphics{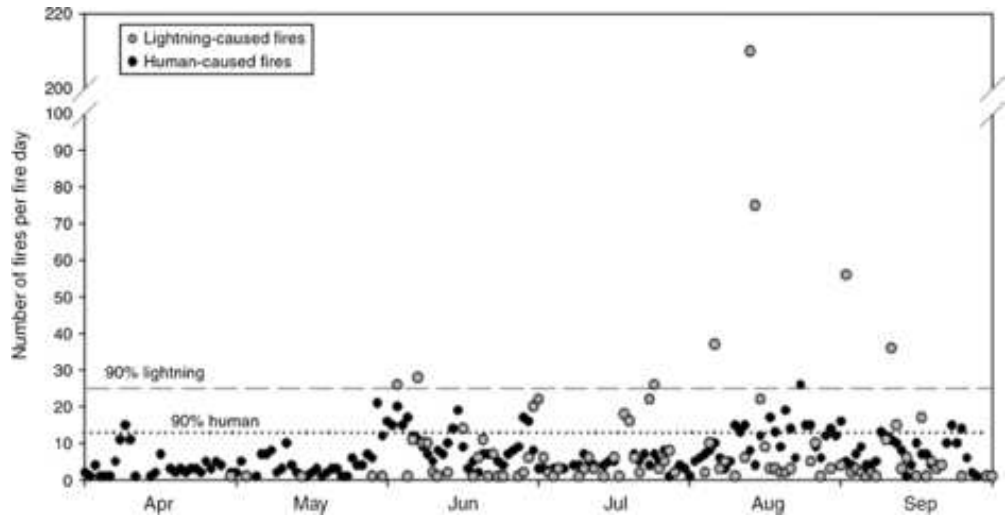}

\caption{The raw daily number of human- and lightning-caused fires in
British Columbia, Canada, during the 1986 fire season observed on fire
days, defined as a day during which at least one fire is observed. The 20-year
90th percentile thresholds are indicated (Magnussen and
Taylor, \citeyear{MagnussenEtAl-2012-IJWF}). Days with $>$90th percentile number of
lightning fire starts are difficult to predict but severely
challenge the ability of fire management agencies to respond quickly to
all fires.}\label{Fig:NumFiresBC}
\end{figure*}

Interspersed throughout this paper, and especially in the closing
section, are discussions of open challenging wildfire management
questions that we hope will be of interest, stimulating the development
of new tools for this critical area of science. We note that,
personally, our collaborative work with teams of statisticians, fire
scientists and fire managers has proven to be a rich and rewarding
platform for interdisciplinary research and training.

\section{Fire Occurrence}\label{sec2}

Wildland fires are ignited by both people and natural processes.
Natural fires are caused mainly\footnote{A small number of wildfires have also been ascribed to volcanic
activity \citep{Ainsworth-2009-PlantEcology} and meteorites [e.g., the
1908 Tunguska event in Siberia \citep{Svetsov-2002-3P}].} by cloud-to-ground lightning
strokes \citep{Anderson-2002-forestFireResearch} that ignite trees or
organic matter at the base of the tree they strike, while people-caused
fires occur when needle, leaf or grass litter is ignited. Anthropogenic
sources of ignition include machinery (sparks, friction and hot
surfaces), arcing from electrical transmission lines, sparks or
firebrands from escaped campfires, prescribed fires, agricultural and
land clearing fires, and arson. An ignition that leads to sustained
fire spread may be reported and record\-ed by a fire management agency
or, in some cases (e.g., in more northern regions of Canada), it is
detected by satellite-borne sensors. The locations, times and number of
forest fire ignitions appearing in historical fire records are
inherently random. In many cases such records contain truncated or
censored data: only fires that are reported to a fire management agency
appear in the records and in many cases the time of ignition is
estimated.\footnote{Most fire managers and researchers use the term ``occurrence'' to
refer to fires that are detected and reported, although the queueing
theory term ``arrivals'' is also sometimes used to distinguish detected
and reported from nondetected fires.} Fire ignition rates vary drastically over both time and
space and their relative frequency of occurrence depends on locally
observed covariates, including the intensity of the ignition process.
There is often greater variability in the daily number of lightning
than anthropogenic ignitions (Figure~\ref{Fig:NumFiresBC}). This is
because, when lightning storms occur, they can produce thousands of
lighting strikes and tens--hundreds of fire starts in a few hours.

\subsection{Probability of Ignition}\label{sec2.1}

Regardless of the initial source of ignition, if sufficient heat is
produced from combustion, adjacent particles (e.g., needle, leaf, grass
and twig litter, or other \mbox{organic} material) will also be heated to
their ignition temperature, resulting in sustained fire spread. The
probability of ignition is related mainly to the physical properties of
dead organic matter and its moisture content, which varies by day and
across all spatial scales. Regression methods have been employed to
quantify the probability of sustained ignition under varying
conditions. In some studies, samples of litter or sub-litter fuels
taken in the field are subjected to ignition experiments in the
laboratory (\cite{Frandsen-1997-CJFR}; \cite{PlucinskiEtAl-2008-IJWF}).
In other cases, ignition experiments are conducted directly in the
field. In their logistic regression based reanalysis of experimental
test fires in Canada, \citet{BeverlyEtAl-2007-IJWF} concluded that the
primary driver of sustained flaming ignition from firebrands is the
moisture content of fine fuels. The moisture in more heavily compacted
organic matter below these fine fuels along with relative humidity also
impacted the probability of sustained ignition for some fuel types.
Similar results have been observed in other regions of the world. An
analysis of experimental fires in Tasmanian grasslands, for example,
revealed that sustained ignition was strongly driven by the moisture
content of the dead fuel as well as the amount of dead fuel available
for combustion \citep{Leonard-2009-EnvManag}. Earlier analyses, using
logistic regression and classification trees, for data on Tasmanian
grassland fires also revealed and quantified the interaction between
wind speed and dead fuel moisture: wetter fuels require a higher wind
speed to sustain ignition, otherwise they are more likely to
self-extinguish (Marsden-Smedley, Catchpole and Pyrke, \citeyear{MarsdenSmedley-2001-IJWF}).

\subsection{Fire Occurrence Prediction}\label{sec2.2}

Early fire occurrence prediction related the number of fires per day to
fire danger indices, usually for a single spatial unit or
administrative region. Many models have subsequently been developed using a
variety of modeling approaches and covariates and for a variety of spatial and temporal scales.
Fire occurrence models typically include
variables believed to influence ignition potential (fuels, fuel
moisture, ignition source) in a particular environment, tempered with
practical considerations regarding data availability. In addition to
weather, fuel moisture and fire danger indices, other explanatory
variables have included historic spatial and seasonal trends,
vegetation type, the number and attributes of lightning strikes,
population and road density.

Fire occurrence models for large areas need to accommodate variation in
topographic, fuel, weather and fuel moisture conditions. Advances in
computing, communication and data collection from weather station
networks in near-real time \citep{LeeEtAl-2002-CompAndElec} have
permitted the implementation of sophisticated grid-based fire
occurrence models for larger and more variable geographic areas. In
these models, the weath\-er and fire danger index variables derived from
multiple weather stations are interpolated across the grid units based
on distance and elevation (\cite{KourtzEtAl-1991-ForestryCan}; \cite{ToddEtAl-1991-ForestryCan}). The advent of lightning location systems
\citep{KriderEtAl-1980-AmMetSoc} also facilitated lightning-caused fire
prediction \citep{KourtzEtAl-1991-ForestryCan}. Models of lightning
fire occurrence should have greater temporal and spatial specificity
(correct prediction) than human-caused fires because the ignition
process can be observed.

It is important to note that the probability of ignition differs from
the probability of fire occurrence in the sense that not all fires that
achieve sustained ignition may be detected: fire occurrence data is
left censored. However, fire occurrence prediction models are much more
common than models for the probability of ignition. \citet
{WoolfordEtAl-2011-EnvStats} provided a brief review of fire occurrence
prediction, which focused on the use of logistic generalized additive
models to approximate the covariate-dependent, inhomogeneous intensity
function of a point process model. There, they also discussed how the
response-based sampling used in some of these models is related to
case--control studies. We paraphrase and expand upon that discourse in
what follows; we also summarize a newly developed methodology for
monitoring for temporal trends in historical records on fire
occurrence, motivated by climate change concerns. For a recent and
concise review of fire risk and other forest fire models, see \citet
{PreislerEtAl-2013-EncOfEnviron}.

Given the stochastic nature of fire ignitions, a point-process with a
conditional intensity function is a natural modeling framework. The
first stochastic model for predicting the occurrence of fires appears
to have been developed by \citet{Bruce-1960-FireControlNotes}, who utilized a
negative binomial model that related counts to a fire danger rating
index. Subsequently, \citet{CunninghamEtAl-1973-CJFR} developed a
Poisson model for counts of fires whose nonspatial conditional
intensity function depended on fuel moisture, as measured by the
Canadian Fine Fuel Moisture Code (FFMC) \citep
{VanWagner-1987-CanForestServ}. The FFMC represents the moisture
content of litter fuels on the forest floor---for example, the higher
the FFMC, the drier the needle litter on the forest floor. Data from a
weather station near the center of a fire management unit in
northwestern Ontario were used to predict daily counts of fires within
that region.

In Ontario, Bernoulli processes have been used to model the risk of
forest fire occurrence since the late 1980s. For example, \citet
{MartellEtAl-1987-CJFR} constructed a set of logistic models for the
daily risk of people-caused fires in northern Ontario. These were
marginal models, without spatial or temporal components, fit to
individual ``subseasons'' that partitioned the fire season. Seasonal
trends were subsequently incorporated by \citet{MartellEtAl-1989-CJFR}
through periodic functions. The seasonality of fire occurrence is of
interest to fire management for planning purposes, although the strong
seasonal variation in Ontario's boreal is not universally observed in
other regions. Moreover, such seasonal trends are not spatially
homogeneous, as illustrated in the site-specific fire risk curves
presented by \citet{WoolfordEtAl-2009-Geomatica} who also explored for
spatial patterns using a singular-value decomposition approach,
somewhat analogous to regression on principal component scores.

Some modeling efforts have quantified ignition and occurrence risk,
such as the site-specific models for fire ignition and occurrence of
\citet{WottonEtAl-2005-CJFR}. Logistic methods have the advantage that
locally observed covariates can be related to each individual fire,
while Poisson-based models connect counts to averages of such
covariates over a larger region. Moreover, overdispersion is of concern
when Poisson-based methods are used to model counts. Overdispersion is
of less concern when logistic models are fit to binary data; however,
this is not true when temporal and/or spatial correlation needs to be
incorporated.

Relatively little work has been done to explore the use of
point-process methods for analyzing the occurrences of forest fires in
space--time. However, some recent advances in point-pattern software
hold promise in this regard [see \citet{Turner-2009-EnvEcoStats} for
an example]. Nonparametric tests for investigating the separability of
a spatio-temporal marked point process are described and compared in
\citeauthor{Schoenberg-2004-Biometrics}\break  (\citeyear{Schoenberg-2004-Biometrics}), where a Cramer--von Mises-type test
is de\-monstrated to be powerful at detecting gradual departures from
separability, while a residual test based on randomly rescaling the
process is powerful at detecting nonseparable clustering or inhibition
of the marks. An application to Los Angeles County wildfire data shows
that the separability hypotheses are invalidated largely due to
clustering of fires of similar sizes within periods of up to about 4
years. In more recent work, \citet{XuEtAl-2011-AnnalsAppStats} showed
that the Burning Index, produced by the US Fire Danger Rating System,
and commonly used in forecasting the hazard of wildfire activity, is
less effective at predicting wildfires in Los Angeles County than
simple point process models incorporating raw meteorological
information. Their point process models incorporate seasonal wildfire
trends, daily and lagged weather variables, and historical spatial burn
patterns. \citet{NicholsEtAl-2011-TimeSeries} developed a method for
summarizing repeated realizations of a space--time marked point process,
called prototyping, and applied this technique to databases of
wildfires in California to produce more precise summaries of patterns
in the spatio-temporal distribution of wildfires within each wildfire season.

The importance of Poisson processes in modeling the risk of wildfire
occurrence was described by \citet{BrillingerEtAl-2003-IMSlecture}, who
focused on the underlying spatio-temporal conditional intensity
function and described methods for approximating the corresponding
likelihood. They advocated partitioning the space--time domain into a
set of space--time voxels $(x, x + dx] \times(y, y + dy] \times(t, t +
dt]$, where $(x, y)$ are spatial location covariates and $t$ indexes
time. The spatio-temporal point process of interest, $N(x, y, t)$,
counts the number of fires in a voxel and has conditional intensity function
\[
\lambda(x, y, t) = \frac{\operatorname{Pr}\{ dN(x, y, t) = 1| H_t\}}{dx\,dy\,dt},
\]
where the $\sigma$-algebra $H_t$ denotes the history of $N(x,\break
y,  t)$ over $(0, t]$, which consists of the set of observed points in
space--time up to time $t$.

If the underlying intensity function depends on a parameter $\theta=
\theta(\mathbf{x})$, where $\mathbf{x}$ denotes a vector of
locally observed covariates, the log-likelihood of the process is
\begin{eqnarray*}
L(\theta) &=& \int_0^T \int
_x \int_y \operatorname{log}\bigl[
\lambda(x, y, t | \theta )\bigr]\,dN(x, y, t)\\
&&\hspace*{37pt}{} - \int_0^T
\int_x \int_y \operatorname{log}
\bigl[\lambda(x, y, t | \theta)\bigr]\,dx\,dy\,dt.
\end{eqnarray*}

\citet{BrillingerEtAl-2003-IMSlecture} listed three practical
approaches to approximation of this log-likelihood that could be used
for model fitting. (Note that although both terms in the above equation
cover large regions of both space and time, it is the second term which
is challenging to evaluate.) Their first approach outlined a method for
approximating the expected value of the log-likelihood. However, that
does not appear to be widely employed in forestry applications.
Instead, their two other approaches, related to binomial approximations
to the Poisson, are more commonly used. In such approximations, the
number of binomial trials may be very large especially if a set of
voxels on a very fine spatio-temporal scale, such as 1 km$^2 \times
\mbox{daily cells}$, is used. On this scale fires are very rare events
and only presence/absence is recorded. Then the underlying rate $\lambda
_{x,y,t} = \lambda(x,y,t|\theta)$ is approximately the Bernoulli
probability of observing a fire in that given space--time region,
leading to the Bernoulli approximation to the log-likelihood:
\begin{eqnarray*}
&&\sum_{x, y, t} N_{x, y, t}\operatorname{log}(
\lambda_{x, y, t})
\\
&&\quad{}+ \sum_{x,y,t} (1 -
N_{x,y,t}) \operatorname{log}(1 - \lambda_{x,y,t}).
\end{eqnarray*}

Therefore, a generalized linear model with the linear
predictor $\operatorname{logit}\{ \lambda[x, y, t|\theta(\mathbf{x})] \} =
\mathbf{x}\bolds{\beta}$, where $\mathbf{x}$ denotes a vector of
covariates and $\bolds{\beta}$ denotes the corresponding vector of
parameters, can be used to approximate the underlying process and, more
importantly, quantify the probability of fire occurrence as a function
of locally observed covariates. Generalized additive models (GAMs) have
been employed to incorporate potential nonlinear relationships for the
explanatory variables (\cite{PreislerEtAl-2004-IJWF}; \cite{PreislerEtAl-2007-AppMeteorAndClimat};
\cite{VilarEtAl-2010-IJWF}; \cite{WoolfordEtAl-2011-EnvStats}).
For example, periodic seasonal effects may
be incorporated into the linear predictor using locally weighted
regression or penalized spline smoothers \citep{wood-IntroToGAMs}. Thin
plate splines have also been used to add a spatial term as a surrogate
for unobservable human land use patterns or unobserved vegetation patterns.

The Bernoulli approximation of the likelihood function induces
computational difficulties since for any practical study, the
cardinality of the set of voxels explodes to such a large size that
model fitting is not computationally convenient/feasible.
Response-based stratified sampling schemes are employed to deal with
this issue: data from voxels where a fire is present are kept, but only
a random sample of the zero-fire voxels are retained for the analysis.

The response-based sampling of the voxel data is analogous to study
designs from logistic retrospective case--control studies. This induces
a deterministic offset of
$\operatorname{log}(1/\pi_{{st}})$ in the logistic GAM, where $\pi_{{st}}$ denotes
the inclusion probability for the observation at site $s$ at time $t$.
Note that the use of an offset in the linear predictor to account for
the response-based sampling only works when modeling on the logit scale
and not when other link functions, such as the probit or the
complementary log--log, are employed in a binomial GAM. \citet
{GarciaEtAl-1995-IJWF} appear to be the first to use response-based
sampling in a logistic model for fire occurrence. More recently, it has
been employed in logistic GAMs which incorporate temporal and spatial
effects (e.g., \citeauthor{BrillingerEtAl-2003-IMSlecture},
\citeyear{BrillingerEtAl-2003-IMSlecture,BrillingerEtAl-2006-Environmetrics};
\cite{PreislerEtAl-2004-IJWF}; \cite{VilarEtAl-2010-IJWF}; \cite{WoolfordEtAl-2011-EnvStats}).

Let $Y$ denote the fire occurrence indicator, $P(Y=1|\mathbf{x}) =
p_\mathbf{x}$, and assume $\operatorname{logit}(p_\mathbf{x}) = \alpha+ \mathbf
{x}\bolds{\beta}$, where $\mathbf{x}$ is a row vector of
covariates and $\bolds{\beta}$ is a column vector of parameters.
This logistic framework implies that the relative risk corresponding to
two voxels with corresponding explanatory variables $\mathbf{x}_1$ and
$\mathbf{x}_2$ is $\operatorname{exp}\{(\mathbf{x}_1- \mathbf{x}_2) \bolds
{\beta}\}$. Similar formulations hold for a logistic GAM because the
nonlinear relationships on the link scale are modeled as a linear
combination of basis functions. In that context, $\operatorname{exp}\{
f_m(x_{m1}) - f_m(x_{m2})\}$ is the associated change in risk when the
covariate in the $m$th additive nonlinear partial effect $f_m$ in a GAM
changes from $x_{m1}$ to $x_{m2}$. This framework is the same as a
prospective analysis in medical studies when whether or not an
individual will develop a disease is not known in advance. In contrast,
with a case--control study, subjects are selected based on their disease
status (here, fire or nonfire voxel is the analogy) and their exposure
or treatment (here, covariate vector) is determined retrospectively. In
this context, the covariate values are viewed as random. However, it
has been shown that inferences surrounding relative risk can be
obtained using the same logistic model as in the prospective study
\citep{BreslowEtAl-1978-Biometrics}. Letting $\delta$ denote an
indicator for whether or not an individual is sampled, the
corresponding inclusion probabilities can be stratified by response:
$\pi_1 = \operatorname{Pr}\{\delta= 1 | Y = 1\}$ and $\pi_0 = \operatorname{Pr}\{
\delta= 1 | Y = 0\}$. Usually a case ($Y=1$) is a rare event, relative
to the population size. In the fire study analogy, all cases are
included ($\pi_1$ is 1) and $\pi_0$ is usually fairly small. Through a
Bayes argument, it is straightforward to show that such
response-dependent sampling induces a deterministic offset into the
model. Specifically, the intercept changes by an additive factor of
$\operatorname{log}(\pi_1/\pi_0)$. Since the sampling probabilities depend only
on the observed disease (fire) status and not on covariates, the
covariate effects are identical to those from a prospective analysis.
The analyses of the fire occurrence data where all fire events are
retained for the analysis and only a sample of the nonfire events are
included is identical to the above case--control formulation.

\begin{figure}
\centering
\begin{tabular}{@{}c@{}}

\includegraphics{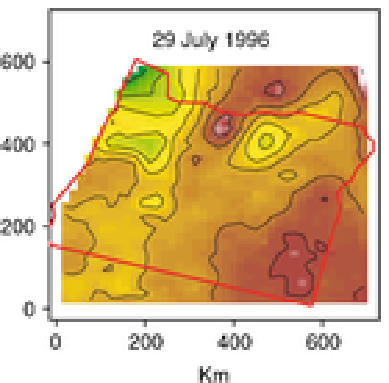}
\\
\footnotesize{(a)}\\[3pt]

\includegraphics{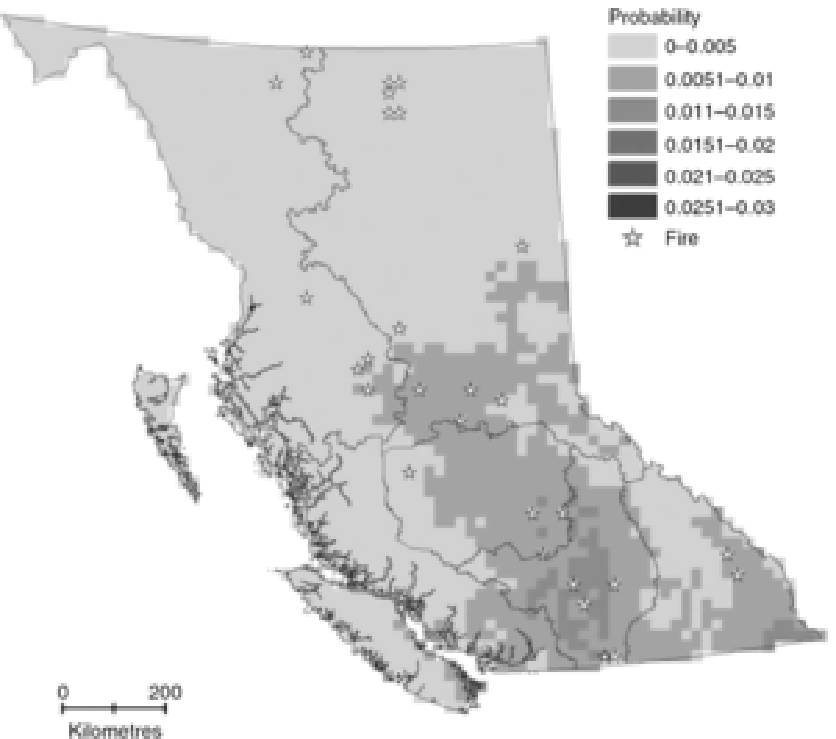}
\\
\footnotesize{(b)}
\end{tabular}
\caption{\textup{(a)} The probability of a large
fire given ignition in 1 km cells in Oregon on 29 July 1996 (Preisler
et~al., \citeyear
{PreislerEtAl-2004-IJWF}). \textup{(b)} The probability of more than
person-caused fire occurring in $\sim$9000 400~km$^2$ cells and
observed fires on 21 May 1985 in British Columbia, Canada (Magnussen and
Taylor, \citeyear
{MagnussenEtAlb-2012-IJWF}).} \label{Fig3}\vspace*{-3pt}
\end{figure}

Over the past decade, there have been multiple studies using
response-specific sampling in logistic GAMs for fire occurrence. \citet
{BrillingerEtAl-2003-IMSlecture} quantified ``baseline'' spatial and
temporal effects for wildfire occurrence in federal lands in Oregon,
U.S.A. \citet{PreislerEtAl-2004-IJWF} extended that work, incorporating
partial effects of other locally observed fire-weather covariates, and
proposed modeling the risk of a large fire event conditional on a fire
occurrence being present [Figure~\ref{Fig3}(a)]. Similar
models for California were presented by \citet
{BrillingerEtAl-2006-Environmetrics}, who also assessed wheth\-er random
effects should be included. Other related work includes \citet
{VilarEtAl-2010-IJWF} and \citet{WoolfordEtAl-2011-EnvStats}, who
modeled people-caused wildfire risk in Madrid, Spain and a region of
boreal forest in northeastern Ontario, Canada, respectively. Both of
those studies explored how locally observed anthropogenic variables
(e.g., density of roads in the cell, distance to the nearest railroad
line, population density, etc.) impacted the probability of fire
occurrence. These types of models have been extended to produce one
month ahead forecasts for the probability of large fires (\cite
{PreislerEtAl-2007-AppMeteorAndClimat}; \cite{PreislerEtAl-2008-IJWF}) and have
been used to quantify spatially explicit risk forecasts for large fires
and to estimate suppression costs \citep{PreislerEtAl-2011-IJWF}. We
elaborate on these latter developments when we discuss burn probability models.

Recently, \citet{MagnussenEtAlb-2012-IJWF} developed a set of six
models to predict daily lightning and person-caused fire occurrence for the
province of British Columbia, Canada, at 20-km (400~km$^2$)
resolution [Figure~\ref{Fig3}(b)]. Their methodology employs an ensemble of
annual logistic models for predicting the risk of fires being present
in a given cell. Piecewise linear predictors were incorporated to
handle nonlinear relationships on the logit scale and separate annual
models were fit. Those models connected linear segments together at
sets of knots which form a partition over the range\vadjust{\goodbreak} of the predictor to
produce a piecewise linear function. This piecewise linear framework
had the advantage that the placement of knots could be done using
domain knowledge, rather than the penalized spline approach where many
knots are employed and the likelihood is penalized to prevent
overfitting of the data. Separate annual models were fit because of
known variability in parameter effects from year to year. This
permitted (1)~leave-one-out cross-validation assessment of predictive
ability (e.g., \cite{wood-IntroToGAMs}) and (2) the quantification of
unbiased estimators of the regression parameters, and corresponding
standard errors, without explicitly stating the structure of
year-to-year random effects. This allowed for the joint fitting of a
province-wide model, rather than separate marginal models over a
partition of a province, such as Wotton and Martell's
(\citeyear{WottonEtAl-2005-CJFR}) lightning occurrence models for
the
province of Ontario,\vadjust{\goodbreak} Canada. \citet{MagnussenEtAlb-2012-IJWF} coupled
the results from their logistic models to zero-truncated Poisson models
in order to model the daily number of fires, conditional on fires being
present in a given cell. They also developed models for predicting
medium-term (i.e., 2--14 days ahead) lightning fire occurrence using
an atmospheric stability index (determined from the mesoscale ensemble
weather model output) as a proxy for future lightning activity. While
this model is less accurate than those including observed lighting
strikes, forecasts over this time period are important for fire
management planning.

It is desirable to model fire occurrence risk on a fine scale, so the
probability that a fire will occur can be related to locally observed
conditions, rather than some average value. Then, fitted values can be
aggregated to ``scale up'' to a coarser resolution. However, not all
such logistic GAMs use a fine scale. Large scale models must often be
at coarser resolution because of data availability and computational
limitations. \citet{KrawchukEtAl-2009-PlosOne} investigated
spatio-temporal patterns in fire activity in a global sense, by
modeling on a coarser 100-km (10,000~km$^2$) $\times$ decadal scale. Climate
scenarios were then used to forecast future changes in fire activity.
Their work found increases in future fire activity in certain regions
and decreases in other regions.

Recently, researchers have been exploring methods for monitoring
long-term trends in forest fire occurrence through analyses driven by
historical fire records, focusing on natural, lightning-caused forest
fires (e.g., \cite{AlbertGreenEtAl-2013-CJFR}; \citeauthor{woolford-2010-Environmetrics},
\citeyear{woolford-2010-Environmetrics,woolfordEtAl-2013-Submitted}).
 \citet
{woolford-2010-Environmetrics} looked for changes to inter and
intra-annual trends in lightning-caused fire occurrences in a region of
Boreal forest in Ontario, Canada. They compared a set of nested
logistic generalized additive mixed models that had fixed effects for
seasonality components, annual trends and their interactions, along
with annual random effects to account for year-to-year variability, and
an autoregressive component to account for daily serial correlation.
Their final model employed a bivariate smoother of the ordered pair
(day of year, year) and suggested that the probability of fires being
present in this region was increasing over time and that the effective
length of the fire season appeared to be lengthening.

One feature of the \citet{woolford-2010-Environmetrics} model was that
the local seasonal behavior within a given year could be impacted by
neighboring years due to the functional form\vadjust{\goodbreak} of the specified signal
component. In arid regions, a wet growing season may result in higher
grass biomass and more fire activity in a subsequent dry year \citep
{GreenvilleEtAl-2009-IJWF}. However, except in cases of extreme drought
at the end of a fire season and low winter precipitation, there is
usually enough wintertime precipitation in temperate and boreal forests
to saturate surface organic fuels \citep{LawsonEtAl-2008-CanForServ}
such that fire seasons are essentially independent. \citet
{AlbertGreenEtAl-2013-CJFR} addressed this concern in the boreal forest
by estimating the historical seasonal trends in fire occurrence risk as
a single risk curve (i.e., a univariate smoother of time over the
entire study period). When annual slices of those curves were explored,
it appeared that the fire season's length was changing by starting
earlier and/or ending later each year. A second stage to their analysis
tested for trends in the lengthening of the fire season. The fire
season was defined as the time between the first upcrossing and last
downcrossing of a pre-specified fire risk threshold each year.
Confidence bands associated with the estimate smoother were used to
find a range of dates that were plausible for each given crossing that
defined the start and end of each year's fire season, so uncertainty in
these estimates was incorporated in testing for trends. They found that
the lightning-caused fire season appeared to be both starting earlier
and ending later in Alberta, Canada, and ending later in Ontario.

A difficulty with historical analyses such as in \citet
{woolford-2010-Environmetrics} or \citet{AlbertGreenEtAl-2013-CJFR} is
the potential confounding effects of changes in fire detection system
effectiveness. For example, \citet{woolford-2010-Environmetrics} noted
that the median size at detection of lightning-caused fires decreased
during 1963--2004. Lightning fires occurring in remote areas may take
longer to detect (and so grow in size) than person-caused fires, which
tend to be concentrated near populated places. Smaller lightning fire
sizes at detection suggested that detection may have become more
effective, which is a potential confounder with any changes due to a
warming climate.

These and further complications to the analysis of data from such
historical records have led to more complicated approaches, such as the
use of mixture models for analyzing trends in historical fire risk.
Three dominant characteristics are observed in records of
lightning-caused fire occurrence in Ontario: regular seasonal patterns
and large departures above or below this pattern, where many more
fires
are observed than normal, or so-called zero-heavy behavior\vadjust{\goodbreak} when no
fires are present on the landscape. Letting $X_t$ denote the number of
fire days during time period $t$, and letting $0$, $R$ and $E$ denote
the zero-heavy, regular and extreme behavior components, \citet
{woolfordEtAl-2013-Submitted} proposed the use of a mixture of logistic
GAMs to model weekly counts of fire days:
\begin{eqnarray*}
X_t &\sim&\pi_0(y) \operatorname{bin}\bigl(7,
p_0(w) = 0\bigr) + \pi_R(y) \operatorname{Bin}\bigl(7,
p_R(w)\bigr) \\
&&{}+ \pi_E(y) \operatorname{Bin}\bigl(7,
p_E(w)\bigr),
\end{eqnarray*}
where $w$ and $y$ index the week and year, respectively. The
binomial probabilities for the nondegenerate component are modeled
using penalized spline smoothers (e.g., \cite{wood-IntroToGAMs}) and
the mixing probabilities are parameterized to test for shifts away from
zero-heavy behavior toward regular or extreme behavior by the
multinomial regression of the log-odds against year ($y$):
\begin{eqnarray*}
\operatorname{logit} \biggl(\frac{\pi_j(y)}{\pi_0(y)} \biggr) = \alpha_j +
\beta _j y,\quad j = R, E.
\end{eqnarray*}
When used to analyze lightning-caused forest fire occurrences
in a region of northwestern Ontario, \citet
{woolfordEtAl-2013-Submitted} found a dramatic decline in the
probability of zero-heavy behavior, which was offset by shifts toward
increased chances membership in the regular seasonal or extreme
behavior components. Their model corroborated that the probability of
fire occurrence, especially the length of elevated risk, has been
increasing over time in that region. Moreover, through a second-stage
analysis they found a significant association with temperature
anomalies and fire-weather indices, which suggests that the increased
likelihood of seeing more fire on the landscape than during ``regular''
years was related to a warming climate. Their work also quantified the
power of three hypothesis tests (Wald, score and permutation) for
testing for trends, as well as the length of historical record which
would be required for achieving high power when testing for trends.
They found that the permutation test had the highest power and that the
power of such tests would dramatically increase as the sample size
(i.e., length of the study period) increased beyond 40 years of data
for this region. Investigating the length of historical records
required to test for trends with a specified power has been overlooked
in these sorts of analyses.

\begin{figure*}

\includegraphics{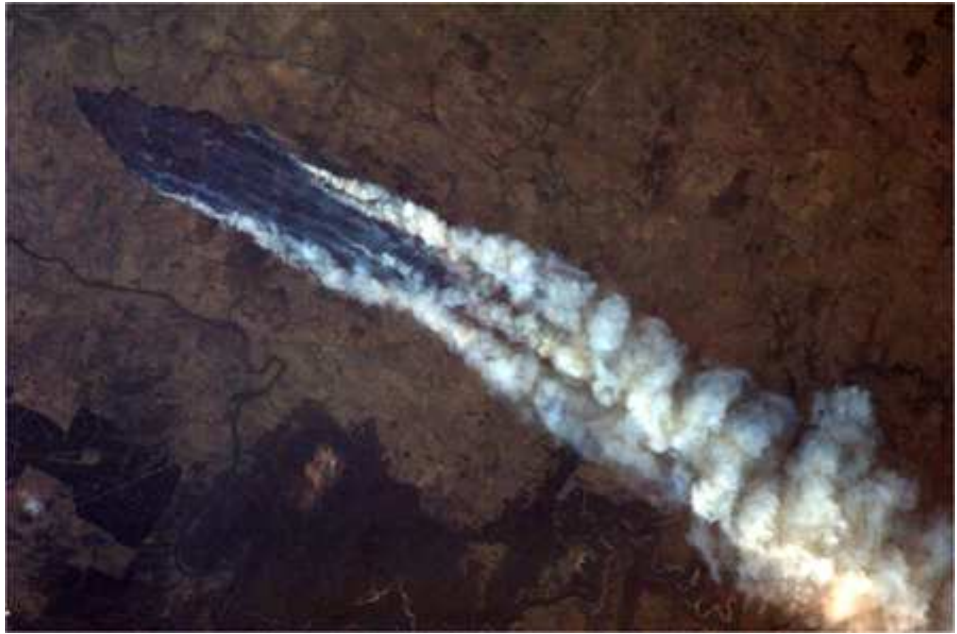}

\caption{The Cobbler Road Fire near Yass, New South Wales, Australia,
on 2 January 2013 (Photo: Chris Hadfield/NASA). At the time the
photograph was taken, the fire was approximately 18 km long and was
spreading through fully-cured grass and open woodland under the
influence of an $\sim$50 km/h wind (Cruz and
Alexander, \citeyear
{CruzEtAl-2013-EnvModelingSoftware}). The maximum flame zone depth and
intensity occurs at the head of the fire in the lower right, and
decreases around the perimeter toward the origin in the upper
left.}\label{Subfig:CobblerRd}
\end{figure*}

\section{Fire Growth}\label{sec3}

After a fire has been ignited, it will continue to spread as long as
sufficient heat is produced by the fire front to\vadjust{\goodbreak} ignite adjacent dead
or live organic matter, if available. The rate of fire spread (ROS) is
determined by the rate at which heat is transferred from burning to
unburned fuel, which is captured in the fundamental equation of spread
\citep{Weber-2001}:
\begin{eqnarray*}
\mbox{{Rate of spread}} &= &\frac{\mbox{{Heat flux from
active combustion}}}{\mbox{{Heat required for fuel ignition}}} \\
&=& \frac{q}{\rho Q_{\mathrm{ig}}},
\end{eqnarray*}
where $q$ is the heat flux from active combustion, $\rho$ is
the fuel density, and $Q_{\mathrm{ig}}$ is the enthalpy per unit mass required
for ignition. ROS is influenced by many environmental factors (e.g.,
moisture content of fine fuels, air temperature and wind speed) and
characteristics of the fuel complex (surface area/\break volume, void space,
depth, temperature).

Fires spread horizontally in surface fuels in two dimensions---with and
parallel to the wind direction at the head of the fire, but also, at a
decreasing rate, laterally and against the wind direction around the
flanks and back of the fire (Figure~\ref{Subfig:CobblerRd}). However,
in coniferous forests and shrub vegetation, fires can also spread from
the ground surface to and in the vegetation canopy if sufficient heat
is produced by the surface fire to heat the crown foliage to ignition
temperature \citep{VanWagner-1977-CJFR}. When a fire ``crowns,'' ROS
increases substantially as the flame zone becomes exposed to the
ambient wind above the vegetation canopy. As a fire continues to grow
in size, firebrands may be lofted ahead and start new fires; as the
smoke plume extends to greater heights in the atmosphere, it may
develop a three-dimensional circulation with stronger upper level
winds.

ROS and fireline intensity (energy release per unit timer per unit of
fire front length) have a diurnal cycle associated with daily variation
in temperature, relative humidity and wind speed---typically following
a sine-wave pattern with a pre-dawn minimum and late afternoon peak
\citep{BeckEtAl-2002-IJWF} which is compounded by stochastic variation
in wind speed over seconds--minutes. Thus, ROS can vary over more than
2 orders of magnitude from less than 1 m min$^{-1}$ to 100--200$+$ m
min$^{-1}$ within and between days during a single fire event, as well
as between fires due to variation in the environment.\footnote{\label{foo5}Sustained ROS of 110$+$ m min$^{-1}$ has been observed in crown
fires in conifer forests in North America, while ROS of 250$+$~m~min$^{-1}$
has been documented in grass fires in Australia \citep{CheneyEtAl-1998-IJWF}.}

Fire duration (the time from ignition to extinguishment) varies from
shorter than 1 day to many weeks or even months. Within this period a
fire may only exhibit significant spread for a period of minutes to
hours within a single day or during a number of burning periods on
multiple days. Variation in wind direction also influences fire growth.
In the extreme case, an abrupt 90$^\circ$ shift in the surface wind
direction (which commonly precedes a cold front) can turn a long fire
flank (e.g., Figure~\ref{Subfig:CobblerRd}) into the head, greatly
increasing fire growth. Thus, variation in the number, magnitude and
direction of spread events results in fire sizes\footnote{The size of the largest recorded individual fire is an unsettled
question. Among the largest documented is the Great Black Dragon fire,
which coalesced from several fires to ultimately burn $1.3 \times 10^5$~km$^2$ in northern China during May 1987 \citep
{CahoonEtAl-1994-GeophysicalResearch}.}
from 10$^{-3}$--10$^4$ km$^2$.

The simplicity of the fundamental equation of fire spread belies the
significant challenge of developing models that provide useful
estimates of wildfire spread and growth over a range of weather
conditions, vegetation types and time periods. \citet
{Show-1919-JournalOfForestry} carried out the first known field
research on wildland fire spread, summarizing fire perimeter\break  growth in
relation to fuel moisture content and wind speed, while \citet
{Fons-1946-AgriResearch} proposed the first physical model of wildfire
spread. Subsequently, spread modeling has followed these two divergent
approaches, which are commonly classified as (a) empirical or
(b)~physical and quasi-physical (\citeauthor{Sullivan-2009a-IJWF},
\citeyear{Sullivan-2009a-IJWF},\break  \citeyear{Sullivan-2009b-IJWF}). Empirical models are based on statistical
relationships between environmental factors and ROS, while physical
models are based on physical and chemical principles; quasi-physical
conserve energy, but do not differentiate between modes of heat
transfer. At least 30 empirical and 40 physical/quasi-physical models
of fire spread have been developed [see reviews by
\cite{Weber-1991-EnergyAndComb}; \cite{PastorEtAl-2003-EnergyAndComb};
\citeauthor{Sullivan-2009a-IJWF},  \citeyear{Sullivan-2009a-IJWF,Sullivan-2009b-IJWF},
and \citeauthor{AlexanderEtAl-2013-ForestryChron}\break  (\citeyear{AlexanderEtAl-2013-ForestryChron})].

\subsection{Spread Rate Models}\label{sec3.1}

ROS models express fire growth as the simple one-dimensional linear
progression of the head, back or flank of the fire at 0, 180 and
90$^\circ$ to the wind direction, respectively (e.g., in units of
m$\cdot$min$^{-1}$ or km$\cdot$hr$^{-1}$).

Empirical approaches have used regression methods to predict ROS as
function of wind speed, fuel moisture content, fuel weight and ground
slope. Models are typically developed for different vegetation
conditions such as conifer and hardwood forests,\break  grassland, shrub and
heathland fuels, and logging slash based on field and laboratory
experiments (e.g., Figure~\ref{Subfig:SuppFig1b}), administrative fire
reports and observations of wildfires.

The Canadian Forest Fire Behavior Prediction\break  (FBP) System \citep
{ForestryResearchGroup-1992} is an example of a well-developed
empirical fire behavior system, where likelihood approaches were used
to estimate the parameters $a$ and $b$ in the Chapman--Richards equation:
\begin{eqnarray*}
\mathrm{ROS} = a \times \bigl[1-e^{(-b \times \mathit{ISI})} \bigr]^c.
\end{eqnarray*}

\begin{figure}

\includegraphics{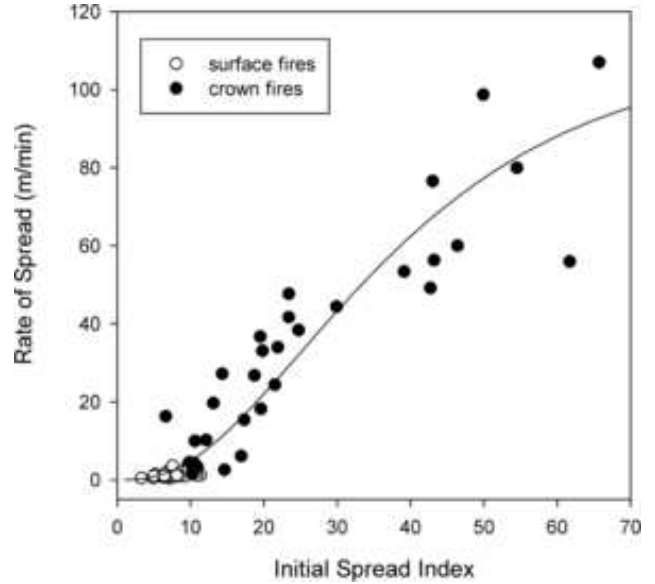}

\caption{Observed (points) and predicted (line) rate of fire spread in
lodgepole and jack pine forests in relation to the Initial Spread Index
of the FWI System (redrawn from Forestry Canada Fire Danger
Group, \citeyear{ForestryResearchGroup-1992}).
The predicted curve is derived from MLE of parameters of the
Chapman--Richards equation. The surface fire observations are from
experimental fires, while a number of the crown fires include wildfire
observations.}\label{Fig:ObservedPredicted}
\end{figure}

The parameter $c$ represents the asymptote and was set for each fuel
type as the maximum fire spread rates observed in coniferous forests
and grasslands (see footnote \ref{foo5}). The Initial Spread Index ($\mathit{ISI}$) is an index that
is based on wind speed and fine fuel moisture content. FBP model
calibration was based on observations of ROS in experimental fires and
wildfires in 17 forest and grass fuel types. The transition from
surface to crown fire is implicit in the sigmoidal curves relating
initial spread index to ROS (Figure~\ref{Fig:ObservedPredicted}),
except in one vegetation type where it is based on physical
considerations \citep{VanWagner-1977-CJFR}. In reanalyses of the FBP
System data, \citet{CruzEtAl-2003-ForestryChron} estimated the
probability of crown fire using a logistic model, and developed a model
of crown fire ROS as a function of crown bulk density, wind speed and
moisture content in coniferous forests (Cruz, Alexander and
Wakimoto, \citeyear{CruzEtAl-2005-CJFR}).

An important milestone in translating physical principles to practical
application was \citeauthor{Rothermel-1972-ForestService}'s (\citeyear
{Rothermel-1972-ForestService}) semi-physical implementation of the
fundamental equation of spread as
\[
\mathrm{ROS} = \frac{(I_p)_o(1 + \phi_w + \phi_s)}{\rho\varepsilon Q_{\mathrm{ig}}},
\]
where the propagating heating flux for a zero wind/\break slope
situation $(I_p)_o$, the wind and slope correction factors $\phi_w$ and
$\phi_s$, and the effective heating number $\varepsilon$ were
parameterized for surface fires in laboratory experiments. The fuel
density $\rho$ can be estimated for various fuel types by field
sampling. This equation was incorporated in the BEHAVE model \citep
{Andrews-1986-BEHAVE}, which is widely used to predict surface fire
spread in the United States and elsewhere.

Sources of error in wildfire spread prediction include lack of model
suitability and accuracy, as well as measurement or sampling errors in
data used as input (\cite{Albini-1976-ForestService}; \cite{AlexanderEtAl-2013-ForestryChron}). It may be difficult to decompose
prediction error into these sources when evaluating the accuracy of a
particular spread model against wildfire observations. A major
challenge in this regard is that model inputs such as wind speed vary
over both space and time. For example, because air flow over and within
forest canopies is turbulent (and also may be affected by the fire
dynamics), wind speed varies at a scale of seconds over distances of
10~s metres, making accurate estimates at the fire front difficult \citep
{Sullivan-2001-CJFR}. In a review of the accuracy of ten empirical and
semi-empirical models of fire spread \citep
{CruzEtAl-2013-EnvModelingSoftware}, six of the models had mean
absolute prediction errors (MAPE) of 20--40\% with respect to their
source data sets. MAPE was defined as
\[
\mathrm{MAPE} = \frac{1}{n}\sum_{i=1}^n
\biggl(\frac{|\hat{y}_i - y_i|}{y_i} \biggr)100,
\]
where $y_i$ was the observed rate of spread, $\hat{y}_i$ was
its corresponding predicted value, and $i$ indexed the sample of size $n$.

Those identical ten spread models have been applied in at least
forty-eight independent studies with more than five observations
arising from experimental, prescribed fires and wildfires. Seven
studies comprising mostly experimental fires (which presumably had the most accurate inputs
and spread documentation) had a MAPE of 20--30\%. A further nine, twenty-six, and seven studies with MAPE of
31--50\%, \mbox{51--75\%}, and $>$75\%, respectively, are a mix of wild, experimental,
and prescribed fires. The \citet
{Rothermel-1972-ForestService} spread model was the most widely
ap\-plied---its median MAPE in twenty-eight studies was 57\% (range 20--310\%).
Because there have been few model comparison studies (e.g.,
Sauvagnargues-\break Lesage et~al. (\citeyear{SauvagnarguesLesage-2001-IJWF})) or systematic model evaluation
programs \citep{CruzEtAl-2013-EnvModelingSoftware}, validation data
have only accumulated slowly over time. The accuracy of some models
and/or the accuracy of predictions in some vegetation types is,
unfortunately, not well described.
\begin{figure}
\centering
\begin{tabular}{@{}c@{}}

\includegraphics{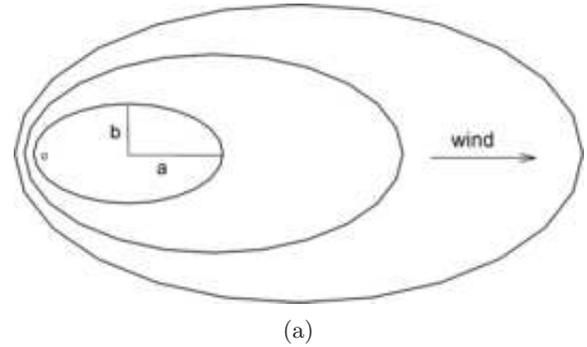}
\\
\footnotesize{(a)}\\[3pt]

\includegraphics{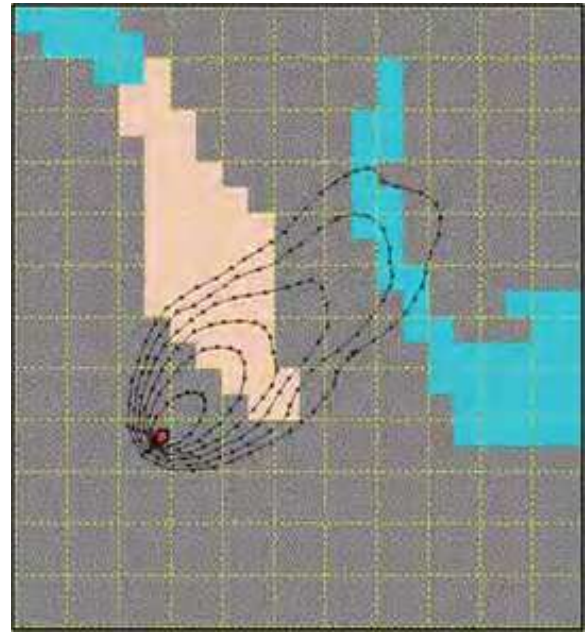}
\\
\footnotesize{(b)}
\end{tabular}
\caption{\textup{(a)} The elliptical fire growth
model is robust in early stages of fire growth. The area and perimeter
can be calculated from the long and short semi-axes, $a$ and $b$,
respectively, which in turn can be calculated from the head, back and
flank fire spread rates and the elapsed time from the fire origin,
denoted by an ``$o$'' in the plot (redrawn from Van Wagner, \citeyear{VanWagner-1969-ForestryChron}). \textup{(b)} Thirty-five-minute
simulation of fire perimeter growth in heterogenous fuels at 25 m
resolution at 5 minute intervals following the wavelet propagation
approach in Prometheus (Tymstra
et~al., \citeyear{TymstraEtAl-2010-NatResourcesCan}). The
colors represent different fuel types (gray is boreal spruce, beige is
grass, and blue is spruce lichen woodland). The red circle represents
the ignition point, and the black dots are the individual vertices
along the fire perimeters.} \label{Fig5}
\end{figure}

\subsection{Fire Growth Models}\label{sec3.2}

Because fire spread rate in empirical and semi-physical models such as
BEHAVE and the FBP System is one dimensional, geometric models have
been developed to project fire growth over time in two dimensions.
\citet{VanWagner-1969-ForestryChron} proposed the use of an elliptical
fire growth model with fire size $A$ and perimeter length $P$:
\begin{eqnarray*}
A &=& \frac{\pi}{2}(v + w)\times u\times t^2; \\
P &\approx&\pi (a+b)
\biggl(1 + \frac{M^2}{4} \biggr),
\end{eqnarray*}
where $u$, $v$ and $w$ are the flank, head and back fire ROS,
respectively, $t$ is the elapsed time since ignition, and $a$ and $b$
are the long and short semi-axes of the ellipse, respectively [which
are related to ROS as $a = (v + w)t/2$ and $b = ut$], and $M = (a - b)
/ (a + b)$. The equation for $P$ is an approximation for the
circumference of an ellipse, truncating an infinite Gauss--Kummer series
at the second term. Other models that extend the idea of an
elliptical-based model have also been proposed \citep
{Anderson-1983-ForestService}. Although critical examination shows that
fire growth, even in uniform conditions, is ellipse-like at best, the
use of a model based on an ellipse nonetheless provides robust
estimates of area and perimeter in the early stages of fire growth
[Figure~\ref{Fig5}(a)].\footnote{It is worth noting that only a thin zone around the fire
perimeter from tens of centimetres--metres in depth is actively
flaming at any time [Figure 5(a)]; this is because the duration of
flaming at any point is in the order of several seconds--minutes, and
flame zone depth${} = \mathrm{ROS} \times$ flaming duration.} The elliptical model has two
useful properties when ROS is constant: (1) the area burned by the fire
at any time is proportional to the square of the time since ignition
(growth in area follows a power function), and (2) the rate of fire
perimeter increase with time is constant \citep{VanWagner-1969-ForestryChron}.

However, where fires spread for periods of hours to days, ROS and
spread direction are influenced by variation in wind speed and
direction, as well as by variation in fuel properties and topographic
conditions. Fire growth simulation models have been developed to
project fire growth in heterogeneous conditions in two dimensions using
one-dimensional ROS equations, often at hourly or sub-hourly intervals,
for periods of hours to days. At least 20 fire growth simulation models
and 22 mathematical analogue models have been developed (\cite
{PastorEtAl-2003-EnergyAndComb}; \cite{Sullivan-2009c-IJWF}); the latter
implement a variety of methods including Markov chains, interacting
particle systems, percolation, cellular automata and differential equations.

\citet{KourtzEtAl-1977-ForestFireResInt} developed one of the first
``contagion'' models of fire growth, implemented in a lattice (grid)
structure, where the spread distance from cell to cell was based on ROS
from the FBP System and wind direction. However, lattice models
constrain the potential spread direction and distance in each time
period. \citeauthor{Richards-1995-IJWF} (\citeyear{Richards-1995-IJWF,Richards-1990-NumMethods}) developed
an algorithm to project the increase in fire perimeter based on
Huygens' principle of wave propagation that overcomes this constraint.
The fire perimeter is discretized into a polygon of vertices joined by
line segments. Fire spread from each vertex is then projected as an
elliptical wavelet of dimensions calculated from ROS equations, and the
new perimeter is formed as the outer hull of the projected points
(removing interior knots, overlaps and crossovers that may evolve)
[Fig-\break ure~\ref{Fig5}(b)]. This method was implemented in the
fire growth simulators FARSITE \citep{Finney-1998-FARSITE}  and\break
Prometheus \citep{TymstraEtAl-2010-NatResourcesCan}; the spread
distance of wavelets in each iteration is calculated using BEHAVE and
the FBP System in the former and latter models, respectively. Minimum
travel time\break  methods have subsequently been implemented in\break  FARSITE \citep
{Finney-2002-CJFR}.

Hybrid empirical--physical approaches have also been used, coupling
empirical surface fire growth with atmospheric fluid dynamics models in
order to represent the complex interactions between large fires and the
atmosphere (\cite{ClarkEtAl-1997-IJWF}; \cite{ClarkEtAl-2004-IJWF}).

More recently, physical models have been developed which allow for fine
scale representations of fuel structures\vadjust{\goodbreak} and fire growth in a
three-dimensional lattice. Examples of these are FIRETEC \citep
{LinnEtAl-2002-IJWF} and the Fire Dynamics Simulator \citep
{MellEtAl-2007-IJWF}. FIRETEC has also been linked to a fluid dynamics
model in order to represent interactions with the atmosphere. Fire
growth is implicit in these physical models, although it is limited to
relatively short time periods and small areas for computational
reasons, while head, back or flank fire spread rates are derived
quantities. Furthermore, replicating the behavior of full scale fires
with physical models remains very challenging (\cite{MellEtAl-2007-IJWF}; \cite{LinnEtAl-2012-CJFR}).

Fire growth prediction errors may also arise due to a lack of model
suitability, accuracy limitations of the given model (e.g., due to the
scale on which predictions are made) and noisy input data. Model
performance has been assessed using various measures that compare
observed and predicted results, including the difference in the radial
distance from the fire origin to points around the perimeter \citep
{Fujioka-2002-IJWF}; difference in fire spread distance
(Duff, Chong and Tolhurst, \citeyear{DuffEtAl-2013-EnvModellingSoftware});
association between predicted and
observed burn perimeters [using Cohen's Kappa coefficient, Sorensen's
coefficient \citet{ArcaEtAl-2007-IJWF} and a Shape Deviation Index
\citep{CuiEtAl-2010-EnviroInfo}]; and agreement in final fire size
distributions [using the Kullback--Leibler divergence \citep
{CouceEtAl-2010-Proceedings}] without regard to spatial association.
However, a major challenge is that validation data from wildfires are
often of poor quality and/or at a coarser spatio-temporal resolution
than model simulations. Weather data inputs may be obtained from a
single station many kilometres distant from the fire location or
interpolated from a number of distant stations, or estimated from a
numerical weather prediction model \citep{JonesEtAl-2003-Proceedings}.
Furthermore, fire perimeters are not usually mapped more frequently
than daily in fire operations. It then becomes problematic when the
interval between observations is several times the model time step
because of error accumulation, particularly in fire spread direction
and head fire location. Importantly, note that after analysis of
twenty-five fires, \citet{Finney-2000-EnvManagement} concluded that it
was not possible to determine growth model performance or error without
controlling or quantifying uncertainty in the input data. On the other
hand, analysis of a large number of fire growth predictions should
reveal model biases if data input errors are unbiased.

The accuracy of both empirical and physical fire spread models, as well
as of fire growth simulation models, is limited by imperfect
understanding of and ability to represent the physical processes over
appropriate scales, variation in atmospheric conditions such as wind
speed and direction that affect spread but which cannot be precisely
known or forecasted, and variation in vegetation and topographic
conditions that is imperfectly represented in models.

However, uncertainty in data inputs has only be incorporated into fire
growth models in a few cases. \citet{WiitalaEtAl-1994-proceedings}
estimated the probability of a free-burning fire in wilderness areas
spreading over a period of weeks from the probability of a ``spread
event day'' with strong winds and the probability of significant
precipitation determined from climatological records. \citet
{Anderson-2010-IJWF} extended\break  these concepts spatially, combining
estimates of the spatial probability of daily spread and extinguishment
in a probabilistic model of the fire growth over weekly to monthly
periods. \citet{AndersonEtAl-2005-AmMeteorSoc} also demonstrated the
use of ensemble methods from meteorology to represent the effect of
varying weather conditions by introducing random and systematic
perturbations to weather forecast inputs to a fire growth model. \citet
{FinneyEtAl-2011-EnvModelling} also applied ensemble methods to
implement FARSITE in a fire probability simulator by randomly and
systematically perturbing the weather input data. Additional links
could be made to probabilistic methods utilized in meteorology and
climatology. Indeed, medium term ensemble numerical weather model
output, such as from the North American Ensemble Forecast System \citep
{TothEtAl-2005-GeophysicalResearch}, are believed to be well suited to
making probabilistic fire projections over 3--10 day time periods, while
climatological methods may be more suited to longer time periods \citep
{Anderson-2002-forestFireResearch}.

\citet{BoychukEtAl-2009-EnvEcoStats} developed a stochastic fire growth
model using a continuous time Markov chain on a lattice, which also
incorporates a stochastic spotting mechanism. They remarked that while
it is well known that embers can be produced from intense fires, lofted
in the smoke plume and deposited ahead of the fire, where they may
start new fires, these processes are difficult to observe and measure.

\subsection{Fire Size}\label{sec3.3}

\citet{WiitalaEtAl-1994-proceedings} observed that the\break  spread of a
free-burning wildfire over a long period is made up of normal spread
days, punctuated by rare spread events, where major growth occurs---this is particularly true for crown fire regimes, where there can be
almost an order of magnitude increase between surface and crown fire
ROS. They considered that the probability of fire movement at any time
was related to the probability of spread and to the probability of
extinguishment, both of which were calculated from waiting time
distributions for major wind events and fire-ending rainfall.

A number of studies have suggested that fire size distributions follow
an exponential (\cite{B89}), power law (\cite{MalamudEtAl-1998-Science}; \cite{JiangEtAl-2009-IJWF})
or a truncated Pareto distribution (\cite{Cumming-2001-CJFR};
\cite{Schoenberg-2003-Environ};
\cite{CuiEtAl-2008-IJWF};
\cite{HolmesEtAl-2008-EcoOfForestDist}).
Power-law behavior has been argued
based on self-organized criticality (Malamud, Morein and
Turcotte, \citeyear{MalamudEtAl-1998-Science}) or
highly optimized tolerance \citep{MoritzEtAl-2005-Proceedings} arising
in dynamical systems.

\begin{figure}[b]

\includegraphics{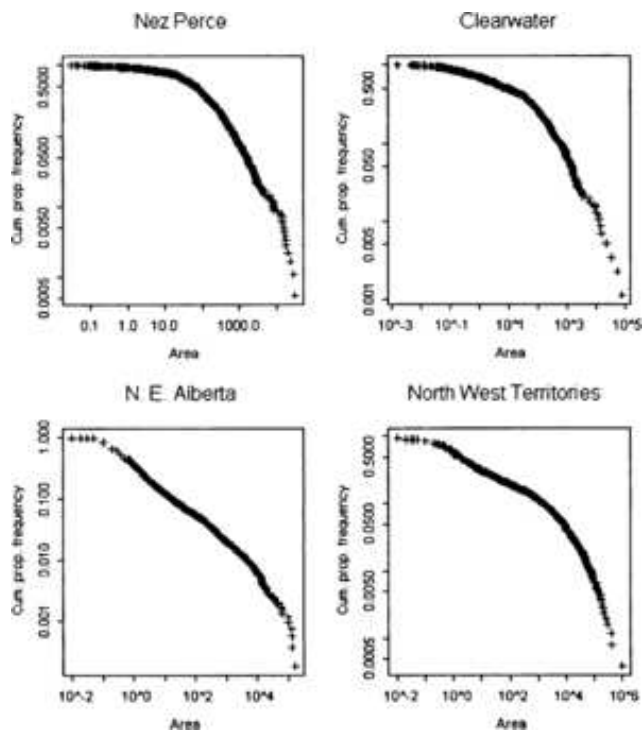}

\caption{Fire size distributions in the Nez Pierce and Clearwater
National Forests, Idaho, USA, and in northern Alberta, and the
Northwest Territories, Canada (Reed and
McKelvey, \citeyear{ReedEtAl-2002-EcoModelling}). Note
the increasing maximum fire sizes in the latter two, larger and less
managed, northern regions.}\label{Fig:FireSizeDistn}
\end{figure}

\citet{ReedEtAl-2002-EcoModelling} provided an important review of
parametric models for fire size distributions. They examined power-law
behavior through the lens of goodness of fit in analyses of several
data sets (Figure~\ref{Fig:FireSizeDistn}) and demonstrated that such
behavior is only approximated over limited ranges of fire sizes. More
importantly, a model is developed which blends both stochastic
processes for growth and extinguishment of fires and is used to develop
an essential model feature, termed the extinguishment growth-rate ratio
(EGRR) from which conditions for power-law behavior are examined in
depth. The growth in area burned is assumed to depend on the current
size of the fire (ignoring spatial aspects, such as the fire's shape),
modeled as a pure birth process whose discrete states represent
regularly spaced, increasing ``markers'' of fire sizes. The
extinguishment of a fire is modeled through a stochastic ``killing
rate'' function, where the probability of extinguishment also depends
on the current size (i.e., state) of the fire. The EGRR is analyzed to
determine general conditions for when a given fire size distribution
follows power-law behavior. For example, power-law behavior over a
given interval of fire sizes would be characterized by a constant EGRR
over that interval; deviations from a constant EGRR suggest departures
from a power-law behavior. Thus, a single power-law distribution for
the size distribution of a given set of fires would be exhibited by a
single, constant EGRR---a rather restrictive condition---while
power-law behavior in the upper tail of a fire size distribution would
be exhibited by an EGRR converging to a positive limit. Special cases
are also considered, for example, when the fire front moves at a fixed
velocity or when the shape of the fire is not regular but fractal, with
area related to length by a power-law relationship and with the fire
front moving at a fixed velocity. None of the special cases were
generally deemed appropriate in practice; most seemed highly
restrictive. Several models are also proposed for fire size, including
a 3-parameter Weibull and a competing hazards model which allows for
competing causes of extinguishment. These are also used to illustrate
that no single model seems superior for the several data sets examined.
Although the power law continues to be used in the literature (\cite
{MalamudEtAl-2005-Proceedings}; \cite{HolmesEtAl-2008-EcoOfForestDist}), \citet
{ZinckEtAl-2009-AmericanNaturalist} emphasized that it is better to
refer to power law-like behavior and to use caution when making
interpretations based on model assumptions.

There are several aspects of fire size modeling\break  which are not well
incorporated into current approaches for analysis. For example, the
amount of effort applied to extinguishing fires varies with a number of
factors, including proximity to settlements, commercial value of timber
and current fire load. Furthermore, fire size is limited by factors
such as fuel continuity, topography and the change of seasons
(especially in regions where snow accompanies the arrival of winter);
the effect of fuel continuity on extinction varies with fire size,
while seasonality effects vary with ignition date.
%

\section{Burned Area and Fire Frequency}\label{sec4}

The annual area burned (BA) in a region is one of the most common
statistics recorded by fire management agencies. It is often used as a
measure of fire season severity, as the risk to timber, air quality and
other values is more closely related to the area burned than the number
of wildfires \citep{Wiitala-1999-proceedings}. Annual BA often varies
by a factor of 10 or more, in a region, with variation in annual
weather and fire danger, and longer term climate cycles \citep
{MeynEtAl-2009-ClimateBio}. A number of different methods have been
used to model the relationship between BA and climate and fire danger
variables, including so-called multivariate adaptive regression splines
\citep{BalshiEtAl-2008-GlobalClimateBio} and general additive models
\citep{KrawchukEtAl-2009-PlosOne}. A surrogate measure of suppression
effectiveness was included in \citet{MartellEtAl-2008-CJFR} along with
fuel and a climatic measure of fire weather in their analysis of BA in
the province of Ontario. Both increases  \citep
{WesterlingEtAl-2006-Science} and decreases\break \citep{MeynEtAl-2010-IJWF}
in BA have been reported in different regions in the past decades,
suggesting that BA is nonstationary in some regions.

Assessing correlation in the number of fires and BA between regions is important for
estimating the collective demand for fire management resources in
larger mutual aid schemes, such as the national resource sharing
systems used in Canada and the Unit\-ed States. \citet
{MagnussenEtAl-2012-IJWF} modeled correlations between regions and
employed Monte Carlo sampling to estimate the likelihood of peaks in BA
between two or more regions occurring within a 14 day period.

\begin{figure}

\includegraphics{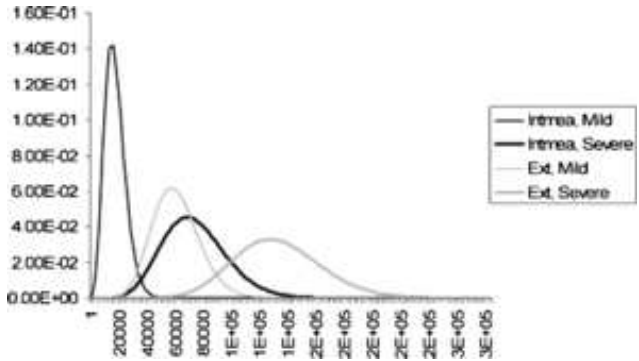}

\caption{Compound Poisson distributions fitted to annual area burned
for four scenarios: mild and severe years in intensively (Intmea.) and
extensively (Ext.) protected forests in Ontario, Canada (Podur, Martell and
Stanford, \citeyear
{PodurEtAl-2010-Environ}). Categorization into mild or severe years was
based on a Seasonal Severity Rating, namely, an average of the Canadian
Forest Fire Weather Index System's Daily Severity Rating.}\label{Fig:CompoundPoisson}
\end{figure}

\citet{Wiitala-1999-proceedings} combined a model for fire size
variability with a Poisson process for fire arrivals to yield the
compound Poisson probability model of BA. However, because of the
difficulties in parameterizing fire size distributions, the risk of BA
exceeding particular values was estimated by discretizing fire sizes
into classes, estimating parameters within classes and calculating
joint probabilities of the number of fires in each class exceeding the
threshold. Drawing on models of aggregate claims in insurance, \citet{PodurEtAl-2010-Environ}
demonstrated that the annual BA could be
estimated as a compound Poisson distribution of the large fire
occurrence rate and expected large fire size. If fire sizes are
exponentially distributed, the total BA is Poisson-exponential and is
distributed as
\[
Fs(s)=e^{-(\lambda+sX)}(2\sqrt{\lambda X/s})I_1(\sqrt{\lambda sX}),\quad s>0,
\]
where $s$ is the annual area burned, $\lambda$ is the annual occurrence rate of
large fires, $X$ the expected fire size, and $I_1$ the modified Bessel
function (Figure~\ref{Fig:CompoundPoisson}). If fire sizes are Weibull-distributed, BA is
Poisson--Weibull and fire size distribution quantities can be calculated
using the lognormal or Pareto approximations.


The annual or average percentage BA has been used as a measure of fire
control success for many years (\cite{ShowEtAl-1941-FireControlNotes};
\cite{Beall-1949-ForestryChron}). \citet{Heinselman-1973-QuatRes} introduced
the term Natural Fire Rotation (NFR) in an ecological context, defined
as the time required to burn an area equal in size to the study area,
\[
\mathrm{NFR} = \frac{A}{A_f}N_y,
\]
where $A$ is the total area of the land, $A_f$ is the total
area burned by all fires (re-burned areas included), and $N_y$ is the
period of observation in years. However, NFR is simply the inverse of
the average annual percent BA, which in turn is equal to the average
probability of a point in the landscape burning \citep
{FallEtAl-1999-EcoSocietyAmerica}, assuming fires occur as a Poisson
process in\vadjust{\goodbreak} space and time. Both percent BA and NFR are calculated using
annual BA, compiled from administrative records (e.g., Figure~\ref{Subfig:SuppFig1a}) or by reconstructing fire boundaries from stand age
maps. Thus, both the size of the sampling area and the length of
observation influence NFR, in as much as they influence the likelihood
of including rare large fire events. Although informal, NFR remains a
popular concept because it is easy to calculate and to communicate.

However, a complete history of burned areas is often not available. In
unmanaged forests of fire origin, the so-called age distribution
depends principally on fire frequency. The age-distribution represents
the distribution of time-since-fire over every point on the landscape.
Conceptually, the statistical problem can be understood as dividing the
study area into a large number of small subunits over a grid and
viewing the resulting survival analysis as a context where time moves
backward, with subunits surviving until they fail through the most
recent past fire occurrence. What is typically available for analysis
is the proportion of the study area that falls within various
time-since-fire classes. Classes are usually determined from forest
stand age maps in decades; where long term maps are available, annual
classes may be used.

Typically, the negative exponential survivorship model is fitted to the
cumulative time since fire data:
\[
A(t) = e^{\lambda t},
\]
where $A(t)$ is the proportion of the landscape surviving to
time $t$, and $\lambda$ is the hazard rate or proportion of area
burned, assuming that fire occurrence in space and time is a Poisson
process (\cite{VanWagner-1969-ForestryChron}; \cite{JohnsonEtAl-1994-EcoRes}).
Sampling areas should be homogeneous with a uniform hazard rate and
larger than the largest fire. The inverse of $\lambda$ has been called
the fire cycle, which is the average stand age of a forest whose age
distribution fits the exponential or Weibull distribution. When age
class data are used, bias may be introduced by the ``missing tail''
\citep{Finney-1995-IJWF}, where very old stands are censored by other
competing hazards (insects, wind, old age).

The key element is the identification of changepoints in fire hazard
rates as well as comparisons of epochs and their hazards over large
scale landscapes globally. Up until the early 1990s, estimation of such
changepoints in the forestry literature was based on identifying
changes through visual inspection of related empirical plots \citep
{Reed-1994-ForScience}. In the late 1990s, likelihood inference emerged
in the forestry literature for estimation of parameters of survivor
functions arising from step-function hazard forms, where changepoints
were specified \citep{ReedEtAl-1998-ForScience}. \citet
{ReedEtAl-1998-ForScience} developed a test for homogeneous hazard
against an alternative of their being a single changepoint.

A substantial shift to more rigorous approaches was initiated by \citeauthor{Reed-2000-CJS}
(\citeyear{Reed-2000-CJS,Reed-2001}), where quasi-likelihood methodology was
employed to obtain estimates of hazards, given $k$ changepoints, while
the number of changepoints was determined through the Bayes Information
Criterion. Using the conceptual framework described earlier where the
study area is divided into $N$ subunits over a grid, the number of
units falling in each time-since-fire class is assumed to follow an
overdispersed multinomial distribution; overdispersion is incorporated
to accommodate spatial correlation in a simple way. The quasi-log
likelihood is
\begin{eqnarray*}
Q& =& \frac{1}{\sigma^2} \sum_{j=1}^m
y_j \operatorname{log}(\theta_j)\\
& =& \frac
{1}{\sigma^2}
\sum_{j=1}^{m-1} \bigl[s_j
\operatorname{log}\bigl\{q^{(j)}\bigr\} + y_j\operatorname
{log}\bigl\{1-q^{(j)} \bigr\} \bigr] ,
\end{eqnarray*}
where
\[
s_j = \sum_{i=j+1}^m
y_i
\]
and
\[
q^{(j)}=e^{-\lambda_jT} .
\]

In the above, $\theta_j$ is the probability that a particular
subunit belongs to time-since-fire class $j$; classes here are $((j -
1)T, jT] (j = 1, \ldots, m - 1)$, while period $m$ is defined
as more than $(m - 1)T$ years ago. As mentioned previously, $T$ is
typically 10 years. Models with $k$ changepoints at prespecified times
$p_1T < \cdots< p_kT$ have hazard rates $\lambda_i$ between $p_{i -
1}T$ and $p_i T$. Estimation of $\lambda_i$ and the overdispersion
parameter $\sigma^2$ is trivially accomplished through quasi-likelihood
estimation. By assigning prior probabilities to models $M_k$ with $k$
changepoints, the Bayes Information Criterion for $M_k$ as well as
posterior probabilities for $M_k$ can be used to guide plausible
choices for $k$. By contrasting changepoint values for a sequence of
models $M_0, M_1, M_2, \ldots,$ and using a sensitivity analysis of
priors, assessments can be made on the consistency of changepoints to
evaluate model choice.\vadjust{\goodbreak}

\begin{figure}

\includegraphics{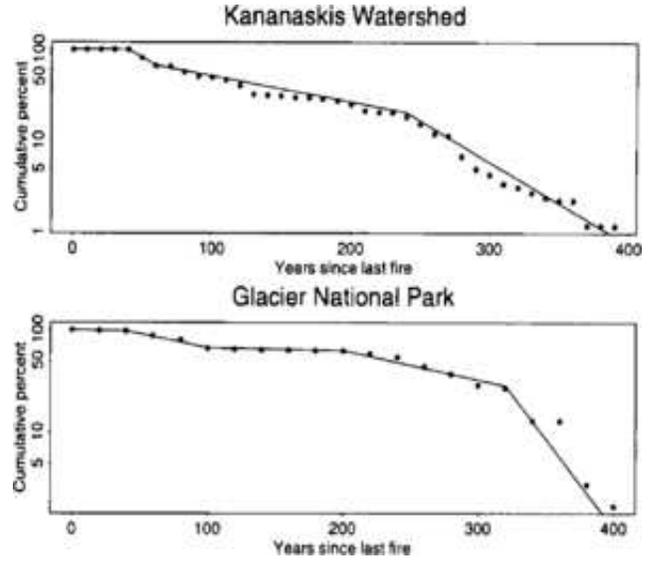}

\caption{Cumulative time-since-fire distributions derived from forest
stand ages in the Kananaskis watershed, Alberta, Canada, and Glacier
National Park, British Columbia, Canada. The line segments extend over
epochs defined by the most plausible change points; epochs are assumed
to have a constant hazard (Reed, \citeyear{Reed-2000-CJS}).}\label{Fig:CumTimeSinceFire}
\end{figure}

\citet{Reed-2000-CJS} applied this methodology to contrast fire epochs
over two major regions, identifying important scientific hypotheses
related to fire regime in these regions (Figure~\ref{Fig:CumTimeSinceFire}). Two major advancements in this methodology
would result from further treatment of spatial correlation using more
modern tools available, as well as incorporation of uncertainty in
estimates arising from the model selection process.

\subsection{Point Frequency}\label{sec4.1}

The interval between fire arrivals at a point in a landscape, as
recorded on fire scars on trees (e.g., Figure~\ref{Subfig:SuppFig1c})
or as charcoal intensity in sediments, is well modeled by a Poisson
process. The challenging problem of estimating fire frequency from
fire-scar data requires essential design and analysis considerations
which take into account that (i) all possible fire-event chronologies
have an equal chance of being chosen, (ii)~not all trees are scarred in
a particular fire, (iii)~methods based on an independent normal
assumption are likely untenable, (iv)~fire frequency intervals change
over large epochs of time. \citet{JohnsonEtAl-1994-EcoRes} discuss
design considerations related to (i) above, including the impact of
choosing trees for fire-scar studies which are easily accessible or
which have the most scars. While this approach can extend estimates of
fire frequency to hundreds and thousands of years (in the case of fire
scars and sediments, resp.), neither trees nor sediments are
perfect recording instruments; some fires may be missed or erased by
other processes.

For many years the traditional approach considered the observed
intervals between scars on all trees in the sample and computed
estimates of the mean time between fires---the mean fire interval (\cite
{ArnoEtAl-1977-ForestService}; \cite{KilgoreEtAl-1979-Ecology}; \cite{Agee-MethodsForFireHistory}) where the confidence intervals come from
the $t$-distribution, assuming that all sites have equal probability of
burning (data are normally distributed and independent). Exploiting
larger numbers of samples, \citet{GrissinoMayer-1999-IJWF} fit two- and
three-parameter Weibull distributions to long fire interval data sets
in Arizona. \citet{ReedEtAl-2004-CJFR} advanced approaches
substantively by developing methods which account for the potential
that fires may not leave scars and, as well, that the independence
assumption is invalid as fires spread spatially. The approach uses
first principles to develop a model whereby a constant hazard rate for
fire occurrence within epochs is combined with an overdispersed
binomial to handle the contagious effect of fire spread; as well, the
probability that a scar-regis\-tering fire leaves a scar is assumed
constant for all objects sampled. By partitioning the probability of
the observed data into a sequence of conditional probabilities, an
overall log likelihood function is constructed. Estimation, however,
proceeds via estimating equations which are a combination of the
maximum likelihood equations for parameters in the mean and a moment
estimator for the dispersion parameter.

\subsection{Burn Probability}\label{sec4.2}

Because data from unmanaged crown-fire dominated forests and fire
scarred trees are restricted to certain environments (and in some cases
are becoming rare within these environments), other methods are needed
to estimate fire frequency at local and landscape scales, the
probability that fires may threaten settlements, infrastructure, timber
and\break  other values at risk, and the influence of climate changes on fire
frequency. In the last decade both simulation and regression-like
approaches have been developed.

\begin{figure}
\centering
\begin{tabular}{@{}c@{}}

\includegraphics{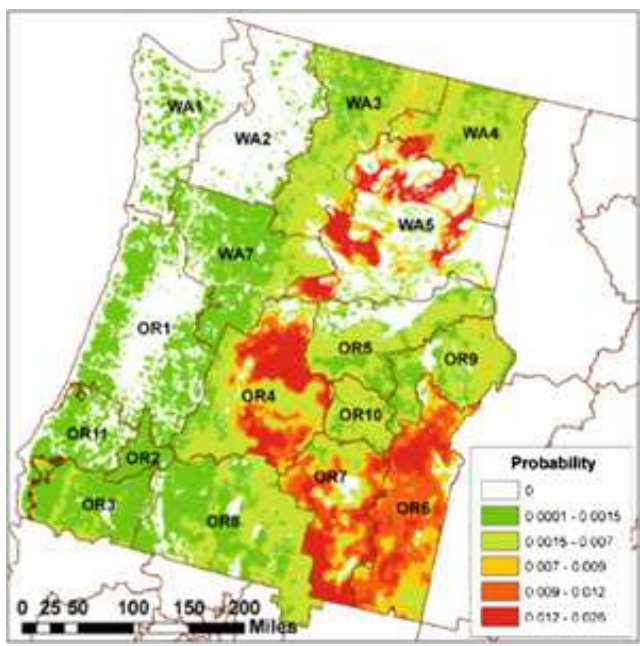}
\\
\footnotesize{(a)}\\[3pt]

\includegraphics{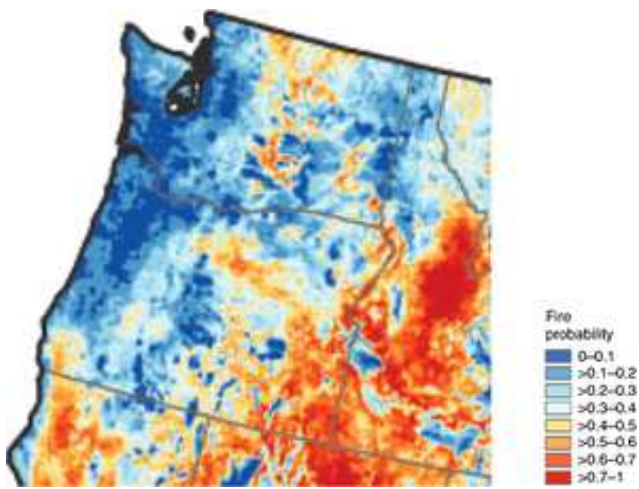}
\\
\footnotesize{(b)}
\end{tabular}
\caption{\textup{(a)} Annual probability of a 270 ha cell burning on federal
lands in Washington and Oregon, USA, estimated from simulation modeling
of potential fire ignitions and spread (Finney
et~al., \citeyear
{FinneyEtAl-2011-StochaticEnvResearch}). \textup{(b)} Relative probability of
burning in the western United States estimated from MaxEnt methods
(Parisien
et~al., \citeyear{ParisienEtAl-2012-IJWF}).} \label{Fig9}
\end{figure}

Monte Carlo approaches implicitly or explicitly combine distributions
of ignitions and spread event days with deterministic fire growth
models to estimate fire sizes, annual area burned and burn probability
or local hazard of burning in a landscape. For example, the
Prometheus\vadjust{\goodbreak}
fire growth model was implemented in software called BurnP3, to
simulate fire spread in landscapes defined by vegetation (fuel type)
and topography grids (slope, aspect), over time periods defined by a
series of daily weather conditions \citep
{ParisienEtAl-2005-NatResourcesCan}. The fire footprints resulting from
many thousands of simulations are ``added up'' to determine the burn
probability or local hazard of burning in a grid cell. Either random or
spatially-explicit ignition probabilities may be used \citep
{BraunEtAl-2010-ProbAndStats}. A similar scheme was used to estimate
burn probability in the contiguous United States of America (i.e.,
excluding the noncontiguous states of Alaska and Hawaii) [Figure~\ref{Fig9}(a)] by implementing the FARSITE growth
modeling in FSIM software \citep{FinneyEtAl-2011-StochaticEnvResearch}.

In a two-stage approach, \citet{PreislerEtAl-2011-IJWF} used a linear
model to estimate mean suppression cost as a function of covariates
(including fire size) and parametric models were developed for the
distribution of fire sizes. Then, a Monte Carlo approach was employed:
spatially explicit probabilities of large fire occurrence were forecast
and then were stochastically mapped to presence/absence of ignition in
a cell. Conditional on large fire ignition being present, a fire size
is simulated and then the projected mean suppression cost is obtained
from the related linear model. This procedure was repeated a large
number of times to produce spatial maps of expected suppression costs
over an upcoming fire season.

\citet{ParisienEtAl-2009-EcoMonographs} applied two tree-based machine
learning algorithms (e.g., \cite{Hastie-dataMining}), MaxEnt (maximum
entropy) and boosted regression trees (BRT), to predict the
environmental space where wildfire can occur in California and in the
contiguous United States of America. The models were fitted to fire map
and large fire occurrence data, including a large suite of
environmental variables such as climate normals, as well as vegetation
and topography covariates, in order to evaluate the contribution of the
individual variables to the susceptibility to fire in a landscape.
\citet{ParisienEtAl-2011-EcoApplications} also used boosted regression
trees to evaluate environmental controls on area burned in the boreal
forest of Canada. MaxEnt methods were used to evaluate a broader set of
environmental variables, including lightning and road density on
wildfire probability in the western United States \citep
{ParisienEtAl-2012-IJWF} [Figure~\ref{Fig9}(b)].

\section{Discussion and Conclusions}\label{sec5}

All events in a wildfire---ignition, growth and extin\-guishment---are
governed by physical principles of conservation of energy, mass,
chemical species and angular momentum \citep{Saito-2001}. While a
number of deterministic physical and empirical models of fire spread
have been developed, wildfire prediction is essentially probabilistic.
This is because, even if we had perfect knowledge of the physical
processes, (1)~human and lightning ignition sources are random, (2)~the
flammability of dead organic fuels and fire spread rates are influenced
by the state of the atmosphere, and this cannot be precisely known over
any period,\footnote{Similar considerations apply to statistical uncertainty in
meteorology \citep{PalmerEtAl-2005-EarthPlanetSci} and climatology
(Von Storch and Zwiers, \citeyear{VonStorch-2002}.)} (3) vegetation characteristics important to fire
behavior vary across the landscape and cannot be precisely represented
in models. While physical models may contribute further understanding,
statistical models and approaches are needed to quantify uncertainties
which are crucial for making decisions with specified precision.
Data are available from a number of sources to support modeling of fire
risk elements over different time periods; these include administrative
and historical records, case studies, laboratory and field experiments,
vegetation proxies (tree rings, stand age, charcoal), remote sensing
and numerical models, as summarized in our article's \hyperref[app]{Appendix}. Each
data source has its own strengths and weaknesses. Administrative
records of fire management organizations have been a primary source for
fire occurrence and size data. However, such records are commonly only
available for decades to a century at most, and in a limited number of
regions. Data quality is variable, and there are few opportunities for
verification of historical records. Furthermore, records collected for
administrative purposes may be at a different resolution or have
missing information that would be important for modeling. For example,
while it is common to record the day a fire starts and its final size,
the dates of control and extinguishment may be missing, and information
on daily fire growth progression is rare and comes mostly from case
studies and historical records. Remote sensing data on fires are
available for the last few decades but at different temporal and
usually coarse spatial scales. Fire frequency can be inferred from
proxy vegetation data over periods of hundreds to thousands of years
but with declining temporal resolution. Censoring is common in all of
these data types. Small fires may be missing (left censored) from
administrative data, vegetation proxy data (tree-rings, age class,
charcoal) and remote sensing data due to incomplete detection;
furthermore, detection effectiveness may vary over time in
administrative data. Right-censoring is common in tree ring and stand
age data because trees can die from other causes. Over long time
series, fire frequency records are nonstationary, due to variation in
climate, fire management strategies and efficiency, patterns of
development and land use practices. Many studies combine more precise,
extensive, physical data on weather or climate covariates with less
accurate, consistent and rigorous fire data (or vice versa) without
accounting for differences in the precision of various data elements.
Different study designs, some perhaps encompassing clustering, repeated
measures, stratification and multi-stage sampling, could be considered.
Hence, substantial data cleaning, in collaboration with forestry
managers and scientists, is required as a first step to any analysis.

The theoretical framework (Poisson process theory) for fire occurrence
modeling is well developed. Further improvements in prediction may come
from both improved data, for model assessment and refinement, and
improved modeling frameworks. Although lightning fire prediction has
been greatly aided by lightning detection system data, strikes are
missed in a \mbox{nonrandom} manner--detection efficiency and spatial
accuracy is related to the proximity to a sensor. If detection system
effectiveness could be quantified (in particular, how the probability
of a fire not being reported has changed over time/space), it could be
incorporated into a logistic model using inclusion probabilities
analogous to the case--control literature. One crucial aspect deserving
of further study is the prediction of sharp peaks where a large number
of lightning fires occur in a very short time period, which can be a
significant fire management problem. A major challenge is the
difficulty in assessing (or predicting) whether lighting storms are
followed by precipitation (which can quench lightning ignitions). This
is because convective precipitation often has a local distribution that
is not measured accurately by sparse weather station networks.
Assimilation schemes that combine data from surface weather stations,
remote sensing, precipitation radar and numerical weather models (e.g., \cite{MahfoufEtAl-2007-AtmosOcean}) may improve the accuracy of
future lightning fire prediction models by providing a better
representation of the spatial distribution of precipitation.

Despite many decades of research and development, fire spread modeling
remains a challenge in some vegetation types. Although more than 70
fire spread models have been developed, only a small number (perhaps
not more than half a dozen) of empirical and quasi-physical fire spread
models are used in fire management;\vadjust{\goodbreak} physical models have limited
ability to replicate the full range of ROS observed in nature. Fire
growth models often have a temporal resolution of seconds--minutes to
match the temporal variability in wind speed. However, most wildfire
data have been obtained by fire management agencies and are often not
recorded more than daily. Detailed data sets on fire spread and growth
at spatio-temporal scales that more closely match model resolution are
needed to facilitate validation and inter-comparison studies in
different vegetation types. However, obtaining good weather data
observations near and during wildfires is difficult, and opportunities
to carry out large free burning experimental fires (where weather can
be closely monitored) are very limited. It may be necessary to monitor
fire growth expressly for validation purposes \citep
{Finney-2000-EnvManagement}, such as with airborne infrared imagers
(e.g., \cite{JonesEtAl-2003-Proceedings}).

Although it is likely that empirical models will continue to be used
for practical applications for many years, it is well recognized that
they have limited flexibility to account for variability in fuel
characteristics and do not explicitly account for interactions with the
mid-atmosphere that may occur in large fires. Representing different
components of fire spread (surface spread, crown fire initiation, crown
fire spread, spotting) in a system of equations may provide a means for
increasing flexibility of empirical models \citep
{CruzEtAl-2008-AustralianForestry}.

Although extinguishment ultimately limits fire\break  growth, it is not well
studied empirically and only rarely included in fire growth models.
Almost all of the physical and empirical fire spread models that have
been developed are deterministic. Methods to represent uncertainty in
fire spread and growth models deserve more attention, as this is
important to decision-making.

Parametric modeling of fire sizes and area burned is difficult due to a
myriad of causes, including spatial heterogeneity and the variable
effects/effective\-ness of fire suppression over the range of fire sizes.
As well, fires that occur late in the year are not as likely to survive
long due to changing weather, while earlier fires have the potential to
last much longer---and hence, grow bigger; seasonality is not accounted
for in current models. Monte Carlo simulation of fire growth can
provide an approximation of fire sizes and area burned; however, it
depends critically on models of fire spread and growth which are
imperfect and often do not account for fire management influences. In
future work, simulation and regression-like approaches might\vadjust{\goodbreak} be used
in a complementary manner, where the latter, for example, may provide
validation of simulation models.

In a warming climate, it will be imperative to improve fire risk
assessment and prediction. This is both a scientific and management
challenge. Systems are needed to predict fire occurrence and frequency
at national and larger scales, including correlation in fire occurrence
between regions. Methods are needed to accommodate nonstationarity.
Model development is constrained in some regions by a lack of long term
fire records. Satellite observations can be used at a coarse scale,
such as for large fire prediction, and will likely become increasingly
important as resolution increases. At present, few fire prediction
models are used by fire managers at national, let alone global scales.

Finally, when the aim of statistical model development is to enhance
fire management decision support systems used by fire managers to
improve their decision-making, it is crucial to consider how such
models can be integrated in management decision support tools while
they are being developed. Complex models need to be implemented in
computer-based fire management information systems in a\break  manner that
provides information (including uncertainty) at the appropriate scale
for the decision problem. It can often take up to 10 years or more from
the development and validation of new models to full implementation in
operational systems and practices. Experience suggests that work is
more likely to influence fire or land management if it involves
collaboration between statisticians with an understanding of the
strengths and limitations of statistical methods as applied for fire
science, and scientists or practitioners with knowledge of the
management questions, the knowledge of limitations of the data, and
sometimes the means to implement new models in practice (e.g., \cite{ReedEtAl-1998-ForScience};
\cite{PreislerEtAl-2004-IJWF};
\cite{WottonEtAl-2005-CJFR}).
Statistical science has an important role in
bringing rigour to fire prediction and risk assessment in both fire
management and fire ecology, and so providing a link between these two
sometimes disparate disciplines.

\begin{appendix}\label{app}
\section*{Appendix: An Overview of Wildfire Data Sources and Limitations}

The types of statistical models that can be developed and analyses that
can be conducted are influenced by the type, resolution and
availability of data. This appendix outlines eight major sources of
quantitative and qualitative\vadjust{\goodbreak} data which have been used to inform
wildfire occurrence, growth, size and frequency models.

\setcounter{figure}{0}
\begin{figure*}

\includegraphics{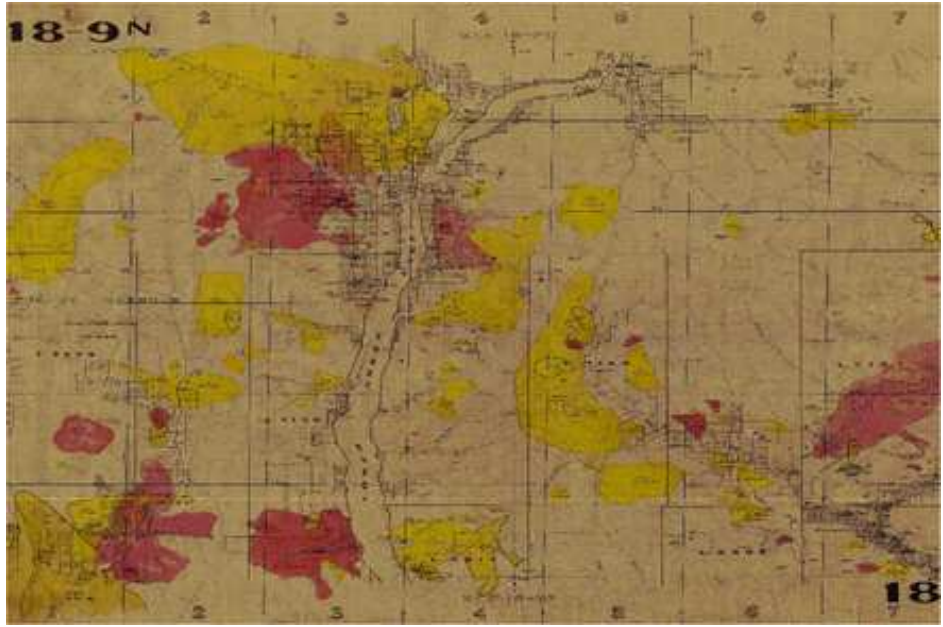}

\caption{Maps and other administrative records are important sources of
data on fire locations, sizes and area burned. Lightning- (yellow) and
person-caused (red) fires along the Columbia River in southern British
Columbia, Canada, during 1920--1945 as recorded on a historical
watercolor-on-linen map.}\label{Subfig:SuppFig1a}
\end{figure*}

1. \textit{Administrative records}: As systematic management
principles began to be applied to wildfire suppression across much of
North America in the early 1900s, foresters in some regions realized
that detailed records would be needed to assess the effectiveness of
fire management efforts---reports of individual fires have been kept in
all national forests in the United States since 1922 \citep
{ShowEtAl-1941-FireControlNotes}. Thus, individual fire reports, maps
of perimeters of significant fires and annual summaries have been
compiled for about 100 years in parts of the United States and Canada,
and more recently in other regions (Figure~\ref{Subfig:SuppFig1a}).
Researchers soon realized that administrative records were a rich data
source. For example, \citet{ShowEtAl-1923} used administrative records
to examine annual fire frequency in California, while \citet
{Abell-1940-ForestService} made inferences about fire spread rates from
individual fire reports. Some common limitations of administrative
records include: (1)~limited accuracy of spatial locations in older
records---fire perimeters, for example, are often from sketched maps;
(2)~data are often left censored, as not all small fires may be
detected; (3)~the observational period may be relatively short in
relation to the return period of extreme events in some regions;
(4)~data collected for administrative purposes may be missing some
information that may be needed to address research questions; (5)~the
fire management agency passes over jurisdiction of some fires to other
agencies (e.g., in the province of Ontario, Canada, the Ontario
Ministry of Natural Resources transfers some fires over to
municipalities). Nevertheless, administrative records continue to be a
primary source of information on fire occurrence, fire size and area
burned in managed forests and other regions where organized fire
management is carried out. It is important to note, however, this is
only a portion of the earth's fire environment.

2. \textit{Historical records}: Though anecdotal, records of
historical wildfires \citep{Plummer-1912-USdeptOfAg} may provide
important information on the occurrence of rare, extreme events. For
example, \citet{HainesEtAl-1970-Weatherwise} reanalyzed the
meteorological conditions of America's deadliest wildfire, the 1871
Peshtigo fire disaster, which killed at least 1500 people in Wisconsin.
That fire had been investigated by \citet{Robinson-1872}, who reported
that embers from the fire landed 7 miles away on the decks of vessels
in Lake Michigan.

\begin{figure*}

\includegraphics{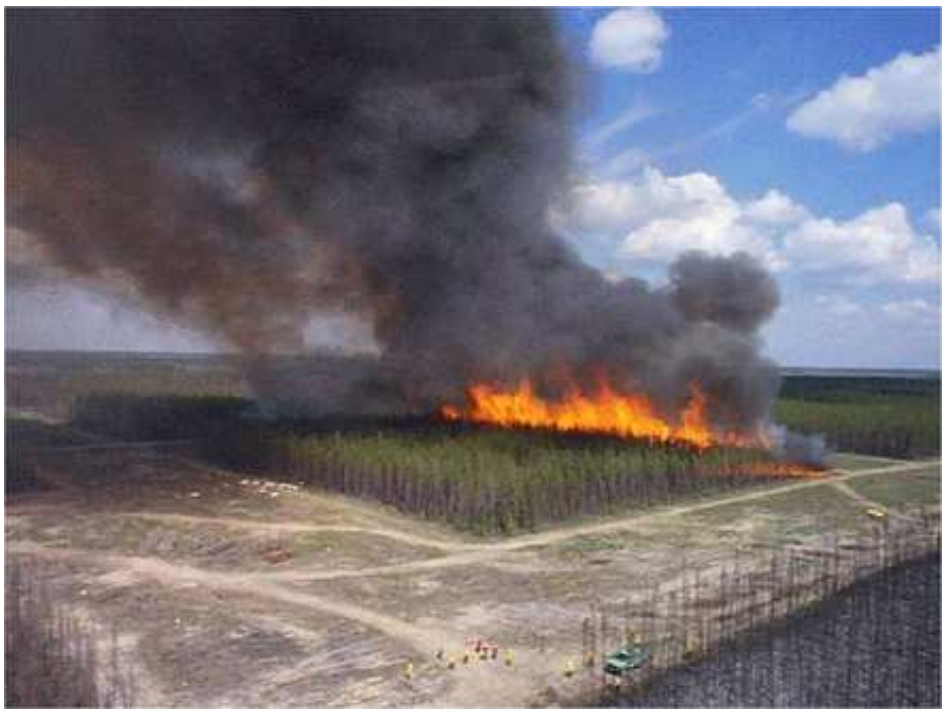}

\caption{Experimental fires allow for detailed observations of fire
behavior. Crown fire in a 3 hectare jackpine stand carried out in the
International Crown Fire Modeling Experiment, Northwest Territories,
Canada (Stocks, Alexander and
Lanoville, \citeyear{StocksEtAl-2004-CJFR}).}\label{Subfig:SuppFig1b}
\end{figure*}

3. \textit{Outdoor experiments}: Fires that were lit under
controlled conditions have provided among the most reliable data on
fire ignition probability and fire behavior under measured
environmental conditions for over half a century \citep
{Curry-1938-AgrResearch}. Because such experiments are logistically
difficult to carry out under severe burning conditions
(Stocks, Alexander and Lanoville, \citeyear{StocksEtAl-2004-CJFR})
(Figure~\ref{Subfig:SuppFig1b}), they have been
limited to fires smaller than 10 hectares in size. However, some
phenomena associated with large fires cannot be readily reproduced in
small experimental fires. These include long-range spotting ahead of
the fire front and the development of smoke plumes reaching and
interacting with winds in the lower and mid-troposphere.

4. \textit{Case studies}: Detailed analyses of significant
wildfires by expert observers (\cite{Gisborne-1927-WeatherReview}; \cite{Olsen-2003-FireManag}) can provide important insights into extreme or
unusual events \citep{AlexanderEtAl-2010-FireManag}. For example, the
report by \citet{KiilEtAl-1969-ABfireWx} remains the most complete
documentation of one of the fastest spreading fires observed in the
northern Hemisphere---a single fire spread event which extended 60 km
in 10~hours; an average sustained rate of spread of 110$+$~m min$^{-1}$.

5. \textit{Laboratory experiments}: The effect of individual
environmental factors, such as fuel moisture \citep{Fons-1946-AgriResearch}
or wind speed (Rothermel, Anderson and
Forest, \citeyear{RothermelEtAl-1966}), on
fire ignition and spread has been examined under controlled laboratory
conditions,\break  which often employ wind tunnels. Such experiments have been
important to parameterize physical and semi-physical models, but are limited to
fires in the order of a metre in size. Fire spread in complex
vegetation structures and phenomena related to vertical development,
such as crown fire initiation, cannot be readily reproduced in the
laboratory.

6. \textit{Numerical and simulation modeling}: Mathematical
models of fire initiation and spread have been implemented in computer
simulations since the 1970s \citep{KourtzEtAl-1971-ForestSci}, allowing
for experi-\break ments in the computer that are not possible in the laboratory
or in nature. Physical models can be computationally intensive, so
simulations of individual fires have typically been of limited size
(several hec\-tares) and duration (minutes). Monte Carlo-like\break  methods
have been used to simulate the growth of many thousands of fires (using
empirical and quasi-physical spread models) on a regional scale to
estimate fire size distributions and burn area probability or fire
frequency (\cite{ParisienEtAl-2005-NatResourcesCan}; \cite{FinneyEtAl-2011-StochaticEnvResearch}). There are still practical limits
on the size of region that can be modeled at high resolution, which can
result in edge matching issues between regions.

\begin{figure*}[t]

\includegraphics{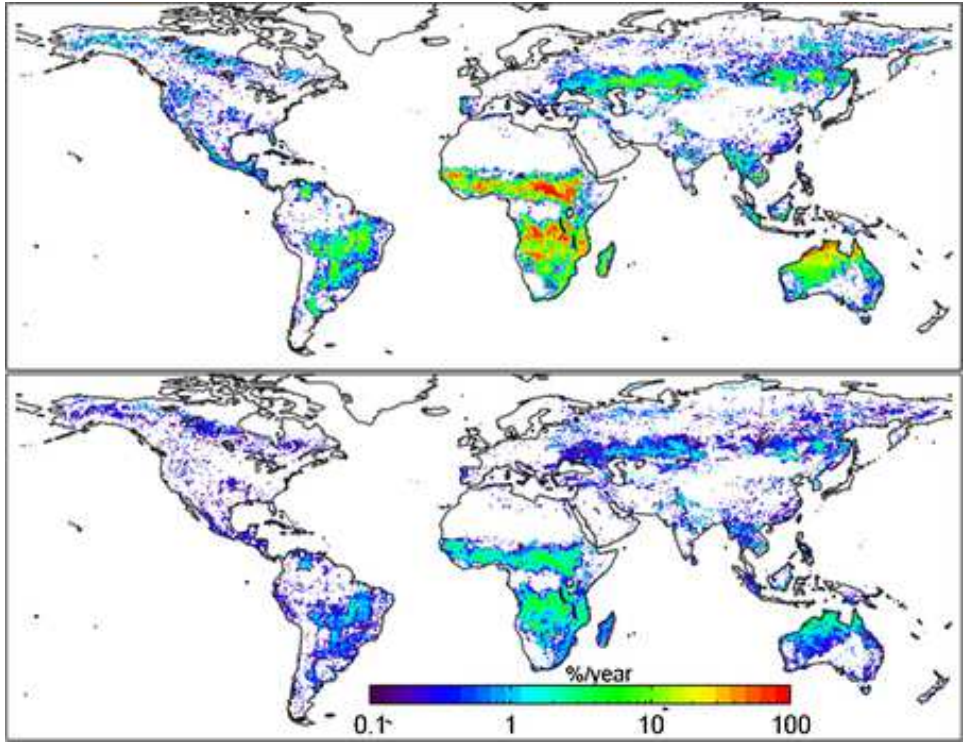}

\caption{Satellite imagery has been used to map fires since the 1970s.
Mean annual global burned area (top) and associated one-sigma
uncertainties (bottom) expressed as a fraction of each grid cell that
burns each year derived from 1997--2008 (Giglio et~al., \citeyear
{Giglio-2009-Biogeosci}).}\label{Subfig:SuppFig1d}
\end{figure*}

7. \textit{Infra-red imaging and other remote sensing}:
Infra-red imaging systems have been used to detect and map forest fires
from aircraft since the late 1960s \citep{Bjornsen-1968-RemoteSensing},
allowing accurate repeated measurements of forest fire perimeters and
growth of large fires over periods of hours or days. Satellite imagery
on the earth's land surface became available in the 1970s, and this has
provided increasingly refined estimates of global burned area. LANDSAT
imagery has been used to map burned areas, particularly in remote
regions, since the late 1970s at 30 m resolution, but with only monthly
sampling frequency. Radiometers such as the Advanced Very High
Resolution Radiometer (AVHRR) and the Moderate Resolution Imaging
Spectroradiometer (MODIS)\break  deployed on the NOAA and the NASA Aqua and
Terra satellites, respectively, have been used to detect and map forest
fires since the 1980s and 2000s (\cite{FlanniganEtAl-1986-CJFR}; \cite{JusticeEtAl-2002-RemoteSensing}). AVHRR and MODIS sensors detect fire
activity at 1000 and 500 m resolution, respectively, several times a
day at a global scale. Global burned area estimates, derived primarily
from MODIS data, are\break  shown in Figure~\ref{Subfig:SuppFig1d}. Since
2002, the European MeteoSat geostationary satellites have detected
fires over Europe and Africa every 15 minutes at 3000 m resolution
\citep{RobertsEtAl-2009-Biogeosciences}. Remote sensing observations
may provide an important source of data for fire occurrence modeling in
regions where administrative records are incomplete.

\begin{figure*}[b]

\includegraphics{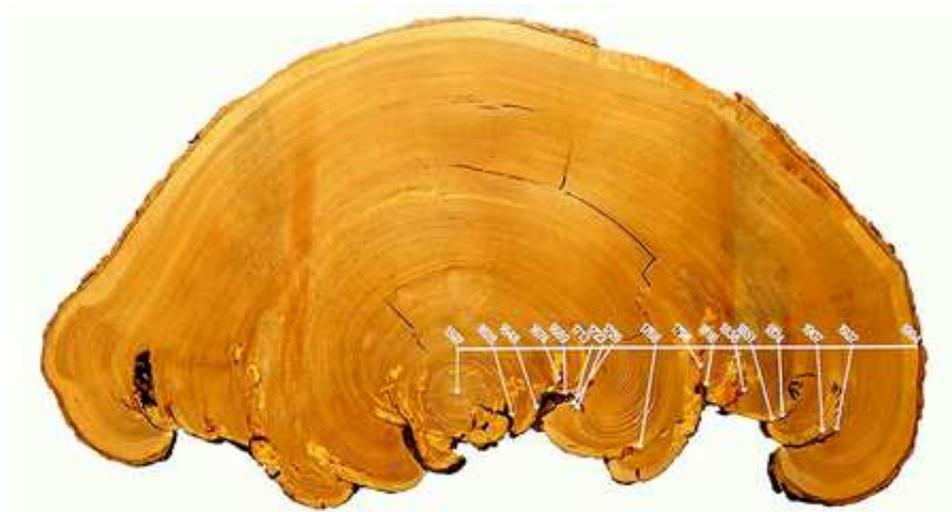}

\caption{Low intensity fires scar living trees, creating a record of
the time between fires at a point over long time periods. Cross section
of a western larch (90 cm diam.) from southern British Columbia with
approximately 10 fire scars over 400 years (NRCan photo).}\label{Subfig:SuppFig1c}
\end{figure*}

8. \textit{Vegetation and charcoal proxies}: Surface fires
often cause nonlethal injuries in tolerant trees that result in ``fire
scars'' observable in the live wood (Figure~\ref{Subfig:SuppFig1c}).
Dating fire scars using tree rings provides a point sample of time
since fire. The frequency of such fires, typically in the order about
10--40 years, was first examined by \citet{Clements-1910}, \citet
{Howe-1915-ForProCan} and other pioneering researchers\break  \citep
{Mcbride-1983-TreeRing}. However, the sampling period for fire scar
records is limited by the lifespan of the tree species---up to several
hundred years for long-lived species such as Ponderosa pine. Thus, the
number of records in a region tends to decrease over time (right
censored) as trees are cut or die from various other causes. Similar
methods have been used to date anomalies in ring growth in eucalypts
(Burrows, Ward and Robinson, \citeyear{BurrowsEtAl-1995-AusForestry}) and Australian grasstrees (\textit
{Xanthorrhoea}) that can survive high intensity fires.

Charcoal resulting from burning of woody vegetation is incorporated
into the soil, while small fragments may be transported and deposited
in lake sediments. Counts of charcoal fragments in soil or lake
sediment cores represent point samples. Combined with carbon dating,
fire frequency has been determined from the time between charcoal
pulses in sediment cores \citep{Swain-1973-QuatResearch}. Although
temporal resolution is coarser than annual tree rings, sampling periods
can extend from centuries to millennia, depending on the geological
history of the sampling area. Laboratory analysis of sediment cores is
time consuming, which limits the sampling intensity. In a review of
data analysis methods, \citet{HigueraEtAl-2011-IJWF} note that fire
frequency over long time periods is usually nonstationary. Indeed, the
association between climate variation and fire risk is often the
motivation for paleo-ecological studies.

Northern temperate and boreal coniferous forests with crown fire
regimes are made up of cohorts of approximately even-aged stands, whose
ages can be used to date the fire initiating events. The age
distribution of stands in a region can be used to estimate the fire
frequency (typically 50 years to several centuries) assuming a
frequency distribution such as the negative exponential (\cite
{Heinselman-1973-QuatRes}; \cite{VanWagner-1978-CJFR}) or the Weibull. However,
the frequency of extreme events may be underestimated because it is
difficult and time consuming to classify the age of all forest stands
except in relatively small areas and such sampling areas are often
small relative to the size of extreme events. As with tree ring data,
older stands may be missed in sampling or lost due to mortality from
other causes (right censored).
\end{appendix}

\section*{Acknowledgments}

This work was supported by Natural Resources Canada and the Natural
Sciences and Engineering Research Council of Canada. Many thanks go to
Haiganoush Preisler (USDA Forest Service), one\break  anonymous referee and
the guest editor, Michel Dek\-king, for their helpful comments on earlier
versions of the paper. Thanks also to A. Albert-Green for technical
assistance with this manuscript.

%

%




\begin{thebibliography}{178}

\bibitem[\protect\citeauthoryear{Abell}{1940}]{Abell-1940-ForestService}
\begin{bmisc}[author]
\bauthor{\bsnm{Abell},~\bfnm{C.}\binits{C.}}
(\byear{1940}).
\bhowpublished{\textit{Rates of Initial Spread of Free-Burning Fires on the National
Forests of California}. California Forest Research Experiment Station, USDA
Forest Service, {Berkeley, CA}.}
\bptok{imsref}%
\end{bmisc}
\endbibitem

\bibitem[\protect\citeauthoryear{Agee}{1996}]{Agee-MethodsForFireHistory}
\begin{binproceedings}[author]
\bauthor{\bsnm{Agee},~\bfnm{J.}\binits{J.}}
(\byear{1996}).
\btitle{Methods for fire history}.
In \bbooktitle{{Fire Ecology of Pacific Northwest Forests.}}
\bpublisher{Island Press}, \blocation{Washington, DC}.
\bptok{imsref}%
\end{binproceedings}
\endbibitem

\bibitem[\protect\citeauthoryear{Ainsworth and
Kauffman}{2009}]{Ainsworth-2009-PlantEcology}
\begin{barticle}[author]
\bauthor{\bsnm{Ainsworth},~\bfnm{A.}\binits{A.}} \AND
\bauthor{\bsnm{Kauffman},~\bfnm{J.~B.}\binits{J.~B.}}
(\byear{2009}).
\btitle{{Response of native Hawaiian woody species to laval-ignited wildfires
in tropical forests and shrublands}}.
\bjournal{Plant Ecology}
\bvolume{201}
\bpages{197--209}.
\bptok{imsref}%
\end{barticle}
\endbibitem

\bibitem[\protect\citeauthoryear{Albert-Green
et~al.}{2013}]{AlbertGreenEtAl-2013-CJFR}
\begin{barticle}[author]
\bauthor{\bsnm{Albert-Green},~\bfnm{A.}\binits{A.}},
\bauthor{\bsnm{Dean},~\bfnm{C.~B.}\binits{C.~B.}},
\bauthor{\bsnm{Martell},~\bfnm{D.~L.}\binits{D.~L.}} \AND
\bauthor{\bsnm{Woolford},~\bfnm{D.~G.}\binits{D.~G.}}
(\byear{2013}).
\btitle{{A methodology for investigating trends in changes in the timing of the
fire season with applications to lightning-caused forest fires in Alberta and
Ontario, Canada}}.
\bjournal{Canadian Journal of Forest Research}
\bvolume{43}
\bpages{39--45}.
\bptok{imsref}%
\end{barticle}
\endbibitem

\bibitem[\protect\citeauthoryear{Albini}{1976}]{Albini-1976-ForestService}
\begin{bbook}[author]
\bauthor{\bsnm{Albini},~\bfnm{F.~A.}\binits{F.~A.}}
(\byear{1976}).
\btitle{{Estimating Wildfire Behavior and Effects.}}
\bpublisher{Intermountain Forest and Range Experiment Station, Forest Service,
US Dept. Agriculture}, \blocation{Ogden, UT}.
\bptok{imsref}%
\end{bbook}
\endbibitem

\bibitem[\protect\citeauthoryear{Albini}{1984}]{Al84}
\begin{barticle}[auto:STB|2013/10/14|10:36:11]
\bauthor{\bsnm{Albini},~\bfnm{F.~A.}\binits{F.~A.}}
(\byear{1984}).
\btitle{Wildland fires: Predicting the behavior of wildland
  fires---among nature's most potent forces---can save
  lives, money, and natural resources}.
\bjournal{American Scientist}
\bvolume{72}
\bpages{590--597}.
\bptok{imsref}%
\end{barticle}
\endbibitem


\bibitem[\protect\citeauthoryear{Alexander and
Cruz}{2013}]{AlexanderEtAl-2013-ForestryChron}
\begin{barticle}[author]
\bauthor{\bsnm{Alexander},~\bfnm{M.~E.}\binits{M.~E.}} \AND
\bauthor{\bsnm{Cruz},~\bfnm{M.~G.}\binits{M.~G.}}
(\byear{2013}).
\btitle{{Limitations on the accuracy of model predictions of wildland fire
behaviour: A~state-of-the-knowledge review}}.
\bjournal{Forestry Chronicle}
\bvolume{89}
\bpages{370--381}.
\bptok{imsref}%
\end{barticle}
\endbibitem

\bibitem[\protect\citeauthoryear{Alexander and
Taylor}{2010}]{AlexanderEtAl-2010-FireManag}
\begin{barticle}[author]
\bauthor{\bsnm{Alexander},~\bfnm{M.~E.}\binits{M.~E.}} \AND
\bauthor{\bsnm{Taylor},~\bfnm{S.~W.}\binits{S.~W.}}
(\byear{2010}).
\btitle{{Wildland fire behaviour case studies and the 1938 Honey Fire
controversy}}.
\bjournal{Fire Management Today}
\bvolume{70}
\bpages{15--25}.
\bptok{imsref}%
\end{barticle}
\endbibitem

\bibitem[\protect\citeauthoryear{Anderson}{1983}]{Anderson-1983-ForestService}
\begin{bbook}[author]
\bauthor{\bsnm{Anderson},~\bfnm{H.~E.}\binits{H.~E.}}
(\byear{1983}).
\btitle{{Predicting Wind-Driven Wild Land Fire Size and Shape.}}
\bpublisher{US Department of Agriculture, Forest Service, Intermountain Forest
and Range Experiment Station}, \blocation{Ogden, UT}.
\bptok{imsref}%
\end{bbook}
\endbibitem

\bibitem[\protect\citeauthoryear{Anderson}{2002}]{Anderson-2002-forestFireRese%
arch}
\begin{binproceedings}[author]
\bauthor{\bsnm{Anderson},~\bfnm{K.~R.}\binits{K.~R.}}
(\byear{2002}).
\btitle{{Fire growth modelling at multiple scales}}.
In \bbooktitle{Forest Fire Research \& Wildland Fire Safety. Proceedings of IV
International Conference on Forest Fire Research/2002 Wildland Fire Safety
Summit}
\bpages{18--23}.
\bpublisher{Milpress}, \blocation{Rotterdam}.
\bptok{imsref}%
\end{binproceedings}
\endbibitem

\bibitem[\protect\citeauthoryear{Anderson}{2010}]{Anderson-2010-IJWF}
\begin{barticle}[author]
\bauthor{\bsnm{Anderson},~\bfnm{K.~R.}\binits{K.~R.}}
(\byear{2010}).
\btitle{{A climatologically based long-range fire growth model}}.
\bjournal{International Journal of Wildland Fire}
\bvolume{19}
\bpages{879--894}.
\bptok{imsref}%
\end{barticle}
\endbibitem

\bibitem[\protect\citeauthoryear{Anderson, Flannigan and
Reuter}{2005}]{AndersonEtAl-2005-AmMeteorSoc}
\begin{binproceedings}[author]
\bauthor{\bsnm{Anderson},~\bfnm{K.~R.}\binits{K.~R.}},
\bauthor{\bsnm{Flannigan},~\bfnm{M.}\binits{M.}} \AND
\bauthor{\bsnm{Reuter},~\bfnm{G.}\binits{G.}}
(\byear{2005}).
\btitle{{Using ensemble techniques in fire-growth modelling}}.
In \bbooktitle{Sixth Symposium on Fire and Forest Meteorology.}
\bpublisher{American Meteorological Society}, \blocation{Boston, MA}.
\bptok{imsref}%
\end{binproceedings}
\endbibitem

\bibitem[\protect\citeauthoryear{Andreae and
Merlet}{2001}]{AndreaeEtAl-2001-Biogeochem}
\begin{barticle}[author]
\bauthor{\bsnm{Andreae},~\bfnm{M.~O.}\binits{M.~O.}} \AND
\bauthor{\bsnm{Merlet},~\bfnm{P.}\binits{P.}}
(\byear{2001}).
\btitle{{Emission of trace gases and aerosols from biomass burning}}.
\bjournal{Global Biogeochemical Cycles}
\bvolume{15}
\bpages{955--966}.
\bptok{imsref}%
\end{barticle}
\endbibitem

\bibitem[\protect\citeauthoryear{Andrews}{1986}]{Andrews-1986-BEHAVE}
\begin{bmisc}[author]
\bauthor{\bsnm{Andrews},~\bfnm{P.~L.}\binits{P.~L.}}
(\byear{1986}).
\bhowpublished{\textit{BEHAVE: Fire Behavior Prediction and Fuel Modeling System-BURN
Subsystem, Part 1}. USDA Forest Service, Ogden, UT}.
\bptok{imsref}%
\end{bmisc}
\endbibitem

\bibitem[\protect\citeauthoryear{Andrews, Finney and
Fischetti}{2007}]{AndrewsEtAl-2007-SciAmer}
\begin{barticle}[author]
\bauthor{\bsnm{Andrews},~\bfnm{P.}\binits{P.}},
\bauthor{\bsnm{Finney},~\bfnm{M.}\binits{M.}} \AND
\bauthor{\bsnm{Fischetti},~\bfnm{M.}\binits{M.}}
(\byear{2007}).
\btitle{{Predicting wildfires}}.
\bjournal{Scientific American}
\bvolume{297}
\bpages{46--55}.
\bptok{imsref}%
\end{barticle}
\endbibitem

\bibitem[\protect\citeauthoryear{Arca et~al.}{2007}]{ArcaEtAl-2007-IJWF}
\begin{barticle}[author]
\bauthor{\bsnm{Arca},~\bfnm{B.}\binits{B.}},
\bauthor{\bsnm{Duce},~\bfnm{P}\binits{P.}},
\bauthor{\bsnm{Laconi},~\bfnm{M.}\binits{M.}},
\bauthor{\bsnm{Pellizaro},~\bfnm{G.}\binits{G.}},
\bauthor{\bsnm{Salis},~\bfnm{M.}\binits{M.}} \AND
\bauthor{\bsnm{Spano},~\bfnm{D.}\binits{D.}}
(\byear{2007}).
\btitle{{Evaluation of FARSITE simulator in Mediterranean maquis}}.
\bjournal{International Journal of Wildland Fire}
\bvolume{16}
\bpages{563--572}.
\bptok{imsref}%
\end{barticle}
\endbibitem

\bibitem[\protect\citeauthoryear{Arno, Sneck and
Forest}{1977}]{ArnoEtAl-1977-ForestService}
\begin{bmisc}[author]
\bauthor{\bsnm{Arno},~\bfnm{S.~F.}\binits{S.~F.}},
\bauthor{\bsnm{Sneck},~\bfnm{K.~M.}\binits{K.~M.}} \AND
\bauthor{\bsnm{Forest},~\bfnm{I.}\binits{I.}}
(\byear{1977}).
\bhowpublished{\textit{A Method for Determining Fire History in Coniferous Forests of
the Mountain West}. Intermountain Forest and Range Experiment Station, Forest
Service, US Dept. Agriculture, Ogden, TU}.
\bptok{imsref}%
\end{bmisc}
\endbibitem

\bibitem[\protect\citeauthoryear{Baker}{1989}]{B89}
\begin{barticle}[auto:STB|2013/10/14|10:36:11]
\bauthor{\bsnm{Baker},~\bfnm{W.~L.}\binits{W.~L.}}
(\byear{1989}).
\btitle{Landscape ecology and nature reserve design in the Boundary Waters
  Canoe Area, Minnesota}.
\bjournal{Ecology}
\bvolume{70}
\bpages{23--35}.
\bptok{imsref}%
\end{barticle}
\endbibitem

\bibitem[\protect\citeauthoryear{Balshi
et~al.}{2008}]{BalshiEtAl-2008-GlobalClimateBio}
\begin{barticle}[author]
\bauthor{\bsnm{Balshi},~\bfnm{M.~S.}\binits{M.~S.}},
\bauthor{\bsnm{McGuire},~\bfnm{A.~D.}\binits{A.~D.}},
\bauthor{\bsnm{Duffy},~\bfnm{P.}\binits{P.}},
\bauthor{\bsnm{Flannigan},~\bfnm{M.}\binits{M.}},
\bauthor{\bsnm{Walsh},~\bfnm{J.}\binits{J.}} \AND
\bauthor{\bsnm{Melillo},~\bfnm{J.}\binits{J.}}
(\byear{2008}).
\btitle{{Assessing the response of area burned to changing climate in western
boreal North Americal using a Multivariate Adaptive Regression Splines (MARS)
approach}}.
\bjournal{Global Change Biology}
\bvolume{15}
\bpages{578--600}.
\bptok{imsref}%
\end{barticle}
\endbibitem

\bibitem[\protect\citeauthoryear{Beall}{1949}]{Beall-1949-ForestryChron}
\begin{barticle}[author]
\bauthor{\bsnm{Beall},~\bfnm{H.}\binits{H.}}
(\byear{1949}).
\btitle{{An outline of forest fire protection standards}}.
\bjournal{Forestry Chronicle}
\bvolume{25}
\bpages{82--106}.
\bptok{imsref}%
\end{barticle}
\endbibitem

\bibitem[\protect\citeauthoryear{Beck et~al.}{2002}]{BeckEtAl-2002-IJWF}
\begin{barticle}[author]
\bauthor{\bsnm{Beck},~\bfnm{J.}\binits{J.}},
\bauthor{\bsnm{Alexander},~\bfnm{M.}\binits{M.}},
\bauthor{\bsnm{Harvey},~\bfnm{S.}\binits{S.}} \AND
\bauthor{\bsnm{Beaver},~\bfnm{A.}\binits{A.}}
(\byear{2002}).
\btitle{{Forecasting diurnal variations in fire intensity to enhance wildland
firefighter safety}}.
\bjournal{International Journal of Wildland Fire}
\bvolume{11}
\bpages{173--182}.
\bptok{imsref}%
\end{barticle}
\endbibitem

\bibitem[\protect\citeauthoryear{Beverly and
Wotton}{2007}]{BeverlyEtAl-2007-IJWF}
\begin{barticle}[author]
\bauthor{\bsnm{Beverly},~\bfnm{J.~L.}\binits{J.~L.}} \AND
\bauthor{\bsnm{Wotton},~\bfnm{B.~M.}\binits{B.~M.}}
(\byear{2007}).
\btitle{{Modelling the probability of sustained flaming: Predictive value of
fire weather index components compared with observations of site weather and
fuel moisture conditions}}.
\bjournal{International Journal of Wildland Fire}
\bvolume{16}
\bpages{161--173}.
\bptok{imsref}%
\end{barticle}
\endbibitem

\bibitem[\protect\citeauthoryear{Bigley and
Roberts}{2001}]{BigleyEtAl-2001-ManagementJournal}
\begin{barticle}[author]
\bauthor{\bsnm{Bigley},~\bfnm{G.~A.}\binits{G.~A.}} \AND
\bauthor{\bsnm{Roberts},~\bfnm{K.~H.}\binits{K.~H.}}
(\byear{2001}).
\btitle{{The indident command system: High-reliability organizing for complex
and volatile task environments}}.
\bjournal{Academy of Management Journal}
\bvolume{44}
\bpages{1281--1299}.
\bptok{imsref}%
\end{barticle}
\endbibitem

\bibitem[\protect\citeauthoryear{Bjornsen}{1968}]{Bjornsen-1968-RemoteSensing}
\begin{binproceedings}[author]
\bauthor{\bsnm{Bjornsen},~\bfnm{R.}\binits{R.}}
(\byear{1968}).
\btitle{{Infrared mapping of large fires}}.
In \bbooktitle{Fifth Symposium on Remote Sensing of the Environment}
\bpages{459--464}.
\bpublisher{Univ. Michigan}, \blocation{Ann Arbor, MI}.
\bptok{imsref}%
\end{binproceedings}
\endbibitem

\bibitem[\protect\citeauthoryear{Bond
et~al.}{2013}]{BondEtAl-2013-GeophysicalResearch}
\begin{barticle}[author]
\bauthor{\bsnm{Bond},~\bfnm{T.}\binits{T.}},
\bauthor{\bsnm{Doherty},~\bfnm{S.}\binits{S.}},
\bauthor{\bsnm{Fahey},~\bfnm{D.}\binits{D.}},
\bauthor{\bsnm{Forster},~\bfnm{P.}\binits{P.}},
\bauthor{\bsnm{Berntsen},~\bfnm{T.}\binits{T.}},
\bauthor{\bsnm{DeAngelo},~\bfnm{B.}\binits{B.}},
\bauthor{\bsnm{Flanner},~\bfnm{M.}\binits{M.}},
\bauthor{\bsnm{Ghan},~\bfnm{S.}\binits{S.}},
\bauthor{\bsnm{Karcher},~\bfnm{B.}\binits{B.}} \AND
\bauthor{\bsnm{Koch},~\bfnm{D.}\binits{D.}}
(\byear{2013}).
\btitle{{Bounding the role of black carbon in the climate system: A scientific
assessment}}.
\bjournal{Journal of Geophysical Research: Atmospheres}
\bvolume{118}
\bpages{1--173}.
\bptok{imsref}%
\end{barticle}
\endbibitem

\bibitem[\protect\citeauthoryear{Bowman et~al.}{2009}]{BowmanEtAl-2009-Science}
\begin{barticle}[author]
\bauthor{\bsnm{Bowman},~\bfnm{D.~M. J.~S.}\binits{D.~M. J.~S.}},
\bauthor{\bsnm{Balch},~\bfnm{J.~K.}\binits{J.~K.}},
\bauthor{\bsnm{Artaxo},~\bfnm{P.}\binits{P.}},
\bauthor{\bsnm{Bond},~\bfnm{W.~J.}\binits{W.~J.}},
\bauthor{\bsnm{Carlson},~\bfnm{J.~M.}\binits{J.~M.}},
\bauthor{\bsnm{Cochrane},~\bfnm{M.~A.}\binits{M.~A.}},
\bauthor{\bsnm{D'Antonio},~\bfnm{C.~M.}\binits{C.~M.}},
\bauthor{\bsnm{DeFries},~\bfnm{R.~S.}\binits{R.~S.}},
\bauthor{\bsnm{Doyle},~\bfnm{J.~C.}\binits{J.~C.}} \AND
\bauthor{\bsnm{Harrison},~\bfnm{S.~P.}\binits{S.~P.}}
(\byear{2009}).
\btitle{{Fire in the Earth system}}.
\bjournal{Science}
\bvolume{324}
\bpages{481--484}.
\bptok{imsref}%
\end{barticle}
\endbibitem

\bibitem[\protect\citeauthoryear{Boychuk
et~al.}{2009}]{BoychukEtAl-2009-EnvEcoStats}
\begin{barticle}[mr]
\bauthor{\bsnm{Boychuk},~\bfnm{Den}\binits{D.}},
\bauthor{\bsnm{Braun},~\bfnm{W.~John}\binits{W.~J.}},
\bauthor{\bsnm{Kulperger},~\bfnm{Reg~J.}\binits{R.~J.}},
\bauthor{\bsnm{Krougly},~\bfnm{Zinovi~L.}\binits{Z.~L.}} \AND
\bauthor{\bsnm{Stanford},~\bfnm{David~A.}\binits{D.~A.}}
(\byear{2009}).
\btitle{A stochastic forest fire growth model}.
\bjournal{Environ. Ecol. Stat.}
\bvolume{16}
\bpages{133--151}.
\bid{doi={10.1007/s10651-007-0079-z}, issn={1352-8505}, mr={2668730}}
\bptok{imsref}%
\end{barticle}
\endbibitem

\bibitem[\protect\citeauthoryear{Braun
et~al.}{2010}]{BraunEtAl-2010-ProbAndStats}
\begin{barticle}[mr]
\bauthor{\bsnm{Braun},~\bfnm{W.~John}\binits{W.~J.}},
\bauthor{\bsnm{Jones},~\bfnm{Bruce~L.}\binits{B.~L.}},
\bauthor{\bsnm{Lee},~\bfnm{Jonathan S.~W.}\binits{J.~S.~W.}},
\bauthor{\bsnm{Woolford},~\bfnm{Douglas~G.}\binits{D.~G.}} \AND
\bauthor{\bsnm{Wotton},~\bfnm{B.~Mike}\binits{B.~M.}}
(\byear{2010}).
\btitle{Forest fire risk assessment: An illustrative example from {O}ntario,
{C}anada}.
\bjournal{J. Probab. Stat.}
\bpages{Art. ID 823018, 26}.
\bid{issn={1687-952X}, mr={2661646}}
\bptok{imsref}%
\end{barticle}
\endbibitem

\bibitem[\protect\citeauthoryear{Breslow and
Powers}{1978}]{BreslowEtAl-1978-Biometrics}
\begin{barticle}[author]
\bauthor{\bsnm{Breslow},~\bfnm{N.}\binits{N.}} \AND
\bauthor{\bsnm{Powers},~\bfnm{W.}\binits{W.}}
(\byear{1978}).
\btitle{{Are there two logistic regressions for retrospecitve studies?}}
\bjournal{Biometrics}
\bvolume{34}
\bpages{100--105}.
\bptok{imsref}%
\end{barticle}
\endbibitem

\bibitem[\protect\citeauthoryear{Brillinger, Preisler and
Benoit}{2003}]{BrillingerEtAl-2003-IMSlecture}
\begin{bincollection}[mr]
\bauthor{\bsnm{Brillinger},~\bfnm{David~R.}\binits{D.~R.}},
\bauthor{\bsnm{Preisler},~\bfnm{Haiganoush~K.}\binits{H.~K.}} \AND
\bauthor{\bsnm{Benoit},~\bfnm{John~W.}\binits{J.~W.}}
(\byear{2003}).
\btitle{Risk assessment: A forest fire example}.
In \bbooktitle{Statistics and Science: A {F}estschrift for {T}erry {S}peed}.
\bseries{Institute of Mathematical Statistics Lecture Notes---Monograph Series}
\bvolume{40}
\bpages{177--196}.
\bpublisher{IMS}, \blocation{Beachwood, OH}.
\bid{doi={10.1214/lnms/1215091142}, mr={2004338}}
\bptok{imsref}%
\end{bincollection}
\endbibitem

\bibitem[\protect\citeauthoryear{Brillinger, Preisler and
Benoit}{2006}]{BrillingerEtAl-2006-Environmetrics}
\begin{barticle}[mr]
\bauthor{\bsnm{Brillinger},~\bfnm{D.~R.}\binits{D.~R.}},
\bauthor{\bsnm{Preisler},~\bfnm{H.~K.}\binits{H.~K.}} \AND
\bauthor{\bsnm{Benoit},~\bfnm{J.~W.}\binits{J.~W.}}
(\byear{2006}).
\btitle{Probabilistic risk assessment for wildfires}.
\bjournal{Environmetrics}
\bvolume{17}
\bpages{623--633}.
\bid{doi={10.1002/env.768}, issn={1180-4009}, mr={2247173}}
\bptok{imsref}%
\end{barticle}
\endbibitem

\bibitem[\protect\citeauthoryear{Bruce}{1960}]{Bruce-1960-FireControlNotes}
\begin{barticle}[author]
\bauthor{\bsnm{Bruce},~\bfnm{D.}\binits{D.}}
(\byear{1960}).
\btitle{{How many fires?}}
\bjournal{Fire Control Notes}
\bvolume{24}
\bpages{45--50}.
\bptok{imsref}%
\end{barticle}
\endbibitem

\bibitem[\protect\citeauthoryear{Burrows, Ward and
Robinson}{1995}]{BurrowsEtAl-1995-AusForestry}
\begin{barticle}[author]
\bauthor{\bsnm{Burrows},~\bfnm{N.}\binits{N.}},
\bauthor{\bsnm{Ward},~\bfnm{B.}\binits{B.}} \AND
\bauthor{\bsnm{Robinson},~\bfnm{A.}\binits{A.}}
(\byear{1995}).
\btitle{{Jarrah forest fire history from stem analysis and anthropological
evidence [Eucalyptus marginata; Western Australia]}}.
\bjournal{Australian Forestry}
\bvolume{58}
\bpages{7--16}.
\bptok{imsref}%
\end{barticle}
\endbibitem

\bibitem[\protect\citeauthoryear{Cahoon
et~al.}{1994}]{CahoonEtAl-1994-GeophysicalResearch}
\begin{barticle}[author]
\bauthor{\bsnm{Cahoon},~\bfnm{D.~R.}\binits{D.~R.}},
\bauthor{\bsnm{Stocks},~\bfnm{B.~J.}\binits{B.~J.}},
\bauthor{\bsnm{Levine},~\bfnm{J.~S.}\binits{J.~S.}},
\bauthor{\bsnm{Cofer},~\bfnm{W.~R.}\binits{W.~R.}} \AND
\bauthor{\bsnm{Pierson},~\bfnm{J.~M.}\binits{J.~M.}}
(\byear{1994}).
\btitle{{Satellite analysis of the severe 1987 forest fires in northern China
and southeastern Siberia}}.
\bjournal{Journal of Geophysical Resarch}
\bvolume{99}
\bpages{18627--18638}.
\bptok{imsref}%
\end{barticle}
\endbibitem

\bibitem[\protect\citeauthoryear{Cheney, Gould and
Catchpole}{1998}]{CheneyEtAl-1998-IJWF}
\begin{barticle}[author]
\bauthor{\bsnm{Cheney},~\bfnm{N.}\binits{N.}},
\bauthor{\bsnm{Gould},~\bfnm{J.}\binits{J.}} \AND
\bauthor{\bsnm{Catchpole},~\bfnm{W.~R.}\binits{W.~R.}}
(\byear{1998}).
\btitle{{Prediction of fire spread in grasslands}}.
\bjournal{International Journal of Wildland Fire}
\bvolume{8}
\bpages{1--13}.
\bptok{imsref}%
\end{barticle}
\endbibitem

\bibitem[\protect\citeauthoryear{Clark, Coen and
Latham}{2004}]{ClarkEtAl-2004-IJWF}
\begin{barticle}[author]
\bauthor{\bsnm{Clark},~\bfnm{T.~L.}\binits{T.~L.}},
\bauthor{\bsnm{Coen},~\bfnm{J.}\binits{J.}} \AND
\bauthor{\bsnm{Latham},~\bfnm{D.}\binits{D.}}
(\byear{2004}).
\btitle{{Description of a coupled atmosphere--fire model}}.
\bjournal{International Journal of Wildland Fire}
\bvolume{13}
\bpages{49--63}.
\bptok{imsref}%
\end{barticle}
\endbibitem

\bibitem[\protect\citeauthoryear{Clark et~al.}{1997}]{ClarkEtAl-1997-IJWF}
\begin{barticle}[author]
\bauthor{\bsnm{Clark},~\bfnm{T.~L.}\binits{T.~L.}},
\bauthor{\bsnm{Jenkins},~\bfnm{M.~A.}\binits{M.~A.}},
\bauthor{\bsnm{Coen},~\bfnm{J.}\binits{J.}} \AND
\bauthor{\bsnm{Packham},~\bfnm{D.~R.}\binits{D.~R.}}
(\byear{1997}).
\btitle{{A coupled atmosphere--fire model: Role of the convective Froude number
and dynamic fingering at the fireline}}.
\bjournal{International Journal of Wildland Fire}
\bvolume{6}
\bpages{177--190}.
\bptok{imsref}%
\end{barticle}
\endbibitem

\bibitem[\protect\citeauthoryear{Clements}{1910}]{Clements-1910}
\begin{bmisc}[author]
\bauthor{\bsnm{Clements},~\bfnm{F.~E.}\binits{F.~E.}}
(\byear{1910}).
\bhowpublished{\textit{The Life History of Lodgepole Burn Forests}. US Dept.
Agriculture, Washington, DC}.
\bptok{imsref}%
\end{bmisc}
\endbibitem

\bibitem[\protect\citeauthoryear{Couce
et~al.}{2010}]{CouceEtAl-2010-Proceedings}
\begin{binproceedings}[author]
\bauthor{\bsnm{Couce},~\bfnm{E.}\binits{E.}},
\bauthor{\bsnm{Knorr},~\bfnm{W.}\binits{W.}},
\bauthor{\bsnm{Perona},~\bfnm{G.}\binits{G.}} \AND
\bauthor{\bsnm{Brebbia},~\bfnm{C.~A.}\binits{C.~A.}}
(\byear{2010}).
\btitle{{Statistical parameter estimation for a cellular auomata wildfire model
based on satellite observations}}.
In \bbooktitle{Second International Conference on Modelling, Monitoring and
Management of Forest Fires.}
\bpages{47--56}.
\bpublisher{WIT Press}, \blocation{Kos, Greece}.
\bptok{imsref}%
\end{binproceedings}
\endbibitem

\bibitem[\protect\citeauthoryear{Cruz, Alexander and
Wakimoto}{2003}]{CruzEtAl-2003-ForestryChron}
\begin{barticle}[author]
\bauthor{\bsnm{Cruz},~\bfnm{M.~G.}\binits{M.~G.}},
\bauthor{\bsnm{Alexander},~\bfnm{M.~E.}\binits{M.~E.}} \AND
\bauthor{\bsnm{Wakimoto},~\bfnm{R.~H.}\binits{R.~H.}}
(\byear{2003}).
\btitle{{Assessing the probability of crown fire initiation based on fire
danger indices}}.
\bjournal{Forestry Chronicle}
\bvolume{79}
\bpages{976--983}.
\bptok{imsref}%
\end{barticle}
\endbibitem

\bibitem[\protect\citeauthoryear{Cruz, Alexander and
Wakimoto}{2005}]{CruzEtAl-2005-CJFR}
\begin{barticle}[author]
\bauthor{\bsnm{Cruz},~\bfnm{M.~G.}\binits{M.~G.}},
\bauthor{\bsnm{Alexander},~\bfnm{M.~E.}\binits{M.~E.}} \AND
\bauthor{\bsnm{Wakimoto},~\bfnm{R.~H.}\binits{R.~H.}}
(\byear{2005}).
\btitle{{Development and testing of models for predicting crown fire rate of
spread in conifer forest stands}}.
\bjournal{Canadian Journal of Forest Research}
\bvolume{35}
\bpages{1628--1639}.
\bptok{imsref}%
\end{barticle}
\endbibitem

\bibitem[\protect\citeauthoryear{Cruz, Alexander and
Fernandes}{2008}]{CruzEtAl-2008-AustralianForestry}
\begin{barticle}[author]
\bauthor{\bsnm{Cruz},~\bfnm{M.~G.}\binits{M.~G.}},
\bauthor{\bsnm{Alexander},~\bfnm{M.~E.}\binits{M.~E.}} \AND
\bauthor{\bsnm{Fernandes},~\bfnm{P.~A.}\binits{P.~A.}}
(\byear{2008}).
\btitle{{Development of a model system to predict wildfire behaviour in pine
plantations}}.
\bjournal{Australian Forestry}
\bvolume{71}
\bpages{113}.
\bptok{imsref}%
\end{barticle}
\endbibitem

\bibitem[\protect\citeauthoryear{Cruz and
Alexander}{2013}]{CruzEtAl-2013-EnvModelingSoftware}
\begin{barticle}[author]
\bauthor{\bsnm{Cruz},~\bfnm{M.~G.}\binits{M.~G.}} \AND
\bauthor{\bsnm{Alexander},~\bfnm{M.~E.}\binits{M.~E.}}
(\byear{2013}).
\btitle{{Uncertainty associated with model predictions of surface and crown
fire rates of spread}}.
\bjournal{Environmental Modeling and Software}
\bvolume{47}
\bpages{16--28}.
\bptok{imsref}%
\end{barticle}
\endbibitem

\bibitem[\protect\citeauthoryear{Cui and Perera}{2008}]{CuiEtAl-2008-IJWF}
\begin{barticle}[author]
\bauthor{\bsnm{Cui},~\bfnm{W.}\binits{W.}} \AND
\bauthor{\bsnm{Perera},~\bfnm{A.~H.}\binits{A.~H.}}
(\byear{2008}).
\btitle{{What do we know about forest fire size distribution, and why is this
knowledge use for forest management}}.
\bjournal{International Journal of Wildland Fire}
\bvolume{17}
\bpages{234--244}.
\bptok{imsref}%
\end{barticle}
\endbibitem

\bibitem[\protect\citeauthoryear{Cui and
Perera}{2010}]{CuiEtAl-2010-EnviroInfo}
\begin{barticle}[author]
\bauthor{\bsnm{Cui},~\bfnm{W.}\binits{W.}} \AND
\bauthor{\bsnm{Perera},~\bfnm{A.~H.}\binits{A.~H.}}
(\byear{2010}).
\btitle{{Quantifying spatio-temporal errors in forest fire spread modelling
explicitly}}.
\bjournal{Journal of Environmental Infomatics}
\bvolume{16}
\bpages{19--26}.
\bptok{imsref}%
\end{barticle}
\endbibitem

\bibitem[\protect\citeauthoryear{Cumming}{2001}]{Cumming-2001-CJFR}
\begin{barticle}[author]
\bauthor{\bsnm{Cumming},~\bfnm{S.}\binits{S.}}
(\byear{2001}).
\btitle{{A parametric model of the fire-size distribution}}.
\bjournal{Canadian Journal of Forest Research}
\bvolume{31}
\bpages{1297--1303}.
\bptok{imsref}%
\end{barticle}
\endbibitem

\bibitem[\protect\citeauthoryear{Cunningham and
Martell}{1973}]{CunninghamEtAl-1973-CJFR}
\begin{barticle}[author]
\bauthor{\bsnm{Cunningham},~\bfnm{A.}\binits{A.}} \AND
\bauthor{\bsnm{Martell},~\bfnm{D.~L.}\binits{D.~L.}}
(\byear{1973}).
\btitle{{A stochastic model for the occurrence of man-caused forest fires}}.
\bjournal{Canadian Journal of Forest Research}
\bvolume{3}
\bpages{282--287}.
\bptok{imsref}%
\end{barticle}
\endbibitem

\bibitem[\protect\citeauthoryear{Curry and Fons}{1938}]{Curry-1938-AgrResearch}
\begin{barticle}[author]
\bauthor{\bsnm{Curry},~\bfnm{J.~R.}\binits{J.~R.}} \AND
\bauthor{\bsnm{Fons},~\bfnm{W.~L.}\binits{W.~L.}}
(\byear{1938}).
\btitle{{Rate of spread of surface fires in the ponderosa pine type of
California}}.
\bjournal{Journal of Agricultural Research}
\bvolume{57}
\bpages{239--267}.
\bptok{imsref}%
\end{barticle}
\endbibitem

\bibitem[\protect\citeauthoryear{Doan and
Martell}{1974}]{DoanEtAl-1974-ForestryChron}
\begin{barticle}[author]
\bauthor{\bsnm{Doan},~\bfnm{G.~E.}\binits{G.~E.}} \AND
\bauthor{\bsnm{Martell},~\bfnm{D.~L.}\binits{D.~L.}}
(\byear{1974}).
\btitle{{The computer based fire weather information system in Ontario}}.
\bjournal{Forestry Chronicle}
\bvolume{50}
\bpages{149--150}.
\bptok{imsref}%
\end{barticle}
\endbibitem

\bibitem[\protect\citeauthoryear{Duff, Chong and
Tolhurst}{2013}]{DuffEtAl-2013-EnvModellingSoftware}
\begin{barticle}[author]
\bauthor{\bsnm{Duff},~\bfnm{T.~J.}\binits{T.~J.}},
\bauthor{\bsnm{Chong},~\bfnm{D.~M.}\binits{D.~M.}} \AND
\bauthor{\bsnm{Tolhurst},~\bfnm{K.~G.}\binits{K.~G.}}
(\byear{2013}).
\btitle{{Quantifying spatio-temporal differences between fire shapes:
Estimating fire travel paths for the improvement of dynamic spread models}}.
\bjournal{Environmental Modelling and Software}
\bvolume{46}
\bpages{33--43}.
\bptok{imsref}%
\end{barticle}
\endbibitem

\bibitem[\protect\citeauthoryear{Fall and
Lertzman}{1999}]{FallEtAl-1999-EcoSocietyAmerica}
\begin{barticle}[author]
\bauthor{\bsnm{Fall},~\bfnm{J.~G.}\binits{J.~G.}} \AND
\bauthor{\bsnm{Lertzman},~\bfnm{K.~P.}\binits{K.~P.}}
(\byear{1999}).
\btitle{{An interactive tutorial on fire frequency analysis (ver. 3.0)}}.
\bjournal{Bulletin of the Ecological Society of America}
\bvolume{80}
\bpages{174--178}.
\bptok{imsref}%
\end{barticle}
\endbibitem

\bibitem[\protect\citeauthoryear{Finlay et~al.}{2012}]{FinlayEtAl-2012-PLoS}
\begin{barticle}[pbm]
\bauthor{\bsnm{Finlay},~\bfnm{Sarah~Elise}\binits{S.~E.}},
\bauthor{\bsnm{Moffat},~\bfnm{Andrew}\binits{A.}},
\bauthor{\bsnm{Gazzard},~\bfnm{Rob}\binits{R.}},
\bauthor{\bsnm{Baker},~\bfnm{David}\binits{D.}} \AND
\bauthor{\bsnm{Murray},~\bfnm{Virginia}\binits{V.}}
(\byear{2012}).
\btitle{Health impacts of wildfires}.
\bjournal{PLoS Curr.}
\bvolume{4}
\bpages{e4f959951cce2c}.
\bid{doi={10.1371/4f959951cce2c}, issn={2157-3999}, pmcid={3492003},
pmid={23145351}}
\bptok{imsref}%
\end{barticle}
\endbibitem

\bibitem[\protect\citeauthoryear{Finney}{1995}]{Finney-1995-IJWF}
\begin{barticle}[author]
\bauthor{\bsnm{Finney},~\bfnm{M.}\binits{M.}}
(\byear{1995}).
\btitle{{The missing tail and other considerations for the use of fire history
models}}.
\bjournal{International Journal of Wildland Fire}
\bvolume{5}
\bpages{197--202}.
\bptok{imsref}%
\end{barticle}
\endbibitem

\bibitem[\protect\citeauthoryear{Finney}{1998}]{Finney-1998-FARSITE}
\begin{bmisc}[author]
\bauthor{\bsnm{Finney},~\bfnm{M.~A.}\binits{M.~A.}}
(\byear{1998}).
\bhowpublished{\textit{FARSITE, Fire Area Simulator---Model Development and
Evaluation}. US Dept. Agriculture, Forest Service, Rocky Mountain Research
Station, Ogden, UT}.
\bptok{imsref}%
\end{bmisc}
\endbibitem

\bibitem[\protect\citeauthoryear{Finney}{2000}]{Finney-2000-EnvManagement}
\begin{bbook}[author]
\bauthor{\bsnm{Finney},~\bfnm{M.~A.}\binits{M.~A.}}
(\byear{2000}).
\btitle{{Efforts at Comparing Simulated and Observed Fire Growth Patterns.}}
\bpublisher{Systems for Environmental Management}, \blocation{Missoula, MT}.
\bptok{imsref}%
\end{bbook}
\endbibitem

\bibitem[\protect\citeauthoryear{Finney}{2002}]{Finney-2002-CJFR}
\begin{barticle}[author]
\bauthor{\bsnm{Finney},~\bfnm{M.~A.}\binits{M.~A.}}
(\byear{2002}).
\btitle{{Fire growth using minimum travel time methods}}.
\bjournal{Canadian Journal of Forest Research}
\bvolume{32}
\bpages{1420--1424}.
\bptok{imsref}%
\end{barticle}
\endbibitem

\bibitem[\protect\citeauthoryear{Finney
et~al.}{2011a}]{FinneyEtAl-2011-EnvModelling}
\begin{barticle}[author]
\bauthor{\bsnm{Finney},~\bfnm{M.~A.}\binits{M.~A.}},
\bauthor{\bsnm{Grenfell},~\bfnm{I.~C.}\binits{I.~C.}},
\bauthor{\bsnm{McHugh},~\bfnm{C.~W.}\binits{C.~W.}},
\bauthor{\bsnm{Seli},~\bfnm{R.~C.}\binits{R.~C.}},
\bauthor{\bsnm{Trethewey},~\bfnm{D.}\binits{D.}},
\bauthor{\bsnm{Stratton},~\bfnm{R.~D.}\binits{R.~D.}} \AND
\bauthor{\bsnm{Brittain},~\bfnm{S.}\binits{S.}}
(\byear{2011}a).
\btitle{{A method for ensemble wildland fire simlation}}.
\bjournal{Environmental Modeling and Assessment}
\bvolume{16}
\bpages{153--167}.
\bptok{imsref}%
\end{barticle}
\endbibitem

\bibitem[\protect\citeauthoryear{Finney
et~al.}{2011b}]{FinneyEtAl-2011-StochaticEnvResearch}
\begin{barticle}[author]
\bauthor{\bsnm{Finney},~\bfnm{M.~A.}\binits{M.~A.}},
\bauthor{\bsnm{McHugh},~\bfnm{C.~W.}\binits{C.~W.}},
\bauthor{\bsnm{Grenfell},~\bfnm{I.~C.}\binits{I.~C.}},
\bauthor{\bsnm{Riley},~\bfnm{K.~L.}\binits{K.~L.}} \AND
\bauthor{\bsnm{Short},~\bfnm{K.~C.}\binits{K.~C.}}
(\byear{2011}b).
\btitle{{A simulation of probabilistic wildfire risk components for the
continental United States}}.
\bjournal{Stoch. Environ. Res. Risk Assess.}
\bvolume{25}
\bpages{973--1000}.
\bptok{imsref}%
\end{barticle}
\endbibitem

\bibitem[\protect\citeauthoryear{Flannigan and
Haar}{1986}]{FlanniganEtAl-1986-CJFR}
\begin{barticle}[author]
\bauthor{\bsnm{Flannigan},~\bfnm{M.~D.}\binits{M.~D.}} \AND
\bauthor{\bsnm{Haar},~\bfnm{T.~H.~V.}\binits{T.~H.~V.}}
(\byear{1986}).
\btitle{{Forest fire monitoring using NOAA satellite AVHRR}}.
\bjournal{Canadian Journal of Forest Research}
\bvolume{16}
\bpages{975--982}.
\bptok{imsref}%
\end{barticle}
\endbibitem

\bibitem[\protect\citeauthoryear{Flannigan
et~al.}{2009}]{FlanniganEtAl-2009-IJWF}
\begin{barticle}[author]
\bauthor{\bsnm{Flannigan},~\bfnm{M.~D.}\binits{M.~D.}},
\bauthor{\bsnm{Krawchuk},~\bfnm{M.~A.}\binits{M.~A.}},
\bauthor{\bparticle{de} \bsnm{Groot},~\bfnm{W.~J.}\binits{W.~J.}},
\bauthor{\bsnm{Wotton},~\bfnm{B.~M.}\binits{B.~M.}} \AND
\bauthor{\bsnm{Gowman},~\bfnm{L.~M.}\binits{L.~M.}}
(\byear{2009}).
\btitle{{Implications of changing climate for global wildland fire}}.
\bjournal{International Journal of Wildland Fire}
\bvolume{18}
\bpages{483--507}.
\bptok{imsref}%
\end{barticle}
\endbibitem

\bibitem[\protect\citeauthoryear{Fons}{1946}]{Fons-1946-AgriResearch}
\begin{barticle}[author]
\bauthor{\bsnm{Fons},~\bfnm{W.~L.}\binits{W.~L.}}
(\byear{1946}).
\btitle{{Analysis of fire spread in light forest fuels}}.
\bjournal{Journal of Agriculture Research}
\bvolume{72}
\bpages{93--121}.
\bptok{imsref}%
\end{barticle}
\endbibitem

\bibitem[\protect\citeauthoryear{Forestry Canada Fire Danger
Group}{1992}]{ForestryResearchGroup-1992}
\begin{bmisc}[author]
\borganization{Forestry Canada Fire Danger Group}
(\byear{1992}).
\bhowpublished{Development and structure of the Canadian forest fire behavior
prediction system. Forestry Canada, Ottawa, ON}.
\bptok{imsref}%
\end{bmisc}
\endbibitem

\bibitem[\protect\citeauthoryear{Frandsen}{1997}]{Frandsen-1997-CJFR}
\begin{barticle}[author]
\bauthor{\bsnm{Frandsen},~\bfnm{W.~H.}\binits{W.~H.}}
(\byear{1997}).
\btitle{{Ignition probability of organic soils}}.
\bjournal{Canadian Journal of Forest Research}
\bvolume{27}
\bpages{1471--1477}.
\bptok{imsref}%
\end{barticle}
\endbibitem

\bibitem[\protect\citeauthoryear{Fujioka}{2002}]{Fujioka-2002-IJWF}
\begin{barticle}[author]
\bauthor{\bsnm{Fujioka},~\bfnm{F.~M.}\binits{F.~M.}}
(\byear{2002}).
\btitle{{A new method for the analysis of fire spread modeling errors}}.
\bjournal{International Journal of Wildland Fire}
\bvolume{11}
\bpages{193--203}.
\bptok{imsref}%
\end{barticle}
\endbibitem

\bibitem[\protect\citeauthoryear{Fujioka
et~al.}{2008}]{FujiokaEtAl-2008-EnvSci}
\begin{barticle}[author]
\bauthor{\bsnm{Fujioka},~\bfnm{F.~M.}\binits{F.~M.}},
\bauthor{\bsnm{Gill},~\bfnm{A.~M.}\binits{A.~M.}},
\bauthor{\bsnm{Viegas},~\bfnm{D.~X.}\binits{D.~X.}} \AND
\bauthor{\bsnm{Wotton},~\bfnm{B.~M.}\binits{B.~M.}}
(\byear{2008}).
\btitle{{Fire danger and fire behavior modeling systems in Australia, Europe,
and North America}}.
\bjournal{Developments in Environmental Science}
\bvolume{11}
\bpages{471--497}.
\bptok{imsref}%
\end{barticle}
\endbibitem

\bibitem[\protect\citeauthoryear{Garcia et~al.}{1995}]{GarciaEtAl-1995-IJWF}
\begin{barticle}[author]
\bauthor{\bsnm{Garcia},~\bfnm{C.~V.}\binits{C.~V.}},
\bauthor{\bsnm{Woodard},~\bfnm{P.}\binits{P.}},
\bauthor{\bsnm{Titus},~\bfnm{S.}\binits{S.}},
\bauthor{\bsnm{Adamowicz},~\bfnm{W.}\binits{W.}} \AND
\bauthor{\bsnm{Lee},~\bfnm{B.}\binits{B.}}
(\byear{1995}).
\btitle{{A logit model for predicting the daily occurrence of human caused
forest-fires}}.
\bjournal{International Journal of Wildland Fire}
\bvolume{5}
\bpages{101--111}.
\bptok{imsref}%
\end{barticle}
\endbibitem

\bibitem[\protect\citeauthoryear{Giglio et~al.}{2009}]{Giglio-2009-Biogeosci}
\begin{barticle}[author]
\bauthor{\bsnm{Giglio},~\bfnm{L.}\binits{L.}},
\bauthor{\bsnm{Randerson},~\bfnm{J.}\binits{J.}},
\bauthor{\bparticle{van~der} \bsnm{Werf},~\bfnm{G.}\binits{G.}},
\bauthor{\bsnm{Kasibhatla},~\bfnm{P.}\binits{P.}},
\bauthor{\bsnm{Collatz},~\bfnm{G.}\binits{G.}},
\bauthor{\bsnm{Morton},~\bfnm{D.}\binits{D.}} \AND
\bauthor{\bsnm{DeFries},~\bfnm{R.}\binits{R.}}
(\byear{2009}).
\btitle{{Assessing variability and long-terms trends in burned area by merging
multiple satellite fire products}}.
\bjournal{Biogeosciences Discussions}
\bvolume{6}
\bpages{11577--11622}.
\bptok{imsref}%
\end{barticle}
\endbibitem

\bibitem[\protect\citeauthoryear{Gisborne}{1927}]{Gisborne-1927-WeatherReview}
\begin{barticle}[author]
\bauthor{\bsnm{Gisborne},~\bfnm{H.}\binits{H.}}
(\byear{1927}).
\btitle{{Meteorological factors in the Quartz Creek forest fire}}.
\bjournal{Monthly Weather Review}
\bvolume{55}
\bpages{56--60}.
\bptok{imsref}%
\end{barticle}
\endbibitem

\bibitem[\protect\citeauthoryear{Greenville
et~al.}{2009}]{GreenvilleEtAl-2009-IJWF}
\begin{barticle}[author]
\bauthor{\bsnm{Greenville},~\bfnm{A.~C.}\binits{A.~C.}},
\bauthor{\bsnm{Dickman},~\bfnm{C.~R.}\binits{C.~R.}},
\bauthor{\bsnm{Wardle},~\bfnm{G.~M.}\binits{G.~M.}} \AND
\bauthor{\bsnm{Letnic},~\bfnm{M.}\binits{M.}}
(\byear{2009}).
\btitle{{The fire history of an arid grassland: The influence of antecedent
rainfall and ENSO}}.
\bjournal{International Journal of Wildland Fire}
\bvolume{18}
\bpages{631--639}.
\bptok{imsref}%
\end{barticle}
\endbibitem

\bibitem[\protect\citeauthoryear{Grissino-Mayer}{1999}]{GrissinoMayer-1999-IJW%
F}
\begin{barticle}[author]
\bauthor{\bsnm{Grissino-Mayer},~\bfnm{H.~D.}\binits{H.~D.}}
(\byear{1999}).
\btitle{{Modeling fire interval data from the American Southwest with the
Weibull distribution}}.
\bjournal{International Journal of Wildland Fire}
\bvolume{9}
\bpages{37--50}.
\bptok{imsref}%
\end{barticle}
\endbibitem

\bibitem[\protect\citeauthoryear{Haines and
Kuehnast}{1970}]{HainesEtAl-1970-Weatherwise}
\begin{barticle}[author]
\bauthor{\bsnm{Haines},~\bfnm{D.~A.}\binits{D.~A.}} \AND
\bauthor{\bsnm{Kuehnast},~\bfnm{E.~L.}\binits{E.~L.}}
(\byear{1970}).
\btitle{{When the Midwest burned}}.
\bjournal{Weatherwise}
\bvolume{23}
\bpages{112--119}.
\bptok{imsref}%
\end{barticle}
\endbibitem

\bibitem[\protect\citeauthoryear{Hansen
et~al.}{2006}]{HansenEtAl-2006-Proceedings}
\begin{binproceedings}[author]
\bauthor{\bsnm{Hansen},~\bfnm{J.}\binits{J.}},
\bauthor{\bsnm{Sato},~\bfnm{M.}\binits{M.}},
\bauthor{\bsnm{Ruedy},~\bfnm{R.}\binits{R.}},
\bauthor{\bsnm{Lo},~\bfnm{K.}\binits{K.}},
\bauthor{\bsnm{Lea},~\bfnm{D.~W.}\binits{D.~W.}} \AND
\bauthor{\bsnm{Medina-Elizade},~\bfnm{M.}\binits{M.}}
(\byear{2006}).
\btitle{{Global temperature change}}.
\bbooktitle{Proc. Natl. Acad. Sci. USA}
\bvolume{103}
\bpages{14288--14293}.
\bptok{imsref}%
\end{binproceedings}
\endbibitem

\bibitem[\protect\citeauthoryear{Hardy and Hardy}{2007}]{HardyEtAl-2007-IJWF}
\begin{barticle}[author]
\bauthor{\bsnm{Hardy},~\bfnm{C.~C.}\binits{C.~C.}} \AND
\bauthor{\bsnm{Hardy},~\bfnm{C.~E.}\binits{C.~E.}}
(\byear{2007}).
\btitle{{Fire danger rating in the United States of America: An evolution since
1916}}.
\bjournal{International Journal of Wildland Fire}
\bvolume{16}
\bpages{217--231}.
\bptok{imsref}%
\end{barticle}
\endbibitem

\bibitem[\protect\citeauthoryear{Hastie, Tibshirani and
Friedman}{2009}]{Hastie-dataMining}
\begin{bbook}[mr]
\bauthor{\bsnm{Hastie},~\bfnm{Trevor}\binits{T.}},
\bauthor{\bsnm{Tibshirani},~\bfnm{Robert}\binits{R.}} \AND
\bauthor{\bsnm{Friedman},~\bfnm{Jerome}\binits{J.}}
(\byear{2009}).
\btitle{The Elements of Statistical Learning: Data Mining, Inference, and
Prediction},
\bedition{2nd} ed.
\bpublisher{Springer}, \blocation{New York}.
\bid{doi={10.1007/978-0-387-84858-7}, mr={2722294}}
\bptok{imsref}%
\end{bbook}
\endbibitem

\bibitem[\protect\citeauthoryear{Heinselman}{1973}]{Heinselman-1973-QuatRes}
\begin{barticle}[author]
\bauthor{\bsnm{Heinselman},~\bfnm{M.~L.}\binits{M.~L.}}
(\byear{1973}).
\btitle{{Fire in the virgin forests of the Boundary Waters Canoe Area,
Minnesota}}.
\bjournal{Quaternary Research}
\bvolume{3}
\bpages{329--382}.
\bptok{imsref}%
\end{barticle}
\endbibitem

\bibitem[\protect\citeauthoryear{Higuera et~al.}{2011}]{HigueraEtAl-2011-IJWF}
\begin{barticle}[author]
\bauthor{\bsnm{Higuera},~\bfnm{P.~E.}\binits{P.~E.}},
\bauthor{\bsnm{Gavin},~\bfnm{D.~G.}\binits{D.~G.}},
\bauthor{\bsnm{Bartlein},~\bfnm{P.~J.}\binits{P.~J.}} \AND
\bauthor{\bsnm{Hallett},~\bfnm{D.~J.}\binits{D.~J.}}
(\byear{2011}).
\btitle{{Peak detection in sediment-charcoal records: Impacts of alternative
data analysis methods on fire-history interpretations}}.
\bjournal{International Journal of Wildland Fire}
\bvolume{19}
\bpages{996--1014}.
\bptok{imsref}%
\end{barticle}
\endbibitem

\bibitem[\protect\citeauthoryear{Holmes, Hugget and
Westerling}{2008}]{HolmesEtAl-2008-EcoOfForestDist}
\begin{binproceedings}[author]
\bauthor{\bsnm{Holmes},~\bfnm{T.~P.}\binits{T.~P.}},
\bauthor{\bsnm{Hugget},~\bfnm{R.~J.}\binits{R.~J.}} \AND
\bauthor{\bsnm{Westerling},~\bfnm{A.~L.}\binits{A.~L.}}
(\byear{2008}).
\btitle{{Statistical analysis of large wildfires}}.
In \bbooktitle{The Economics of Forest Disturbances}
\bpages{59--77}.
\bpublisher{Springer}, \blocation{New York, NY}.
\bptok{imsref}%
\end{binproceedings}
\endbibitem

\bibitem[\protect\citeauthoryear{Howe}{1915}]{Howe-1915-ForProCan}
\begin{binproceedings}[author]
\bauthor{\bsnm{Howe},~\bfnm{C.~D.}\binits{C.~D.}}
(\byear{1915}).
\btitle{{The effect of repeated forest fires upon the reproduction of
commercial species in Peterborough County, Ontario}}.
In \bbooktitle{Forest Protection in Canada: 1913--1914}
\bpages{162--211}.
\bpublisher{Canada Commission on Conservation}, \blocation{Ottawa, ON}.
\bptok{imsref}%
\end{binproceedings}
\endbibitem

\bibitem[\protect\citeauthoryear{Jiang et~al.}{2009}]{JiangEtAl-2009-IJWF}
\begin{barticle}[author]
\bauthor{\bsnm{Jiang},~\bfnm{Y.}\binits{Y.}},
\bauthor{\bsnm{Zhuang},~\bfnm{Q.}\binits{Q.}},
\bauthor{\bsnm{Flannigan},~\bfnm{M.}\binits{M.}} \AND
\bauthor{\bsnm{Little},~\bfnm{J.~M.}\binits{J.~M.}}
(\byear{2009}).
\btitle{{Characterization of wildfire regimes in Canadian boreal terrestrial
ecosystems}}.
\bjournal{International Journal of Wildland Fire}
\bvolume{18}
\bpages{992--1002}.
\bptok{imsref}%
\end{barticle}
\endbibitem

\bibitem[\protect\citeauthoryear{Johnson and
Gutsell}{1994}]{JohnsonEtAl-1994-EcoRes}
\begin{barticle}[author]
\bauthor{\bsnm{Johnson},~\bfnm{E.~A.}\binits{E.~A.}} \AND
\bauthor{\bsnm{Gutsell},~\bfnm{S.~L.}\binits{S.~L.}}
(\byear{1994}).
\btitle{{Fire frequency models, methods and interpretations}}.
\bjournal{Advances in Ecological Research}
\bvolume{25}
\bpages{239--287}.
\bptok{imsref}%
\end{barticle}
\endbibitem

\bibitem[\protect\citeauthoryear{Jones
et~al.}{2003}]{JonesEtAl-2003-Proceedings}
\begin{binproceedings}[author]
\bauthor{\bsnm{Jones},~\bfnm{C.}\binits{C.}},
\bauthor{\bsnm{Dennison},~\bfnm{P.}\binits{P.}},
\bauthor{\bsnm{Fujioka},~\bfnm{F.}\binits{F.}},
\bauthor{\bsnm{Weise},~\bfnm{D.}\binits{D.}} \AND
\bauthor{\bsnm{Benoit},~\bfnm{J.}\binits{J.}}
(\byear{2003}).
\btitle{{Analysis of space/time characteristics of errors in an integrated
weather/fire spread simulation}}.
In \bbooktitle{Proceedings of the 5th Symposium on Fire and Forest
Meteorology}.
\bpublisher{American Meteorological Society}, \blocation{Orlando, FL}.
\bptok{imsref}%
\end{binproceedings}
\endbibitem

\bibitem[\protect\citeauthoryear{Justice
et~al.}{2002}]{JusticeEtAl-2002-RemoteSensing}
\begin{barticle}[author]
\bauthor{\bsnm{Justice},~\bfnm{C.}\binits{C.}},
\bauthor{\bsnm{Giglio},~\bfnm{L.}\binits{L.}},
\bauthor{\bsnm{Korontzi},~\bfnm{S.}\binits{S.}},
\bauthor{\bsnm{Owens},~\bfnm{J.}\binits{J.}},
\bauthor{\bsnm{Morisette},~\bfnm{J.}\binits{J.}},
\bauthor{\bsnm{Roy},~\bfnm{D.}\binits{D.}},
\bauthor{\bsnm{Descloitres},~\bfnm{J.}\binits{J.}},
\bauthor{\bsnm{Alleaume},~\bfnm{S.}\binits{S.}},
\bauthor{\bsnm{Petitcolin},~\bfnm{F.}\binits{F.}} \AND
\bauthor{\bsnm{Kaufman},~\bfnm{Y.}\binits{Y.}}
(\byear{2002}).
\btitle{{The MODIS fire products}}.
\bjournal{Remote Sensing of Environment}
\bvolume{83}
\bpages{244--262}.
\bptok{imsref}%
\end{barticle}
\endbibitem

\bibitem[\protect\citeauthoryear{Kiil and
Grigel}{1969}]{KiilEtAl-1969-ABfireWx}
\begin{bbook}[author]
\bauthor{\bsnm{Kiil},~\bfnm{A.}\binits{A.}} \AND
\bauthor{\bsnm{Grigel},~\bfnm{J.~E.}\binits{J.~E.}}
(\byear{1969}).
\btitle{{The May 1968 Forest Conflagrations in Central Alberta: A Review of
Fire Weather, Fuels and Fire Behavior}}.
\bpublisher{Canada Dept. Forestry}, \blocation{Calgary, AB}.
\bptok{imsref}%
\end{bbook}
\endbibitem

\bibitem[\protect\citeauthoryear{Kilgore and
Taylor}{1979}]{KilgoreEtAl-1979-Ecology}
\begin{barticle}[author]
\bauthor{\bsnm{Kilgore},~\bfnm{B.~M.}\binits{B.~M.}} \AND
\bauthor{\bsnm{Taylor},~\bfnm{D.}\binits{D.}}
(\byear{1979}).
\btitle{{Fire history of a sequoia-mixed conifer forest}}.
\bjournal{Ecology}
\bvolume{60}
\bpages{129--142}.
\bptok{imsref}%
\end{barticle}
\endbibitem

\bibitem[\protect\citeauthoryear{Kourtz, Nozaki and
O'Regan}{1977}]{KourtzEtAl-1977-ForestFireResInt}
\begin{bbook}[author]
\bauthor{\bsnm{Kourtz},~\bfnm{P.}\binits{P.}},
\bauthor{\bsnm{Nozaki},~\bfnm{S.}\binits{S.}} \AND
\bauthor{\bsnm{O'Regan},~\bfnm{W.~G.}\binits{W.~G.}}
(\byear{1977}).
\btitle{{Forest Fires in the Computer---A Model to Predict the Perimeter
Location of a Forest Fire.}}
\bpublisher{Fish. Environ. Can., Forest Fire Research Institute},
\blocation{Ottawa, ON}.
\bptok{imsref}%
\end{bbook}
\endbibitem

\bibitem[\protect\citeauthoryear{Kourtz and
O'Regan}{1971}]{KourtzEtAl-1971-ForestSci}
\begin{barticle}[author]
\bauthor{\bsnm{Kourtz},~\bfnm{P.}\binits{P.}} \AND
\bauthor{\bsnm{O'Regan},~\bfnm{W.~G.}\binits{W.~G.}}
(\byear{1971}).
\btitle{{A model for a small forest fire\ldots to simulate burned and burning
areas for use in a detection model}}.
\bjournal{Forest Science}
\bvolume{17}
\bpages{163--169}.
\bptok{imsref}%
\end{barticle}
\endbibitem

\bibitem[\protect\citeauthoryear{Kourtz and
Todd}{1991}]{KourtzEtAl-1991-ForestryCan}
\begin{bbook}[author]
\bauthor{\bsnm{Kourtz},~\bfnm{P.}\binits{P.}} \AND
\bauthor{\bsnm{Todd},~\bfnm{B.}\binits{B.}}
(\byear{1991}).
\btitle{{Predicting the Daily Occurrence of Lightning-Caused Forest Fires.}}
\bpublisher{Forestry Canada}, \blocation{Ottawa, ON}.
\bptok{imsref}%
\end{bbook}
\endbibitem

\bibitem[\protect\citeauthoryear{Krawchuk
et~al.}{2009}]{KrawchukEtAl-2009-PlosOne}
\begin{barticle}[pbm]
\bauthor{\bsnm{Krawchuk},~\bfnm{Meg~A.}\binits{M.~A.}},
\bauthor{\bsnm{Moritz},~\bfnm{Max~A.}\binits{M.~A.}},
\bauthor{\bsnm{Parisien},~\bfnm{Marc-Andr{\'{e}}}\binits{M.-A.}},
\bauthor{\bsnm{Dorn},~\bfnm{Jeff~Van}\binits{J.~V.}} \AND
\bauthor{\bsnm{Hayhoe},~\bfnm{Katharine}\binits{K.}}
(\byear{2009}).
\btitle{Global pyrogeography: The current and future distribution of wildfire}.
\bjournal{PLoS ONE}
\bvolume{4}
\bpages{e5102}.
\bid{doi={10.1371/journal.pone.0005102}, issn={1932-6203}, pmcid={2662419},
pmid={19352494}}
\bptok{imsref}%
\end{barticle}
\endbibitem

\bibitem[\protect\citeauthoryear{Krider
et~al.}{1980}]{KriderEtAl-1980-AmMetSoc}
\begin{barticle}[author]
\bauthor{\bsnm{Krider},~\bfnm{E.}\binits{E.}},
\bauthor{\bsnm{Noggle},~\bfnm{R.}\binits{R.}},
\bauthor{\bsnm{Pifer},~\bfnm{A.}\binits{A.}} \AND
\bauthor{\bsnm{Vance},~\bfnm{D.}\binits{D.}}
(\byear{1980}).
\btitle{{Lightning direction-finding systems for forest fire detection}}.
\bjournal{Bulletin of the American Meteorological Society}
\bvolume{61}
\bpages{980--986}.
\bptok{imsref}%
\end{barticle}
\endbibitem

\bibitem[\protect\citeauthoryear{Lawson and
Armitage}{2008}]{LawsonEtAl-2008-CanForServ}
\begin{bbook}[author]
\bauthor{\bsnm{Lawson},~\bfnm{B.~D.}\binits{B.~D.}} \AND
\bauthor{\bsnm{Armitage},~\bfnm{O.}\binits{O.}}
(\byear{2008}).
\btitle{{Weather Guide for the Canadian Forest Fire Danger Rating System.}}
\bpublisher{Nat. Resour. Can., Can. For. Serv., North. For. Cent.},
\blocation{Edmonton, AB}.
\bptok{imsref}%
\end{bbook}
\endbibitem

\bibitem[\protect\citeauthoryear{Lee et~al.}{2002}]{LeeEtAl-2002-CompAndElec}
\begin{barticle}[author]
\bauthor{\bsnm{Lee},~\bfnm{B.}\binits{B.}},
\bauthor{\bsnm{Alexander},~\bfnm{M.}\binits{M.}},
\bauthor{\bsnm{Hawkes},~\bfnm{B.}\binits{B.}},
\bauthor{\bsnm{Lynham},~\bfnm{T.}\binits{T.}},
\bauthor{\bsnm{Stocks},~\bfnm{B.}\binits{B.}} \AND
\bauthor{\bsnm{Englefield},~\bfnm{P.}\binits{P.}}
(\byear{2002}).
\btitle{{Information systems in support of wildland fire management decisions
making in Canada}}.
\bjournal{Computers and Electronics in Agriculture}
\bvolume{37}
\bpages{185--198}.
\bptok{imsref}%
\end{barticle}
\endbibitem

\bibitem[\protect\citeauthoryear{Leonard}{2009}]{Leonard-2009-EnvManag}
\begin{barticle}[pbm]
\bauthor{\bsnm{Leonard},~\bfnm{Steven}\binits{S.}}
(\byear{2009}).
\btitle{Predicting sustained fire spread in Tasmanian native grasslands}.
\bjournal{Environ. Manage.}
\bvolume{44}
\bpages{430--440}.
\bid{doi={10.1007/s00267-009-9340-6}, issn={1432-1009}, pmid={19597866}}
\bptok{imsref}%
\end{barticle}
\endbibitem

\bibitem[\protect\citeauthoryear{Linn et~al.}{2002}]{LinnEtAl-2002-IJWF}
\begin{barticle}[author]
\bauthor{\bsnm{Linn},~\bfnm{R.}\binits{R.}},
\bauthor{\bsnm{Reisner},~\bfnm{J.}\binits{J.}},
\bauthor{\bsnm{Colman},~\bfnm{J.~J.}\binits{J.~J.}} \AND
\bauthor{\bsnm{Winterkamp},~\bfnm{J.}\binits{J.}}
(\byear{2002}).
\btitle{{Studying wildfire behavior using FIRETEC}}.
\bjournal{International Journal of Wildland Fire}
\bvolume{11}
\bpages{233--246}.
\bptok{imsref}%
\end{barticle}
\endbibitem

\bibitem[\protect\citeauthoryear{Linn et~al.}{2012}]{LinnEtAl-2012-CJFR}
\begin{barticle}[author]
\bauthor{\bsnm{Linn},~\bfnm{R.}\binits{R.}},
\bauthor{\bsnm{Anderson},~\bfnm{K.}\binits{K.}},
\bauthor{\bsnm{Winterkamp},~\bfnm{J.}\binits{J.}},
\bauthor{\bsnm{Brooks},~\bfnm{A.}\binits{A.}},
\bauthor{\bsnm{Wotton},~\bfnm{B.~M.}\binits{B.~M.}},
\bauthor{\bsnm{Dupuy},~\bfnm{J.~L.}\binits{J.~L.}},
\bauthor{\bsnm{Pimont},~\bfnm{F.}\binits{F.}} \AND
\bauthor{\bsnm{Edminster},~\bfnm{C.}\binits{C.}}
(\byear{2012}).
\btitle{{Incorporating field wind data into FIRETEC simulations of the
International Crown Fire Modeling Experiment (ICFME): Preliminary lessons
learned}}.
\bjournal{Canadian Journal of Forest Research}
\bvolume{42}
\bpages{879--898}.
\bptok{imsref}%
\end{barticle}
\endbibitem

\bibitem[\protect\citeauthoryear{Magnussen and
Taylor}{2012a}]{MagnussenEtAlb-2012-IJWF}
\begin{barticle}[author]
\bauthor{\bsnm{Magnussen},~\bfnm{S.}\binits{S.}} \AND
\bauthor{\bsnm{Taylor},~\bfnm{S.~W.}\binits{S.~W.}}
(\byear{2012}a).
\btitle{{Prediction of daily lightning- and human-caused fires in British
Columbia}}.
\bjournal{International Journal of Wildland Fire}
\bvolume{21}
\bpages{342--356}.
\bptok{imsref}%
\end{barticle}
\endbibitem

\bibitem[\protect\citeauthoryear{Magnussen and
Taylor}{2012b}]{MagnussenEtAl-2012-IJWF}
\begin{barticle}[author]
\bauthor{\bsnm{Magnussen},~\bfnm{S.}\binits{S.}} \AND
\bauthor{\bsnm{Taylor},~\bfnm{S.~W.}\binits{S.~W.}}
(\byear{2012}b).
\btitle{{Inter- and intra-annual profiles of fire regmines in the managed
forests of Canada and implications for resource sharing}}.
\bjournal{International Journal of Wildland Fire}
\bvolume{21}
\bpages{328--341}.
\bptok{imsref}%
\end{barticle}
\endbibitem

\bibitem[\protect\citeauthoryear{Mahfouf, Brasnett and
Gagnon}{2007}]{MahfoufEtAl-2007-AtmosOcean}
\begin{barticle}[author]
\bauthor{\bsnm{Mahfouf},~\bfnm{J.~F.}\binits{J.~F.}},
\bauthor{\bsnm{Brasnett},~\bfnm{B.}\binits{B.}} \AND
\bauthor{\bsnm{Gagnon},~\bfnm{S.}\binits{S.}}
(\byear{2007}).
\btitle{{A~Canadian precipitation analysis (CaPA) project: Description and
preliminary results}}.
\bjournal{Atmosphere--Ocean}
\bvolume{45}
\bpages{1--17}.
\bptok{imsref}%
\end{barticle}
\endbibitem

\bibitem[\protect\citeauthoryear{Malamud, Millington and
Perry}{2005}]{MalamudEtAl-2005-Proceedings}
\begin{barticle}[author]
\bauthor{\bsnm{Malamud},~\bfnm{B.~D.}\binits{B.~D.}},
\bauthor{\bsnm{Millington},~\bfnm{J.~D.~A.}\binits{J.~D.~A.}} \AND
\bauthor{\bsnm{Perry},~\bfnm{G.~L.~W.}\binits{G.~L.~W.}}
(\byear{2005}).
\btitle{{Characterizing wildfire regimes in the United States}}.
\bjournal{Proc. Natl. Acad. Sci. USA}
\bvolume{102}
\bpages{4694--4699}.
\bptok{imsref}%
\end{barticle}
\endbibitem

\bibitem[\protect\citeauthoryear{Malamud, Morein and
Turcotte}{1998}]{MalamudEtAl-1998-Science}
\begin{barticle}[pbm]
\bauthor{\bsnm{Malamud},~\bfnm{B.~D.}\binits{B.~D.}},
\bauthor{\bsnm{Morein},~\bfnm{G.}\binits{G.}} \AND
\bauthor{\bsnm{Turcotte},~\bfnm{D.~L.}\binits{D.~L.}}
(\byear{1998}).
\btitle{Forest fires: An example of self-organized critical behavior}.
\bjournal{Science}
\bvolume{281}
\bpages{1840--1842}.
\bid{issn={1095-9203}, pmid={9743494}}
\bptok{imsref}%
\end{barticle}
\endbibitem

\bibitem[\protect\citeauthoryear{Marsden-Smedley, Catchpole and
Pyrke}{2001}]{MarsdenSmedley-2001-IJWF}
\begin{barticle}[author]
\bauthor{\bsnm{Marsden-Smedley},~\bfnm{J.~B.}\binits{J.~B.}},
\bauthor{\bsnm{Catchpole},~\bfnm{W.~R.}\binits{W.~R.}} \AND
\bauthor{\bsnm{Pyrke},~\bfnm{A.}\binits{A.}}
(\byear{2001}).
\btitle{{Fire modelling in Tasmanian buttongrass moorlands. IV. Sustaining
versus non-sustaining fires}}.
\bjournal{International Journal of Wildland Fire}
\bvolume{10}
\bpages{255--262}.
\bptok{imsref}%
\end{barticle}
\endbibitem

\bibitem[\protect\citeauthoryear{Martell}{1982}]{Martell-1982-CJFR}
\begin{barticle}[author]
\bauthor{\bsnm{Martell},~\bfnm{D.~L.}\binits{D.~L.}}
(\byear{1982}).
\btitle{{A review of operational research studies in forest fire management}}.
\bjournal{Canadian Journal of Forest Research}
\bvolume{12}
\bpages{119--140}.
\bptok{imsref}%
\end{barticle}
\endbibitem

\bibitem[\protect\citeauthoryear{Martell, Bevilacqua and
Stocks}{1989}]{MartellEtAl-1989-CJFR}
\begin{barticle}[author]
\bauthor{\bsnm{Martell},~\bfnm{D.~L.}\binits{D.~L.}},
\bauthor{\bsnm{Bevilacqua},~\bfnm{E.}\binits{E.}} \AND
\bauthor{\bsnm{Stocks},~\bfnm{B.~J.}\binits{B.~J.}}
(\byear{1989}).
\btitle{{Modelling seaonal variation in daily people-caused forest fire
occurrence in Ontario}}.
\bjournal{Canadian Journal of Forest Research}
\bvolume{19}
\bpages{1555--1563}.
\bptok{imsref}%
\end{barticle}
\endbibitem

\bibitem[\protect\citeauthoryear{Martell, Otukol and
Stocks}{1987}]{MartellEtAl-1987-CJFR}
\begin{barticle}[author]
\bauthor{\bsnm{Martell},~\bfnm{D.~L.}\binits{D.~L.}},
\bauthor{\bsnm{Otukol},~\bfnm{S.}\binits{S.}} \AND
\bauthor{\bsnm{Stocks},~\bfnm{B.~J.}\binits{B.~J.}}
(\byear{1987}).
\btitle{{A logistic model for predicting daily people-caused forest fire
occurrence in Ontario}}.
\bjournal{Canadian Journal of Forest Research}
\bvolume{17}
\bpages{394--401}.
\bptok{imsref}%
\end{barticle}
\endbibitem

\bibitem[\protect\citeauthoryear{Martell and Sun}{2008}]{MartellEtAl-2008-CJFR}
\begin{barticle}[author]
\bauthor{\bsnm{Martell},~\bfnm{D.~L.}\binits{D.~L.}} \AND
\bauthor{\bsnm{Sun},~\bfnm{H.}\binits{H.}}
(\byear{2008}).
\btitle{{The impact of fire suppression, vegetation, and weather on the area
burned by lightning-cause forest fires in Ontario}}.
\bjournal{Canadian Journal of Forest Research}
\bvolume{38}
\bpages{1547--1563}.
\bptok{imsref}%
\end{barticle}
\endbibitem

\bibitem[\protect\citeauthoryear{McBride}{1983}]{Mcbride-1983-TreeRing}
\begin{barticle}[author]
\bauthor{\bsnm{McBride},~\bfnm{J.~R.}\binits{J.~R.}}
(\byear{1983}).
\btitle{{Analysis of tree rings and fire scars to establish fire history}}.
\bjournal{Tree-Ring Bulletin}
\bvolume{43}
\bpages{51--67}.
\bptok{imsref}%
\end{barticle}
\endbibitem

\bibitem[\protect\citeauthoryear{Mell et~al.}{2007}]{MellEtAl-2007-IJWF}
\begin{barticle}[author]
\bauthor{\bsnm{Mell},~\bfnm{W.}\binits{W.}},
\bauthor{\bsnm{Jenkins},~\bfnm{M.~A.}\binits{M.~A.}},
\bauthor{\bsnm{Gould},~\bfnm{J.}\binits{J.}} \AND
\bauthor{\bsnm{Cheney},~\bfnm{P.}\binits{P.}}
(\byear{2007}).
\btitle{{A physics-based approach to modelling grassland fires}}.
\bjournal{International Journal of Wildland Fire}
\bvolume{16}
\bpages{1--22}.
\bptok{imsref}%
\end{barticle}
\endbibitem

\bibitem[\protect\citeauthoryear{Meyn et~al.}{2009}]{MeynEtAl-2009-ClimateBio}
\begin{barticle}[author]
\bauthor{\bsnm{Meyn},~\bfnm{A.}\binits{A.}},
\bauthor{\bsnm{Taylor},~\bfnm{S.~W.}\binits{S.~W.}},
\bauthor{\bsnm{Flannigan},~\bfnm{M.~D.}\binits{M.~D.}},
\bauthor{\bsnm{Thonicke},~\bfnm{K.}\binits{K.}} \AND
\bauthor{\bsnm{Cramer},~\bfnm{W.}\binits{W.}}
(\byear{2009}).
\btitle{{Relationship between fire, climate oscillations, and drought in
British Columbia, Canada, 1920--2000}}.
\bjournal{Global Change Biology}
\bvolume{16}
\bpages{977--989}.
\bptok{imsref}%
\end{barticle}
\endbibitem

\bibitem[\protect\citeauthoryear{Meyn et~al.}{2010}]{MeynEtAl-2010-IJWF}
\begin{barticle}[author]
\bauthor{\bsnm{Meyn},~\bfnm{A.}\binits{A.}},
\bauthor{\bsnm{Schmidtlein},~\bfnm{S.}\binits{S.}},
\bauthor{\bsnm{Taylor},~\bfnm{S.~W.}\binits{S.~W.}},
\bauthor{\bsnm{Girardin},~\bfnm{M.~P.}\binits{M.~P.}},
\bauthor{\bsnm{Thonicke},~\bfnm{K.}\binits{K.}} \AND
\bauthor{\bsnm{Cramer},~\bfnm{W.}\binits{W.}}
(\byear{2010}).
\btitle{{Spatial variation of trends in wildfire and summer drought in British
Columbia, Canada, 1920--2000}}.
\bjournal{International Journal of Wildland Fire}
\bvolume{19}
\bpages{272--283}.
\bptok{imsref}%
\end{barticle}
\endbibitem

\bibitem[\protect\citeauthoryear{Moritz
et~al.}{2005}]{MoritzEtAl-2005-Proceedings}
\begin{barticle}[pbm]
\bauthor{\bsnm{Moritz},~\bfnm{Max~A.}\binits{M.~A.}},
\bauthor{\bsnm{Morais},~\bfnm{Marco~E.}\binits{M.~E.}},
\bauthor{\bsnm{Summerell},~\bfnm{Lora~A.}\binits{L.~A.}},
\bauthor{\bsnm{Carlson},~\bfnm{J.~M.}\binits{J.~M.}} \AND
\bauthor{\bsnm{Doyle},~\bfnm{John}\binits{J.}}
(\byear{2005}).
\btitle{Wildfires, complexity, and highly optimized tolerance}.
\bjournal{Proc. Natl. Acad. Sci. USA}
\bvolume{102}
\bpages{17912--17917}.
\bid{doi={10.1073/pnas.0508985102}, issn={0027-8424}, pii={0508985102},
pmcid={1312407}, pmid={16332964}}
\bptok{imsref}%
\end{barticle}
\endbibitem

\bibitem[\protect\citeauthoryear{Nichols
et~al.}{2011}]{NicholsEtAl-2011-TimeSeries}
\begin{barticle}[mr]
\bauthor{\bsnm{Nichols},~\bfnm{Kevin}\binits{K.}},
\bauthor{\bsnm{Schoenberg},~\bfnm{Frederic~Paik}\binits{F.~P.}},
\bauthor{\bsnm{Keeley},~\bfnm{Jon~E.}\binits{J.~E.}},
\bauthor{\bsnm{Bray},~\bfnm{Andrew}\binits{A.}} \AND
\bauthor{\bsnm{Diez},~\bfnm{David}\binits{D.}}
(\byear{2011}).
\btitle{The application of prototype point processes for the summary and
description of {C}alifornia wildfires}.
\bjournal{J. Time Series Anal.}
\bvolume{32}
\bpages{420--429}.
\bid{doi={10.1111/j.1467-9892.2011.00734.x}, issn={0143-9782}, mr={2857337}}
\bptok{imsref}%
\end{barticle}
\endbibitem

\bibitem[\protect\citeauthoryear{Olsen}{2003}]{Olsen-2003-FireManag}
\begin{barticle}[author]
\bauthor{\bsnm{Olsen},~\bfnm{C.~F.}\binits{C.~F.}}
(\byear{2003}).
\btitle{{An analysis of the Honey Fire}}.
\bjournal{Fire Management Today}
\bvolume{29}
\bpages{28--41}.
\bptok{imsref}%
\end{barticle}
\endbibitem

\bibitem[\protect\citeauthoryear{Palmer
et~al.}{2005}]{PalmerEtAl-2005-EarthPlanetSci}
\begin{bincollection}[mr]
\bauthor{\bsnm{Palmer},~\bfnm{T.~N.}\binits{T.~N.}},
\bauthor{\bsnm{Shutts},~\bfnm{G.~J.}\binits{G.~J.}},
\bauthor{\bsnm{Hagedorn},~\bfnm{R.}\binits{R.}},
\bauthor{\bsnm{Doblas-Reyes},~\bfnm{F.~J.}\binits{F.~J.}},
\bauthor{\bsnm{Jung},~\bfnm{T.}\binits{T.}} \AND
\bauthor{\bsnm{Leutbecher},~\bfnm{M.}\binits{M.}}
(\byear{2005}).
\btitle{Representing model uncertainty in weather and climate prediction}.
\bbooktitle{Annual Review of Earth and Planetary Sciences}
\bvolume{33}
\bpages{163--193}.
\bid{mr={2153320}}
\bptok{imsref}%
\end{bincollection}
\endbibitem

\bibitem[\protect\citeauthoryear{Parisien and
Moritz}{2009}]{ParisienEtAl-2009-EcoMonographs}
\begin{barticle}[author]
\bauthor{\bsnm{Parisien},~\bfnm{M.~A.}\binits{M.~A.}} \AND
\bauthor{\bsnm{Moritz},~\bfnm{M.~A.}\binits{M.~A.}}
(\byear{2009}).
\btitle{{Environmental control on the distribution of wildfire at multiple
spatial scales}}.
\bjournal{Ecological Monographs}
\bvolume{79}
\bpages{127--153}.
\bptok{imsref}%
\end{barticle}
\endbibitem

\bibitem[\protect\citeauthoryear{Parisien
et~al.}{2005}]{ParisienEtAl-2005-NatResourcesCan}
\begin{bbook}[author]
\bauthor{\bsnm{Parisien},~\bfnm{M.~A.}\binits{M.~A.}},
\bauthor{\bsnm{Kafka},~\bfnm{V.}\binits{V.}},
\bauthor{\bsnm{Hirsch},~\bfnm{K.}\binits{K.}},
\bauthor{\bsnm{Todd},~\bfnm{J.}\binits{J.}},
\bauthor{\bsnm{Lavoie},~\bfnm{S.}\binits{S.}} \AND
\bauthor{\bsnm{Maczek},~\bfnm{P.}\binits{P.}}
(\byear{2005}).
\btitle{{Mapping Wildfire Susceptibility with the BURN-P3 Simulation Model.}}
\bpublisher{Nat. Resour. Can., Can. For. Serv., North. For. Cent.},
\blocation{Edmonton, AB}.
\bptok{imsref}%
\end{bbook}
\endbibitem

\bibitem[\protect\citeauthoryear{Parisien
et~al.}{2011}]{ParisienEtAl-2011-EcoApplications}
\begin{barticle}[pbm]
\bauthor{\bsnm{Parisien},~\bfnm{Marc-Andr{\'{e}}}\binits{M.-A.}},
\bauthor{\bsnm{Parks},~\bfnm{Sean~A.}\binits{S.~A.}},
\bauthor{\bsnm{Krawchuk},~\bfnm{Meg~A.}\binits{M.~A.}},
\bauthor{\bsnm{Flannigan},~\bfnm{Mike~D.}\binits{M.~D.}},
\bauthor{\bsnm{Bowman},~\bfnm{Lynn~M.}\binits{L.~M.}} \AND
\bauthor{\bsnm{Moritz},~\bfnm{Max~A.}\binits{M.~A.}}
(\byear{2011}).
\btitle{Scale-dependent controls on the area burned in the boreal forest of
Canada, 1980--2005}.
\bjournal{Ecol. Appl.}
\bvolume{21}
\bpages{789--805}.
\bid{issn={1051-0761}, pmid={21639045}}
\bptok{imsref}%
\end{barticle}
\endbibitem

\bibitem[\protect\citeauthoryear{Parisien
et~al.}{2012}]{ParisienEtAl-2012-IJWF}
\begin{barticle}[author]
\bauthor{\bsnm{Parisien},~\bfnm{M.~A.}\binits{M.~A.}},
\bauthor{\bsnm{Snetsinger},~\bfnm{S.}\binits{S.}},
\bauthor{\bsnm{Greenberg},~\bfnm{J.~A.}\binits{J.~A.}},
\bauthor{\bsnm{Nelson},~\bfnm{C.~R.}\binits{C.~R.}},
\bauthor{\bsnm{Schoennagel},~\bfnm{T.}\binits{T.}},
\bauthor{\bsnm{Dobrowski},~\bfnm{S.~Z.}\binits{S.~Z.}} \AND
\bauthor{\bsnm{Moritz},~\bfnm{M.~A.}\binits{M.~A.}}
(\byear{2012}).
\btitle{{Spatial variabilty in wildfire probability across the western United
States}}.
\bjournal{International Journal of Wildland Fire}
\bvolume{21}
\bpages{313--327}.
\bptok{imsref}%
\end{barticle}
\endbibitem

\bibitem[\protect\citeauthoryear{Pastor
et~al.}{2003}]{PastorEtAl-2003-EnergyAndComb}
\begin{barticle}[author]
\bauthor{\bsnm{Pastor},~\bfnm{E.}\binits{E.}},
\bauthor{\bsnm{Zarate},~\bfnm{L.}\binits{L.}},
\bauthor{\bsnm{Planas},~\bfnm{E.}\binits{E.}} \AND
\bauthor{\bsnm{Arnaldos},~\bfnm{J.}\binits{J.}}
(\byear{2003}).
\btitle{{Mathematical models and calculation systems for the study of wildland
behaviour}}.
\bjournal{Progress in Energy and Combusion Science}
\bvolume{29}
\bpages{139--153}.
\bptok{imsref}%
\end{barticle}
\endbibitem

\bibitem[\protect\citeauthoryear{Plucinski and
Anderson}{2008}]{PlucinskiEtAl-2008-IJWF}
\begin{barticle}[author]
\bauthor{\bsnm{Plucinski},~\bfnm{M.~P.}\binits{M.~P.}} \AND
\bauthor{\bsnm{Anderson},~\bfnm{W.~R.}\binits{W.~R.}}
(\byear{2008}).
\btitle{{Laboratory determination of factors influencing successful point
ignition in the litter layer of shrubland vegetation}}.
\bjournal{International Journal of Wildland Fire}
\bvolume{17}
\bpages{628--637}.
\bptok{imsref}%
\end{barticle}
\endbibitem

\bibitem[\protect\citeauthoryear{Plummer}{1912}]{Plummer-1912-USdeptOfAg}
\begin{bmisc}[author]
\bauthor{\bsnm{Plummer},~\bfnm{F.~G.}\binits{F.~G.}}
(\byear{1912}).
\bhowpublished{\textit{Forest Fires: Their Causes, Extent and Effects, With a Summary of
Recorded Destruction and Loss}.
US Dept. Agriculture, Forest Service, Washington, DC.}
\bptok{imsref}%
\end{bmisc}
\endbibitem

\bibitem[\protect\citeauthoryear{Podur, Martell and
Knight}{2002}]{PodurEtAl-2002-CJFR}
\begin{barticle}[author]
\bauthor{\bsnm{Podur},~\bfnm{J.}\binits{J.}},
\bauthor{\bsnm{Martell},~\bfnm{D.~L.}\binits{D.~L.}} \AND
\bauthor{\bsnm{Knight},~\bfnm{K.}\binits{K.}}
(\byear{2002}).
\btitle{{Statistical quality control analysis of forest fire activity in
Canada}}.
\bjournal{Canadian Journal of Forest Research}
\bvolume{32}
\bpages{195--205}.
\bptok{imsref}%
\end{barticle}
\endbibitem

\bibitem[\protect\citeauthoryear{Podur, Martell and
Stanford}{2010}]{PodurEtAl-2010-Environ}
\begin{barticle}[mr]
\bauthor{\bsnm{Podur},~\bfnm{Justin~J.}\binits{J.~J.}},
\bauthor{\bsnm{Martell},~\bfnm{David~L.}\binits{D.~L.}} \AND
\bauthor{\bsnm{Stanford},~\bfnm{David}\binits{D.}}
(\byear{2010}).
\btitle{A~compound {P}oisson model for the annual area burned by forest fires
in the province of {O}ntario}.
\bjournal{Environmetrics}
\bvolume{21}
\bpages{457--469}.
\bid{doi={10.1002/env.996}, issn={1180-4009}, mr={2842261}}
\bptok{imsref}%
\end{barticle}
\endbibitem

\bibitem[\protect\citeauthoryear{Preisler and
Ager}{2013}]{PreislerEtAl-2013-EncOfEnviron}
\begin{bmisc}[author]
\bauthor{\bsnm{Preisler},~\bfnm{H.~K.}\binits{H.~K.}} \AND
\bauthor{\bsnm{Ager},~\bfnm{A.~A.}\binits{A.~A.}}
(\byear{2013}).
\bhowpublished{Forest-Fire Models. Encyclopedia of Environmetrics.}
\bptok{imsref}%
\end{bmisc}
\endbibitem

\bibitem[\protect\citeauthoryear{Preisler and
Westerling}{2007}]{PreislerEtAl-2007-AppMeteorAndClimat}
\begin{barticle}[author]
\bauthor{\bsnm{Preisler},~\bfnm{H.~K.}\binits{H.~K.}} \AND
\bauthor{\bsnm{Westerling},~\bfnm{A.~L.}\binits{A.~L.}}
(\byear{2007}).
\btitle{{Statistical model for forecasting monthly large wildfire events in
western United States}}.
\bjournal{Journal of Applied Meteorology and Climatology}
\bvolume{46}
\bpages{1020--1030}.
\bptok{imsref}%
\end{barticle}
\endbibitem

\bibitem[\protect\citeauthoryear{Preisler
et~al.}{2004}]{PreislerEtAl-2004-IJWF}
\begin{barticle}[author]
\bauthor{\bsnm{Preisler},~\bfnm{H.~K.}\binits{H.~K.}},
\bauthor{\bsnm{Brillinger},~\bfnm{D.}\binits{D.}},
\bauthor{\bsnm{Burgan},~\bfnm{R.~E.}\binits{R.~E.}} \AND
\bauthor{\bsnm{Benoit},~\bfnm{J.~W.}\binits{J.~W.}}
(\byear{2004}).
\btitle{{Probability based models for estimation of wildfire risk}}.
\bjournal{International Journal of Wildland Fire}
\bvolume{13}
\bpages{133--142}.
\bptok{imsref}%
\end{barticle}
\endbibitem

\bibitem[\protect\citeauthoryear{Preisler
et~al.}{2008}]{PreislerEtAl-2008-IJWF}
\begin{barticle}[author]
\bauthor{\bsnm{Preisler},~\bfnm{H.~K.}\binits{H.~K.}},
\bauthor{\bsnm{Chen},~\bfnm{S.~C.}\binits{S.~C.}},
\bauthor{\bsnm{Fujioka},~\bfnm{F.}\binits{F.}},
\bauthor{\bsnm{Benoit},~\bfnm{J.~W.}\binits{J.~W.}} \AND
\bauthor{\bsnm{Westerling},~\bfnm{A.~L.}\binits{A.~L.}}
(\byear{2008}).
\btitle{{Wildland fire probabilities estimated from weather model-deduced
monthly mean fire danger indices}}.
\bjournal{International Journal of Wildland Fire}
\bvolume{17}
\bpages{305--316}.
\bptok{imsref}%
\end{barticle}
\endbibitem

\bibitem[\protect\citeauthoryear{Preisler
et~al.}{2011}]{PreislerEtAl-2011-IJWF}
\begin{barticle}[author]
\bauthor{\bsnm{Preisler},~\bfnm{H.~K.}\binits{H.~K.}},
\bauthor{\bsnm{Westerling},~\bfnm{A.~L.}\binits{A.~L.}},
\bauthor{\bsnm{Gebert},~\bfnm{K.~M.}\binits{K.~M.}},
\bauthor{\bsnm{Munoz-Arriola},~\bfnm{F.}\binits{F.}} \AND
\bauthor{\bsnm{Holmes},~\bfnm{T.~P.}\binits{T.~P.}}
(\byear{2011}).
\btitle{{Spatially explicit forecasts of large wildland fire probability and
suppression costs for California}}.
\bjournal{International Journal of Wildland Fire}
\bvolume{20}
\bpages{508--517}.
\bptok{imsref}%
\end{barticle}
\endbibitem

\bibitem[\protect\citeauthoryear{Reed}{1994}]{Reed-1994-ForScience}
\begin{barticle}[author]
\bauthor{\bsnm{Reed},~\bfnm{W.~J.}\binits{W.~J.}}
(\byear{1994}).
\btitle{{Estimating the historic probability of stand-replacement fire using
the age--class distribution of undisturbed forest}}.
\bjournal{Forest Science}
\bvolume{40}
\bpages{104--119}.
\bptok{imsref}%
\end{barticle}
\endbibitem

\bibitem[\protect\citeauthoryear{Reed}{2000}]{Reed-2000-CJS}
\begin{barticle}[mr]
\bauthor{\bsnm{Reed},~\bfnm{William~J.}\binits{W.~J.}}
(\byear{2000}).
\btitle{Reconstructing the history of forest fire frequency: Identifying hazard
rate change points using the {B}ayes information criterion}.
\bjournal{Canad. J. Statist.}
\bvolume{28}
\bpages{353--365}.
\bid{doi={10.2307/3315984}, issn={0319-5724}, mr={1791689}}
\bptok{imsref}%
\end{barticle}
\endbibitem

\bibitem[\protect\citeauthoryear{Reed}{2001}]{Reed-2001}
\begin{bincollection}[author]
\bauthor{\bsnm{Reed},~\bfnm{W.~J.}\binits{W.~J.}}
(\byear{2001}).
\btitle{Statistical inference for historical fire frequency using
spatial mosiac. Chapter~12.}
In \bbooktitle{{Forest Fires: Behavior and Ecological Effects}}
(\beditor{\bfnm{E.}\binits{E.}~\bsnm{Johnson}} \AND
\beditor{\bfnm{K.}\binits{K.}~\bsnm{Miyanshi}}, eds.).
\bpublisher{Academic Press}, \blocation{San Diego, CA}.
\bptok{imsref}%
\end{bincollection}
\endbibitem

\bibitem[\protect\citeauthoryear{Reed and Johnson}{2004}]{ReedEtAl-2004-CJFR}
\begin{barticle}[author]
\bauthor{\bsnm{Reed},~\bfnm{W.~J.}\binits{W.~J.}} \AND
\bauthor{\bsnm{Johnson},~\bfnm{E.~A.}\binits{E.~A.}}
(\byear{2004}).
\btitle{{Statistical methods for estimating historical fire frequency from
multiple fire-scar data}}.
\bjournal{Canadian Journal of Forest Research}
\bvolume{34}
\bpages{2306--2313}.
\bptok{imsref}%
\end{barticle}
\endbibitem

\bibitem[\protect\citeauthoryear{Reed and
McKelvey}{2002}]{ReedEtAl-2002-EcoModelling}
\begin{barticle}[author]
\bauthor{\bsnm{Reed},~\bfnm{W.~J.}\binits{W.~J.}} \AND
\bauthor{\bsnm{McKelvey},~\bfnm{K.~S.}\binits{K.~S.}}
(\byear{2002}).
\btitle{{Power-law behaviour and parametric models for the size-distribution of
forest fires}}.
\bjournal{Ecological Modelling}
\bvolume{150}
\bpages{239--254}.
\bptok{imsref}%
\end{barticle}
\endbibitem

\bibitem[\protect\citeauthoryear{Reed et~al.}{1998}]{ReedEtAl-1998-ForScience}
\begin{barticle}[author]
\bauthor{\bsnm{Reed},~\bfnm{W.}\binits{W.}},
\bauthor{\bsnm{Larson},~\bfnm{C.}\binits{C.}},
\bauthor{\bsnm{Johnson},~\bfnm{E.}\binits{E.}} \AND
\bauthor{\bsnm{MacDonald},~\bfnm{G.}\binits{G.}}
(\byear{1998}).
\btitle{{Estimation of temporal variations in historical fire frequency from
time-since-fire map data}}.
\bjournal{Forest Science}
\bvolume{44}
\bpages{465--475}.
\bptok{imsref}%
\end{barticle}
\endbibitem

\bibitem[\protect\citeauthoryear{Richards}{1990}]{Richards-1990-NumMethods}
\begin{barticle}[author]
\bauthor{\bsnm{Richards},~\bfnm{G.~D.}\binits{G.~D.}}
(\byear{1990}).
\btitle{{An elliptical growth model of forest fire fronts and its numerical
solution}}.
\bjournal{Internat. J. Numer. Methods Engrg.}
\bvolume{30}
\bpages{1163--1179}.
\bptok{imsref}%
\end{barticle}
\endbibitem

\bibitem[\protect\citeauthoryear{Richards}{1995}]{Richards-1995-IJWF}
\begin{barticle}[author]
\bauthor{\bsnm{Richards},~\bfnm{G.}\binits{G.}}
(\byear{1995}).
\btitle{{A general mathematical framework for modeling two-dimensional wildland
fire spread}}.
\bjournal{International Journal of Wildland Fire}
\bvolume{5}
\bpages{63--72}.
\bptok{imsref}%
\end{barticle}
\endbibitem

\bibitem[\protect\citeauthoryear{Roberts, Wooster and
Lagoudakis}{2009}]{RobertsEtAl-2009-Biogeosciences}
\begin{barticle}[author]
\bauthor{\bsnm{Roberts},~\bfnm{G.}\binits{G.}},
\bauthor{\bsnm{Wooster},~\bfnm{M.}\binits{M.}} \AND
\bauthor{\bsnm{Lagoudakis},~\bfnm{E.}\binits{E.}}
(\byear{2009}).
\btitle{{Annual and diurnal african biomass burning temporal dynamics}}.
\bjournal{Biogeosciences}
\bvolume{6}
\bpages{849--866}.
\bptok{imsref}%
\end{barticle}
\endbibitem

\bibitem[\protect\citeauthoryear{Robinson}{1872}]{Robinson-1872}
\begin{binproceedings}[author]
\bauthor{\bsnm{Robinson},~\bfnm{C.~D.}\binits{C.~D.}}
(\byear{1872}).
\btitle{{Account of the Great Peshtigo fire of 1871}}.
In \bbooktitle{Report on Forestry to the Commisioner of Agriculture}
(\beditor{\bfnm{F.~B.}\binits{F.~B.}~\bsnm{Hough}}, ed.)
\bpages{231--242}.
\bpublisher{U.S. Government Printing Office}, \blocation{Washington, DC}.
\bptok{imsref}%
\end{binproceedings}
\endbibitem

\bibitem[\protect\citeauthoryear{Rothermel}{1972}]{Rothermel-1972-ForestServic%
e}
\begin{bmisc}[author]
\bauthor{\bsnm{Rothermel},~\bfnm{R.~C.}\binits{R.~C.}}
(\byear{1972}).
\bhowpublished{\textit{A Mathematical Model for Predicting Fire Spread in
Wildland Fuels}. Intermountain Forest \& Range Experiment Station, Forest
Service, US Dept. Agriculture, Washington, DC}.
\bptok{imsref}%
\end{bmisc}
\endbibitem

\bibitem[\protect\citeauthoryear{Rothermel, Anderson and
Forest}{1966}]{RothermelEtAl-1966}
\begin{bmisc}[author]
\bauthor{\bsnm{Rothermel},~\bfnm{R.~C.}\binits{R.~C.}},
\bauthor{\bsnm{Anderson},~\bfnm{H.~E.}\binits{H.~E.}} \AND
\bauthor{\bsnm{Forest},~\bfnm{I.}\binits{I.}}
(\byear{1966}).
\bhowpublished{\textit{Fire Spread Characteristics Determined in the
Laboratory}. Intermountain Forest \& Range Experiment Station, Forest
Service, US Dept. Agriculture, Washington, DC}.
\bptok{imsref}%
\end{bmisc}
\endbibitem

\bibitem[\protect\citeauthoryear{Saito}{2001}]{Saito-2001}
\begin{bincollection}[author]
\bauthor{\bsnm{Saito},~\bfnm{K.}\binits{K.}}
(\byear{2001}).
\btitle{{Flames. Chapter~2}}.
In \bbooktitle{{Forest Fires, Behavior and Ecological Effects}}
(\beditor{\bfnm{K.~M.~E.}\binits{K.~M.~E.}~\bsnm{Johnson}}, ed.).
\bpublisher{Academic Press}, \blocation{San Diego, CA}.
\bptok{imsref}%
\end{bincollection}
\endbibitem

\bibitem[\protect\citeauthoryear{Sauvagnargues-Lesage
et~al.}{2001}]{SauvagnarguesLesage-2001-IJWF}
\begin{barticle}[author]
\bauthor{\bsnm{Sauvagnargues-Lesage},~\bfnm{S.}\binits{S.}},
\bauthor{\bsnm{Dusserre},~\bfnm{G.}\binits{G.}},
\bauthor{\bsnm{Robert},~\bfnm{F.}\binits{F.}},
\bauthor{\bsnm{Dray},~\bfnm{G.}\binits{G.}} \AND
\bauthor{\bsnm{Pearson},~\bfnm{D.}\binits{D.}}
(\byear{2001}).
\btitle{{Experimental validation in Mediterranean shrubs fuel of seven wildland
fire rate of spread models}}.
\bjournal{International Journal of Wildland Fire}
\bvolume{10}
\bpages{15--22}.
\bptok{imsref}%
\end{barticle}
\endbibitem

\bibitem[\protect\citeauthoryear{Schoenberg}{2004}]{Schoenberg-2004-Biometrics}
\begin{barticle}[mr]
\bauthor{\bsnm{Schoenberg},~\bfnm{Frederic~Paik}\binits{F.~P.}}
(\byear{2004}).
\btitle{Testing separability in spatial-temporal marked point processes}.
\bjournal{Biometrics}
\bvolume{60}
\bpages{471--481}.
\bid{doi={10.1111/j.0006-341X.2004.00192.x}, issn={0006-341X}, mr={2066282}}
\bptok{imsref}%
\end{barticle}
\endbibitem

\bibitem[\protect\citeauthoryear{Schoenberg, Peng and
Woods}{2003}]{Schoenberg-2003-Environ}
\begin{barticle}[author]
\bauthor{\bsnm{Schoenberg},~\bfnm{F.~P.}\binits{F.~P.}},
\bauthor{\bsnm{Peng},~\bfnm{R.}\binits{R.}} \AND
\bauthor{\bsnm{Woods},~\bfnm{J.}\binits{J.}}
(\byear{2003}).
\btitle{{On the distribution of wildfire sizes}}.
\bjournal{Environmetrics}
\bvolume{14}
\bpages{583--592}.
\bptok{imsref}%
\end{barticle}
\endbibitem

\bibitem[\protect\citeauthoryear{Show}{1919}]{Show-1919-JournalOfForestry}
\begin{barticle}[author]
\bauthor{\bsnm{Show},~\bfnm{S.}\binits{S.}}
(\byear{1919}).
\btitle{{Climate and forest fires in northern California}}.
\bjournal{Journal of Forestry}
\bvolume{17}
\bpages{965--979}.
\bptok{imsref}%
\end{barticle}
\endbibitem

\bibitem[\protect\citeauthoryear{Show and Kotok}{1923}]{ShowEtAl-1923}
\begin{bmisc}[author]
\bauthor{\bsnm{Show},~\bfnm{S.~B.}\binits{S.~B.}} \AND
\bauthor{\bsnm{Kotok},~\bfnm{E.~I.}\binits{E.~I.}}
(\byear{1923}).
\bhowpublished{Forest fires in California 1911--1920: An analytical study.
Department Circular 243, United States Department of Agriculture,
Washington, DC.}
\bptok{imsref}%
\end{bmisc}
\endbibitem

\bibitem[\protect\citeauthoryear{Show
et~al.}{1941}]{ShowEtAl-1941-FireControlNotes}
\begin{barticle}[author]
\bauthor{\bsnm{Show},~\bfnm{S.}\binits{S.}},
\bauthor{\bsnm{Abell},~\bfnm{C.}\binits{C.}},
\bauthor{\bsnm{Deering},~\bfnm{R.}\binits{R.}} \AND
\bauthor{\bsnm{Hanson},~\bfnm{P.}\binits{P.}}
(\byear{1941}).
\btitle{{A~planning basis for adequate fire control on the southern California
national forests}}.
\bjournal{Fire Control Notes}
\bvolume{5}
\bpages{1--59}.
\bptok{imsref}%
\end{barticle}
\endbibitem

\bibitem[\protect\citeauthoryear{Simard}{1991}]{Simard-1991-IJWF}
\begin{barticle}[author]
\bauthor{\bsnm{Simard},~\bfnm{A.~J.}\binits{A.~J.}}
(\byear{1991}).
\btitle{{Fire severity, changing scales, and how things hang together}}.
\bjournal{International Journal of Wildland Fire}
\bvolume{1}
\bpages{23--34}.
\bptok{imsref}%
\end{barticle}
\endbibitem

\bibitem[\protect\citeauthoryear{Stocks, Alexander and
Lanoville}{2004}]{StocksEtAl-2004-CJFR}
\begin{barticle}[author]
\bauthor{\bsnm{Stocks},~\bfnm{M.}\binits{M.}},
\bauthor{\bsnm{Alexander},~\bfnm{M.}\binits{M.}} \AND
\bauthor{\bsnm{Lanoville},~\bfnm{R.}\binits{R.}}
(\byear{2004}).
\btitle{{Overview of the International Crown Fire Modelling Experiment
(ICFME)}}.
\bjournal{Canadian Journal of Forest Research}
\bvolume{34}
\bpages{1543--1547}.
\bptok{imsref}%
\end{barticle}
\endbibitem

\bibitem[\protect\citeauthoryear{Sullivan}{2009a}]{Sullivan-2009a-IJWF}
\begin{barticle}[author]
\bauthor{\bsnm{Sullivan},~\bfnm{A.~L.}\binits{A.~L.}}
(\byear{2009}a).
\btitle{{Wildland surface fire spread modelling, 1990--2007. 1: Physical and
quasi-physical models}}.
\bjournal{International Journal of Wildland Fire}
\bvolume{18}
\bpages{349--368}.
\bptok{imsref}%
\end{barticle}
\endbibitem

\bibitem[\protect\citeauthoryear{Sullivan}{2009b}]{Sullivan-2009b-IJWF}
\begin{barticle}[author]
\bauthor{\bsnm{Sullivan},~\bfnm{A.~L.}\binits{A.~L.}}
(\byear{2009}b).
\btitle{{Wildland surface fire spread modelling, 1990--2007. 2: Empirical and
quasi-empirical models}}.
\bjournal{International Journal of Wildland Fire}
\bvolume{18}
\bpages{369--386}.
\bptok{imsref}%
\end{barticle}
\endbibitem

\bibitem[\protect\citeauthoryear{Sullivan}{2009c}]{Sullivan-2009c-IJWF}
\begin{barticle}[author]
\bauthor{\bsnm{Sullivan},~\bfnm{A.~L.}\binits{A.~L.}}
(\byear{2009}c).
\btitle{{Wildland surface fire spread modelling, 1990--2007. 3: Simulation and
mathematical analogue models}}.
\bjournal{International Journal of Wildland Fire}
\bvolume{18}
\bpages{387--403}.
\bptok{imsref}%
\end{barticle}
\endbibitem

\bibitem[\protect\citeauthoryear{Sullivan and
Knight}{2001}]{Sullivan-2001-CJFR}
\begin{barticle}[author]
\bauthor{\bsnm{Sullivan},~\bfnm{A.}\binits{A.}} \AND
\bauthor{\bsnm{Knight},~\bfnm{I.}\binits{I.}}
(\byear{2001}).
\btitle{{Estimating the error in wind speed measurements for experimental
fires}}.
\bjournal{Canadian Journal of Forest Research}
\bvolume{31}
\bpages{401--409}.
\bptok{imsref}%
\end{barticle}
\endbibitem

\bibitem[\protect\citeauthoryear{Svetsov}{2002}]{Svetsov-2002-3P}
\begin{barticle}[author]
\bauthor{\bsnm{Svetsov},~\bfnm{V.~V.}\binits{V.~V.}}
(\byear{2002}).
\btitle{{Comment on ``Extraterrestrial impacts and wildfires.''}}
\bjournal{Palaeogeography, Palaeoclimatology, Palaeoecology}
\bvolume{185}
\bpages{403--405}.
\bptok{imsref}%
\end{barticle}
\endbibitem

\bibitem[\protect\citeauthoryear{Swain}{1973}]{Swain-1973-QuatResearch}
\begin{barticle}[author]
\bauthor{\bsnm{Swain},~\bfnm{A.~M.}\binits{A.~M.}}
(\byear{1973}).
\btitle{{A history of fire and vegetation in northeastern Minnisota as recorded
in lake sediments}}.
\bjournal{Quaternary Research}
\bvolume{3}
\bpages{383--396}.
\bptok{imsref}%
\end{barticle}
\endbibitem

\bibitem[\protect\citeauthoryear{Taylor and
Alexander}{2006}]{TaylorEtAl-2006-IJWF}
\begin{barticle}[author]
\bauthor{\bsnm{Taylor},~\bfnm{S.~W.}\binits{S.~W.}} \AND
\bauthor{\bsnm{Alexander},~\bfnm{M.~E.}\binits{M.~E.}}
(\byear{2006}).
\btitle{{Science, technology, and human factors in fire danger rating: The
Canadian experience}}.
\bjournal{International Journal of Wildland Fire}
\bvolume{15}
\bpages{121--135}.
\bptok{imsref}%
\end{barticle}
\endbibitem

\bibitem[\protect\citeauthoryear{Todd and
Kourtz}{1991}]{ToddEtAl-1991-ForestryCan}
\begin{bbook}[author]
\bauthor{\bsnm{Todd},~\bfnm{B.}\binits{B.}} \AND
\bauthor{\bsnm{Kourtz},~\bfnm{P.~H.}\binits{P.~H.}}
(\byear{1991}).
\btitle{Predicting the Daily Occurence of People-Caused Forest Fires}.
\bpublisher{Forestry Canada}, \blocation{Chalk River, Ontario}.
\bptok{imsref}%
\end{bbook}
\endbibitem

\bibitem[\protect\citeauthoryear{Toth
et~al.}{2005}]{TothEtAl-2005-GeophysicalResearch}
\begin{barticle}[author]
\bauthor{\bsnm{Toth},~\bfnm{Z.}\binits{Z.}},
\bauthor{\bsnm{Desmarais},~\bfnm{J.~G.}\binits{J.~G.}},
\bauthor{\bsnm{Brunet},~\bfnm{G.}\binits{G.}},
\bauthor{\bsnm{Zhu},~\bfnm{Y.}\binits{Y.}},
\bauthor{\bsnm{Verret},~\bfnm{R.}\binits{R.}},
\bauthor{\bsnm{Wobus},~\bfnm{R.}\binits{R.}},
\bauthor{\bsnm{Hogue},~\bfnm{R.}\binits{R.}} \AND
\bauthor{\bsnm{Cui},~\bfnm{B.}\binits{B.}}
(\byear{2005}).
\btitle{{The North American Ensemble Forecast System (NAEFS)}}.
\bjournal{Geophysical Research Abstracts}
\bvolume{7}
\bpages{02501}.
\bptok{imsref}%
\end{barticle}
\endbibitem

\bibitem[\protect\citeauthoryear{Turner}{2009}]{Turner-2009-EnvEcoStats}
\begin{barticle}[mr]
\bauthor{\bsnm{Turner},~\bfnm{Rolf}\binits{R.}}
(\byear{2009}).
\btitle{Point pattern of forest fire locations}.
\bjournal{Environ. Ecol. Stat.}
\bvolume{16}
\bpages{197--223}.
\bid{doi={10.1007/s10651-007-0085-1}, issn={1352-8505}, mr={2668733}}
\bptok{imsref}%
\end{barticle}
\endbibitem

\bibitem[\protect\citeauthoryear{Tymstra
et~al.}{2010}]{TymstraEtAl-2010-NatResourcesCan}
\begin{bbook}[author]
\bauthor{\bsnm{Tymstra},~\bfnm{C.}\binits{C.}},
\bauthor{\bsnm{Bryce},~\bfnm{R.}\binits{R.}},
\bauthor{\bsnm{Wotton},~\bfnm{B.~M.}\binits{B.~M.}},
\bauthor{\bsnm{Taylor},~\bfnm{S.~W.}\binits{S.~W.}} \AND
\bauthor{\bsnm{Armitage},~\bfnm{O.}\binits{O.}}
(\byear{2010}).
\btitle{{Development and Structure of Prometheus: The Canadian Wildland Fire
Growth Simulation Model}}.
\bpublisher{Nat. Resour. Can., Can. For. Serv., North. For. Cent.},
\blocation{Edmonton, AB}.
\bptok{imsref}%
\end{bbook}
\endbibitem

\bibitem[\protect\citeauthoryear{Van~Wagner}{1969}]{VanWagner-1969-ForestryChr%
on}
\begin{barticle}[author]
\bauthor{\bsnm{Van~Wagner},~\bfnm{C.~E.}\binits{C.~E.}}
(\byear{1969}).
\btitle{{A simple fire-growth model}}.
\bjournal{Forestry Chronicle}
\bvolume{45}
\bpages{103--104}.
\bptok{imsref}%
\end{barticle}
\endbibitem

\bibitem[\protect\citeauthoryear{Van~Wagner}{1977}]{VanWagner-1977-CJFR}
\begin{barticle}[author]
\bauthor{\bsnm{Van~Wagner},~\bfnm{C.~E.}\binits{C.~E.}}
(\byear{1977}).
\btitle{{Conditions for the start and spread of crown fire}}.
\bjournal{Canadian Journal of Forest Research}
\bvolume{7}
\bpages{23--34}.
\bptok{imsref}%
\end{barticle}
\endbibitem

\bibitem[\protect\citeauthoryear{Van~Wagner}{1978}]{VanWagner-1978-CJFR}
\begin{barticle}[author]
\bauthor{\bsnm{Van~Wagner},~\bfnm{C.~E.}\binits{C.~E.}}
(\byear{1978}).
\btitle{{Age-class distribution and the forest fire cycle}}.
\bjournal{Canadian Journal of Forest Research}
\bvolume{8}
\bpages{220--227}.
\bptok{imsref}%
\end{barticle}
\endbibitem

\bibitem[\protect\citeauthoryear{Van~Wagner}{1987}]{VanWagner-1987-CanForestSe%
rv}
\begin{bbook}[author]
\bauthor{\bsnm{Van~Wagner},~\bfnm{C.~E.}\binits{C.~E.}}
(\byear{1987}).
\btitle{{Development and Structure of the Canadian Forest Fire Weather Index
System}}.
\bpublisher{Canadian Forest Service}, \blocation{Ottawa}.
\bptok{imsref}%
\end{bbook}
\endbibitem

\bibitem[\protect\citeauthoryear{Viegas et~al.}{1999}]{ViegasEtAl-1999-IJWF}
\begin{barticle}[author]
\bauthor{\bsnm{Viegas},~\bfnm{D.~X.}\binits{D.~X.}},
\bauthor{\bsnm{Bovio},~\bfnm{G.}\binits{G.}},
\bauthor{\bsnm{Ferreira},~\bfnm{A.}\binits{A.}},
\bauthor{\bsnm{Nosenzo},~\bfnm{A.}\binits{A.}} \AND
\bauthor{\bsnm{Sol},~\bfnm{B.}\binits{B.}}
(\byear{1999}).
\btitle{{Comparative study of various mthods of fire danger evaluation in
sourthern Europe}}.
\bjournal{International Journal of Wildland Fire}
\bvolume{9}
\bpages{235--246}.
\bptok{imsref}%
\end{barticle}
\endbibitem

\bibitem[\protect\citeauthoryear{Vilar et~al.}{2010}]{VilarEtAl-2010-IJWF}
\begin{barticle}[author]
\bauthor{\bsnm{Vilar},~\bfnm{L.}\binits{L.}},
\bauthor{\bsnm{Woolford},~\bfnm{D.~G.}\binits{D.~G.}},
\bauthor{\bsnm{Martell},~\bfnm{D.~L.}\binits{D.~L.}} \AND
\bauthor{\bsnm{Martn},~\bfnm{M.~P.}\binits{M.~P.}}
(\byear{2010}).
\btitle{{A model for predicting human-caused wildfire occurrence in the region
of Madrid, Spain}}.
\bjournal{International Journal of Wildland Fire}
\bvolume{19}
\bpages{325--337}.
\bptok{imsref}%
\end{barticle}
\endbibitem

\bibitem[\protect\citeauthoryear{Von~Storch and Zwiers}{2002}]{VonStorch-2002}
\begin{bbook}[author]
\bauthor{\bsnm{Von~Storch},~\bfnm{H.}\binits{H.}} \AND
\bauthor{\bsnm{Zwiers},~\bfnm{F.~W.}\binits{F.~W.}}
(\byear{2002}).
\btitle{{Statistical Analysis in Climate Research}}.
\bpublisher{Cambridge Univ. Press}, \blocation{Cambridge}.
\bptok{imsref}%
\end{bbook}
\endbibitem

\bibitem[\protect\citeauthoryear{Weber}{1991}]{Weber-1991-EnergyAndComb}
\begin{barticle}[author]
\bauthor{\bsnm{Weber},~\bfnm{R.}\binits{R.}}
(\byear{1991}).
\btitle{{Modelling fire spread through fuel beds}}.
\bjournal{Process in Energy and Combustion Science}
\bvolume{17}
\bpages{67--82}.
\bptok{imsref}%
\end{barticle}
\endbibitem

\bibitem[\protect\citeauthoryear{Weber}{2001}]{Weber-2001}
\begin{bbook}[author]
\bauthor{\bsnm{Weber},~\bfnm{R.}\binits{R.}}
(\byear{2001}).
\btitle{{Forest Fires: Behaviour and Ecological Effects}}.
\bpublisher{Academic Press}, \blocation{San Diego, CA}.
\bptok{imsref}%
\end{bbook}
\endbibitem

\bibitem[\protect\citeauthoryear{Westerling
et~al.}{2006}]{WesterlingEtAl-2006-Science}
\begin{barticle}[author]
\bauthor{\bsnm{Westerling},~\bfnm{A.~L.}\binits{A.~L.}},
\bauthor{\bsnm{Hidalgo},~\bfnm{H.~G.}\binits{H.~G.}},
\bauthor{\bsnm{Cayan},~\bfnm{D.~R.}\binits{D.~R.}} \AND
\bauthor{\bsnm{Swetnam},~\bfnm{T.~W.}\binits{T.~W.}}
(\byear{2006}).
\btitle{{Warming and earlier spring increase western US forest wildfire
activity}}.
\bjournal{Science}
\bvolume{313}
\bpages{940--943}.
\bptok{imsref}%
\end{barticle}
\endbibitem

\bibitem[\protect\citeauthoryear{Wiitala}{1999}]{Wiitala-1999-proceedings}
\begin{bmisc}[author]
\bauthor{\bsnm{Wiitala},~\bfnm{M.~R.}\binits{M.~R.}}
(\byear{1999}).
\bhowpublished{Assessing the risk of cumulative burned acreage using the
Poisson probability model. In \textit{Proceedings of the Symp. on Fire
Economics, Planning and Policy: Bottom Lines} 51--58. USDA For. Serv.}
\bptok{imsref}%
\end{bmisc}
\endbibitem

\bibitem[\protect\citeauthoryear{Wiitala and
Carlton}{1994}]{WiitalaEtAl-1994-proceedings}
\begin{bmisc}[author]
\bauthor{\bsnm{Wiitala},~\bfnm{M.~R.}\binits{M.~R.}} \AND
\bauthor{\bsnm{Carlton},~\bfnm{D.~W.}\binits{D.~W.}}
(\byear{1994}).
\bhowpublished{Assessing long-term fire movement risk in wilderness fire
management. In \textit{12th Conf. on Fire and Forest Meteorology} 187--194.
Jekyll Island, GA.}
\bptok{imsref}%
\end{bmisc}
\endbibitem

\bibitem[\protect\citeauthoryear{Wood}{2006}]{wood-IntroToGAMs}
\begin{bbook}[mr]
\bauthor{\bsnm{Wood},~\bfnm{Simon~N.}\binits{S.~N.}}
(\byear{2006}).
\btitle{Generalized Additive Models: An Introduction with \mbox{\texttt{R}}}.
\bpublisher{Chapman \& Hall/CRC}, \blocation{Boca Raton, FL}.
\bid{mr={2206355}}
\bptok{imsref}%
\end{bbook}
\endbibitem

\bibitem[\protect\citeauthoryear{Woolford
et~al.}{2009}]{WoolfordEtAl-2009-Geomatica}
\begin{barticle}[author]
\bauthor{\bsnm{Woolford},~\bfnm{D.~G.}\binits{D.~G.}},
\bauthor{\bsnm{Braun},~\bfnm{W.~J.}\binits{W.~J.}},
\bauthor{\bsnm{Dean},~\bfnm{C.~B.}\binits{C.~B.}} \AND
\bauthor{\bsnm{Martell},~\bfnm{D.~L.}\binits{D.~L.}}
(\byear{2009}).
\btitle{{Site-specific seasonal baselines for forest fire risk in Ontario}}.
\bjournal{Geomatica}
\bvolume{63}
\bpages{356--364}.
\bptok{imsref}%
\end{barticle}
\endbibitem

\bibitem[\protect\citeauthoryear{Woolford
et~al.}{2010}]{woolford-2010-Environmetrics}
\begin{barticle}[mr]
\bauthor{\bsnm{Woolford},~\bfnm{Douglas~G.}\binits{D.~G.}},
\bauthor{\bsnm{Cao},~\bfnm{Jiguo}\binits{J.}},
\bauthor{\bsnm{Dean},~\bfnm{Charmaine~B.}\binits{C.~B.}} \AND
\bauthor{\bsnm{Martell},~\bfnm{David~L.}\binits{D.~L.}}
(\byear{2010}).
\btitle{Characterizing temporal changes in forest fire ignitions: Looking for
climate change signals in a region of the {C}anadian boreal forest}.
\bjournal{Environmetrics}
\bvolume{21}
\bpages{789--800}.
\bid{doi={10.1002/env.1067}, issn={1180-4009}, mr={2838446}}
\bptok{imsref}%
\end{barticle}
\endbibitem

\bibitem[\protect\citeauthoryear{Woolford
et~al.}{2011}]{WoolfordEtAl-2011-EnvStats}
\begin{barticle}[author]
\bauthor{\bsnm{Woolford},~\bfnm{D.~G.}\binits{D.~G.}},
\bauthor{\bsnm{Bellhouse},~\bfnm{D.~R.}\binits{D.~R.}},
\bauthor{\bsnm{Braun},~\bfnm{W.~J.}\binits{W.~J.}},
\bauthor{\bsnm{Dean},~\bfnm{C.~B.}\binits{C.~B.}},
\bauthor{\bsnm{Martell},~\bfnm{D.~L.}\binits{D.~L.}} \AND
\bauthor{\bsnm{Sun},~\bfnm{J.}\binits{J.}}
(\byear{2011}).
\btitle{{A~spatio-temporal model for people-caused forest fire occurrence in
the Romeo Malette Forest}}.
\bjournal{Journal of Environmental Statistics}
\bvolume{2}
\bpages{2--16}.
\bptok{imsref}%
\end{barticle}
\endbibitem

\bibitem[\protect\citeauthoryear{Woolford
et~al.}{2013}]{woolfordEtAl-2013-Submitted}
\begin{bmisc}[author]
\bauthor{\bsnm{Woolford},~\bfnm{D.~G.}\binits{D.~G.}},
\bauthor{\bsnm{Dean},~\bfnm{C.~B.}\binits{C.~B.}},
\bauthor{\bsnm{Martell},~\bfnm{D.~L.}\binits{D.~L.}},
\bauthor{\bsnm{Cao},~\bfnm{J.}\binits{J.}} \AND
\bauthor{\bsnm{Wotton},~\bfnm{B.~M.}\binits{B.~M.}}
(\byear{2013}).
\bhowpublished{Lightning-caused forest fire risk in Northwestern Ontario,
Canada is increasing and associated with anomalies in fire-weather.
Unpublished manuscript.}
\bptok{imsref}%
\end{bmisc}
\endbibitem

\bibitem[\protect\citeauthoryear{Wotton}{2009}]{Wotton-2009-EnvAndEcoStats}
\begin{barticle}[mr]
\bauthor{\bsnm{Wotton},~\bfnm{B.~Mike}\binits{B.~M.}}
(\byear{2009}).
\btitle{Interpreting and using outputs from the {C}anadian forest fire danger
rating system in research applications}.
\bjournal{Environ. Ecol. Stat.}
\bvolume{16}
\bpages{107--131}.
\bid{doi={10.1007/s10651-007-0084-2}, issn={1352-8505}, mr={2668729}}
\bptok{imsref}%
\end{barticle}
\endbibitem

\bibitem[\protect\citeauthoryear{Wotton and
Martell}{2005}]{WottonEtAl-2005-CJFR}
\begin{barticle}[author]
\bauthor{\bsnm{Wotton},~\bfnm{B.~M.}\binits{B.~M.}} \AND
\bauthor{\bsnm{Martell},~\bfnm{D.~L.}\binits{D.~L.}}
(\byear{2005}).
\btitle{{A lightning fire occurrence model for Ontario}}.
\bjournal{Canadian Journal of Forest Research}
\bvolume{35}
\bpages{1389--1401}.
\bptok{imsref}%
\end{barticle}
\endbibitem

\bibitem[\protect\citeauthoryear{Xu and
Schoenberg}{2011}]{XuEtAl-2011-AnnalsAppStats}
\begin{barticle}[mr]
\bauthor{\bsnm{Xu},~\bfnm{Haiyong}\binits{H.}} \AND
\bauthor{\bsnm{Schoenberg},~\bfnm{Frederic~Paik}\binits{F.~P.}}
(\byear{2011}).
\btitle{Point process modeling of wildfire hazard in {L}os {A}ngeles {C}ounty,
{C}alifornia}.
\bjournal{Ann. Appl. Stat.}
\bvolume{5}
\bpages{684--704}.
\bid{doi={10.1214/10-AOAS401}, issn={1932-6157}, mr={2840171}}
\bptok{imsref}%
\end{barticle}
\endbibitem

\bibitem[\protect\citeauthoryear{Zinck and
Grimm}{2009}]{ZinckEtAl-2009-AmericanNaturalist}
\begin{barticle}[pbm]
\bauthor{\bsnm{Zinck},~\bfnm{Richard~D.}\binits{R.~D.}} \AND
\bauthor{\bsnm{Grimm},~\bfnm{Volker}\binits{V.}}
(\byear{2009}).
\btitle{Unifying wildfire models from ecology and statistical physics}.
\bjournal{Am. Nat.}
\bvolume{174}
\bpages{E170--E185}.
\bid{doi={10.1086/605959}, issn={1537-5323}, pmid={19799499}}
\bptok{imsref}%
\end{barticle}
\endbibitem

\end{thebibliography}
\end{document}